\documentclass[final,5p,times,twocolumn]{elsarticle}

\usepackage[T1]{fontenc}
\usepackage{array} 
\usepackage[table,dvipsnames,svgnames,x11names]{xcolor} 
\usepackage{graphicx} 
\usepackage[english]{babel}
\usepackage{hhline} 
\newcommand*\rot{\rotatebox{90}} 
\definecolor{version}{HTML}{DCDCDC}
\definecolor{background}{HTML}{FFFFFF}
\definecolor{C1}{HTML}{FFFFFF}
\definecolor{C2}{HTML}{A0A0A0}
\definecolor{C3}{HTML}{303030}
\usepackage{floatrow}
\newfloatcommand{capbtabbox}{table}[][\FBwidth]

\usepackage{geometry}
\usepackage[many]{tcolorbox}
\usepackage{setspace}
\usepackage{multicol}
\definecolor{main}{HTML}{5989cf}
\definecolor{sub}{HTML}{cde4ff}

\newtcolorbox{boxResults}{
    colback = sub, 
    colframe = main, 
    boxrule = 0pt, 
    leftrule = 4pt 
}

\usepackage{amssymb}
\usepackage{natbib}
\usepackage{nameref}
\usepackage{enumitem,hyperref}
\usepackage{graphicx}
\graphicspath{ {./images/} }
\usepackage[colorinlistoftodos]{todonotes}
\usepackage{ctable}
\usepackage{lscape}
\usepackage{outlines}
\usepackage{comment}
\usepackage{hyperref}
\usepackage{pgfplots}
\usepackage{pgf-pie}
\definecolor{MyBlue}{rgb}{0.25, 0.5, 0.75}
\colorlet{NextBlue}{MyBlue!20}
\colorlet{SecondBlue}{MyBlue!40}
\pgfplotsset{compat=1.9}

\usepackage{tabularx}

\usepackage{makecell}
\usepackage{array}

\usepackage{float,caption}
\restylefloat{table}

\usepackage[outline]{contour}
\contourlength{1pt}

\usepackage[markup=underlined]{changes}
\definechangesauthor[color=CarnationPink]{PAE}
\definechangesauthor[color=NavyBlue]{OKO}
\setcommentmarkup{\todo[color={authorcolor!20},size=\scriptsize]{#3: #1}}

\usepackage{array}
\newcolumntype{P}[1]{>{\centering\arraybackslash}p{#1}}

\usepackage{contour}

\usepackage{pst-text}
\DeclareFixedFont{\RM}{T1}{ptm}{b}{n}{2cm}

\journal{}

\begin{document}

\begin{frontmatter}

\title{Systematic Mapping Study on Requirements Engineering for Regulatory Compliance of Software Systems}

\author[inst1,inst2]{Oleksandr Kosenkov$^*$}
\author[inst1,inst2]{Parisa Elahidoost$^*$}
\author[inst1,inst2]{Tony Gorschek}
\author[inst1,inst4]{Jannik Fischbach}
\author[inst1,inst2]{Daniel Mendez}
\author[inst2]{Michael Unterkalmsteiner}
\author[inst2]{Davide Fucci}
\author[inst3]{Rahul Mohanani}

\affiliation[inst1]
            {organization={fortiss GmbH},
            addressline={Guerickestraße 25}, 
            city={Munich},
            postcode={80805}, 
            country={Germany}}
            
\affiliation[inst2]
            {organization={Software Engineering Research Lab SERL, Blekinge Institute of Technology},
            addressline={Valhallavägen 1},
            city={Karlskrona},
            postcode={371 79},
            country={Sweden}}

\affiliation[inst3]
            {organization={University of Jyväskylä},
            addressline={Seminaarinkatu 15},
            city={Jyväskylä},
            postcode={40014},
            country={Finland}}

\affiliation[inst4]
            {organization={Netlight Consulting GmbH},
            addressline={Prannerstraße 4},
            city={Munich},
            postcode={80333},
            country={Germany}}

\begin{abstract}

\textit{Context:} As the diversity and complexity of regulations affecting Software-Intensive Products and Services (SIPS) is increasing, software engineers need to address the growing regulatory scrutiny. We argue that, as with any other non-negotiable requirements, SIPS compliance should be addressed early in SIPS engineering---i.e., during requirements engineering (RE). 

\textit{Objectives:} In the conditions of the expanding regulatory landscape, existing research offers scattered insights into regulatory compliance of SIPS. This study addresses the pressing need for a structured overview of the state of the art in software RE and its contribution to regulatory compliance of SIPS. 

\textit{Method:} We conducted a systematic mapping study to provide an overview of the current state of research regarding challenges, principles, and practices for regulatory compliance of SIPS related to RE. We focused on the role of RE and its contribution to other SIPS lifecycle process areas. We retrieved 6914 studies published from 2017 (January 1) until 2023 (December 31) from four academic databases, which we filtered down to 280 relevant primary studies. 

\textit{Results:} We identified and categorized the RE-related challenges in regulatory compliance of SIPS and their potential connection to six types of principles and practices addressing challenges. We found that about 13.6\% of the primary studies considered the involvement of both software engineers and legal experts in developing principles and practices. About 20.7\% of primary studies considered RE in connection to other process areas. Most primary studies focused on a few popular regulation fields (privacy, quality) and application domains (healthcare, software development, avionics). Our results suggest that there can be differences in terms of challenges and involvement of stakeholders across different fields of regulation.

\textit{Conclusion:} Our findings highlight the need for an in-depth investigation of stakeholders' roles, relationships between process areas, and specific challenges for distinct regulatory fields to guide research and practice.
\end{abstract}


\end{frontmatter}

\section{Introduction}
\renewcommand*{\thefootnote}{\fnsymbol{footnote}}
\footnotetext[1]{These authors contributed equally to this work}
\renewcommand*{\thefootnote}{\arabic{footnote}}
\label{sec:Intro}
\paragraph*{Motivation} Software engineers are facing an increasing amount of regulations that affect software-intensive products and services (SIPS) either directly by regulating technology (e.g., artificial intelligence) or indirectly by regulating processes along which SIPS are developed and/or in which they operate (e.g., personal data processing). Analysts predict that further regulatory scrutiny of the Information Technology (IT) sector will continue to intensify for various issues, ranging from privacy to new advanced technologies~\cite{hays2019shaping}. Software engineers must respond to multiple regulations that are often concurrently applicable. These regulations can also address different issues (e.g., privacy and security) and different aspects of software systems (e.g., properties of software systems and user behavior~\cite{breaux2008analyzing}) with structured principles and practices. A well-structured approach to implementing and adhering to regulations in software engineering is essential to achieve regulatory compliance and provide evidence of that compliance (i.e., the reproducible implementation of demands contained in the regulations). Failure to implement and demonstrate compliance can result in sanctions ranging from large fines to removing non-compliant SIPS from the market.

Existing scientific studies already provide a significant body of knowledge about the regulatory compliance of SIPS. Due to the steady growth of regulatory demands, SE research and practice would benefit from a more systematic understanding of regulatory compliance and its multiple facets. In particular, such a systematic understanding is essential, especially in (software) requirements engineering (RE), which largely contributes to the compliance of SIPS by translating regulations into manageable (and testable) requirements~\cite{kempe2021regulatory, kempe2021perspectives}. Hence, we argue that RE practices should be systematically guided to contribute to implementing regulatory compliance and raise awareness of specific engineering challenges it needs to address. The connection of RE to other software development life cycle (SDLC) process areas is important~\cite{kempe2021regulatory} to seamlessly and consistently address the implementation of regulations throughout the entire SDLC~\cite{hamou2015regulatory, leite2021impact}. Existing literature focuses on multiple RE principles and practices; however, there is no clear and commonly accepted understanding of the challenges they address, their contribution to subsequent SDLC process areas, and the various stakeholders' involvement and roles. A synthesis of the data available in primary studies on these matters would contribute significantly to systematizing RE for regulatory compliance.

\paragraph*{Objective} Our objective is to structure and synthesize the publication via a secondary study focusing on challenges, principles, and practices in regulatory compliance of SIPS. Our research aims to identify \textbf{the state of reported evidence in RE for regulatory compliance}.
We are mainly focused on the RE-related aspects of regulatory compliance of SIPS related to (1) challenges to the regulatory compliance of software systems, (2) principles and practices, and (3) affected SDLC process areas.
Furthermore, we analyze the involvement of stakeholders in the development of principles and practices. \added{We also extract and analyze data about different regulatory fields and application domains to account for the differences in implementing compliance with different regulations in different} application domains.

\paragraph*{Contribution} 

Our study makes the following core contributions:
\begin{itemize}
    \item categorization of RE-related challenges to regulatory compliance of SIPS into 14 categories;
    \item identification of six main types of principles and practices and their mapping to challenges to the implementation of regulatory compliance;
    \item identification of five stakeholder types that are reported to be involved in the development, application, and validation of principles and practices and their mapping with challenges and fields of regulation;
    \item identification of SDLC process areas and RE processes that are considered in the primary studies we identified;
    \item identification of domains of application, fields of regulation, and regulations reported in primary studies.
\end{itemize}

The contribution of our research is complementary to multiple previous secondary studies that have considered regulatory compliance from different, more isolated perspectives. Some of the secondary studies were mainly focused on the compliance of SE processes~\cite{moyon2020security,castellanos2022compliance}, the application of a single principle or practice such as goal-oriented modeling \cite{fame,goal}, particular compliance tasks such as compliance management~\cite{syed2010emerging}, or compliance to one specific regulation (e.g., GDPR~\cite{9420457}). Other studies have considered regulatory compliance challenges~\cite{syed2010emerging} in general without having the entire software development life-cycle in scope, and those studies considering the regulatory compliance implementation throughout the SDLC did not, in turn, analyze challenges, principles, and practices along the main process areas~\cite{kempe2021regulatory}. Our study aims to close those gaps and provide a synthesized view of the broad publication landscape to the community of researchers and practitioners.

\paragraph*{Outline} 
The rest of the paper is structured as follows---in Section~\ref{sec:background}, we synthesize the theoretical background based on computer science and legal literature. In Section~\ref{sec:related}, we provide an overview of related secondary studies and describe how our study complements existing work. In Section~\ref{sec:SD}, we present the design and the main stages of the systematic mapping study. Section~\ref{sec:results} introduces and discusses the results, structured along our research questions. We analyze the potential threats to the validity of our study in Section~\ref{sec:threats} before concluding, in Section~\ref{sec:conclusions}, our work and suggesting potential directions for future work.

\paragraph*{Data availability}
We publish our dataset of the data extracted from the 280 primary studies we have identified for this study (available on \href{https://zenodo.org/records/13999201}{Zenodo} (DOI: 10.5281/zenodo.13999201) or on the website\footnote{\href{https://regulatory-re.com/sms}{https://regulatory-re.com/sms}}). The dataset includes the following data: study metadata (\added{title, venue, publication year, authors, authors' affiliation, abstract}), challenges to regulatory compliance (direct excerpts from studies), categories of challenges to compliance, principles and practices (direct excerpts from text), categories of principles and practices, types of automation of principles and practices, involved stakeholders (direct excerpts from studies), categories of involved stakeholders, phase of principle and practice life cycle for which involvement of stakeholders was considered, SDLC process areas covered by the study, regulations considered in the study, fields of regulations that were considered, domains of application that were considered and assessment of rigor and relevance.

\section{Fundamentals and Related Work}\label{sec:fundamentals}
Regulatory RE is an interdisciplinary research area requiring the synthesis of legal and SE perspectives. Therefore, we first introduce the basic terms used throughout our study (from the perspective of SE) before reviewing work directly related to our own.

\subsection{Background and Terminology used in this Manuscript}\label{sec:background}
This section provides an overview of the terms and concepts used in this study.

\subsubsection{Regulatory Compliance: Terms and Concepts}\label{sec:terms}
Nowadays, organizations operate in a complex business landscape with multiple obligatory norms of different origins. This makes \textit{compliance}---a state of verifiable adherence to all obligatory pre-defined legally binding norms~\cite{lin2016compliance,engiel2017tool,zdun2012guest,orozco2019systems}---one of the priorities for organizations. Organizations pay special attention to \textit{regulatory compliance}---the state of verifiable adherence to public, general, obligatory norms specified in regulations~\cite{goal, lin}. In this research, we define \textit{regulation} as any official document that is a source of public, general, obligatory norms issued by regulators~\cite{brownsword2009law, koop2017regulation}. We also consider standards to be a type of regulation. While standards are not legally binding, many regulators can use them as a reference point to amend regulations or enforce standards as a source of legally binding requirements for regulatory compliance~\cite{laporte2018software} (e.g., GDPR makes references to approved certification mechanisms as an element to demonstrate compliance to GDPR norms). Moreover, standards are recognized as one of the compliance concerns in SE research~\cite{emmerich1999managing}. 

In this study, we consider \textit{regulators} as any public entities (and other entities empowered by regulators) that have a special authority to issue, monitor, and enforce regulations in the name of public interests~\cite{koop2017regulation, levi2011regulation, oecd2016governance, brownsword2009law}. Regulations are enacted to address a specific \textit{field of regulation} (like the concept of ``field of law'')---a group of social relations that are addressed by the regulation (e.g., GDPR belongs to the personal data protection field of regulation).

\subsubsection{Regulatory Compliance of SIPS: Terms and Concepts}
Based on IEEE 1471-2000~\cite{875998}, we define \textit{software-intensive products and services (SIPS)} as any products and services in which software components contribute to the design, construction, deployment, and evolution of the system as a whole or make an essential contribution to the added value of products and services.

Nowadays, SIPS are an essential element of organizations~\cite{winter2006essential} that should comply with regulations applicable to organizations operating them. Moreover, regulators increasingly enact new regulations specifically addressing SIPS. This creates a challenge of \textit{regulatory compliance of SIPS}. We define regulatory compliance of SIPS as the state of verifiable conformance of functional and non-functional properties of SIPS to criteria laid out in regulations about (1) SIPS as a system, solution, or any other type of a tangible or intangible good, (2) SIPS development process, (3) SIPS stakeholders (end-users, operators, etc.), (4) SIPS operational environment~\cite{zdun2012guest,engiel2017tool}. Regulatory compliance of SIPS should be considered throughout the SIPS life cycle. We identify the following \textit{process areas (PAs)} in the SIPS life cycle based on SWEBOK~\cite{SWEBOK} and ISO/IEC 12207 and deduce their potential contribution to regulatory compliance of SIPS:

\begin{itemize}
\item SIPS Requirements Engineering (RE)---systematic handling of requirements to SIPS that derive from regulations. This main process area includes the following processes: requirements elicitation, requirements analysis and modeling (REA/M), requirements specification (RES), requirements verification and validation (REVV), and requirements management (REM). Regulatory RE is an area of requirements engineering practice and research that contributes to the compliance of SIPS by processing requirements derived from regulations for SE purposes.
\item SIPS design (SD) - the process of defining the internal structure (e.g., architecture, components, interfaces) of SIPS in a way that will enable their regulatory compliance (e.g., assure verifiability, address future evolution of regulations).
\item SIPS development (SDev) refers to the detailed creation of working SIPS through coding, verification, unit testing, integration testing, and debugging. SIPS development process area \added{is} related to implementing requirements derived from regulations.
\item SIPS quality assurance (SQA) consists of the dynamic verification that SIPS provide expected behaviors on a set of test cases addressing the regulators' perspective on compliance by verifying the implementation of all requirements originating from regulations (e.g., \cite{corriveau2014requirements}) and also by using tools and methods applied by regulators for compliance verification.
\item In our study, we define SIPS deployment \added{(SDep)} as a set of activities directed towards generating executable and testable SIPS components, combining related components with a single deployable artefact and putting SIPS into operation~\cite{iso32675}. With the advancement of DevOps and automation practices, it becomes essential to \added{assure} regulatory compliance of the SIPS deployment process and regulatory compliance of the deployed SIPS.
\item SIPS maintenance (SM) is the totality of activities required to provide cost-effective support to SIPS \added{in operation} to ensure that SIPS remains compliant throughout the software and regulatory evolution.
\end{itemize}

\textit{Challenges to the regulatory compliance of SIPS} are general types of SE issues that emerge in achieving the regulatory compliance of SIPS.

\textit{Principles and practices for regulatory compliance of SIPS (PPs)} are any means employed to implement regulatory compliance of SIPS and to tackle the related challenges. These include but are not limited to SE methods, tools, frameworks, solutions, and models.
For this research, we consider that the life cycle of PPs comprises three core stages: 1) development, 2) application, and 3) validation. In each of these stages, specific types of stakeholders can be involved. For example, legal experts can be involved in developing PPs but not in their application and validation.
PPs can be applied in a particular \textit{domain of application} - operational domain in which SIPS are engineered, and PPs for the regulatory compliance of SIPS are applied.

\begin{table*}
\small
\resizebox{\textwidth}{!}{%
\centering
    \begin{tabular}{p{5cm}P{1.2cm}P{1.2cm}P{1.2cm}P{1.2cm}P{1.2cm}P{1.2cm}P{1.2cm}P{1.2cm}P{1.2cm}P{2cm}l} \toprule  
    \thead{Study scope} & \thead{\cite{negri2024understanding}} & \thead{\cite{mubarkoot2023software}} & \thead{\cite{castellanos2022compliance}} & \thead{\cite{kempe2021regulatory}} & \thead{\cite{new}} & \thead{\cite{moyon2020security}} & \thead{\cite{goal}}  & \thead{\cite{Hashmi2017AreWD}} & \thead{Our\\ Study} \\\midrule
    \emph{Year} & 2024 & 2023 & 2022 & 2021 & 2020 & 2020 & 2019 & 2018 & 2024 \\\midrule
    \emph{Aspect of compliance} \\
    • Challenges                    & - & X & X & X & - & - & X & X & X  \\
    • Principles \& Practices       & X & X & X & - & X & X & X & X & X \\
    — Goal/nongoal modelling      & - & - & - & - & - & - & X & - & - \\
    — Automated                   & - & - & X & - & - & - & - & - & - \\
    • Stakeholders involvement      & - & X & - & - & - & - & X & - & X \\
    • Different PAs Considered      & - & - & - & X & - & X & - & - & X \\
    • Domains of application        & - & X & X & - & X & X & X & - & X \\
    • Fields of regulation          & - & X & X & - & - & X & X & - & X \\
    — Security                     & - & - & - & - & - & X & - & - & - \\\midrule
    \emph{Type of compliance} \\
    • Regulatory                    & X & X & X & X & X & X & X & X & X  \\
    • Non-obligatory                & - & X & - & - & - & - & - & X & - \\\midrule
    \emph{Compliance scope} \\
    • Business process              & - & - & - & - & X & - & X & X & - \\
    • Engineering process           & - & X & X & - & X & - & - & X & - \\
    • SIPS                          & X & X & - & X & - & - & X & - & X \\
    \bottomrule
    \end{tabular}}
\caption{Overview of the most recent related secondary studies showing research areas (e.g., regulatory RE principles and practice) and topics (goal and non-goal-modeling principles and practices) covered by each secondary study}
\label{tab:relatedWork}
\end{table*}

\subsection{Related work}\label{sec:related}
Although no systematic mapping study exists on RE for regulatory compliance of SIPS, multiple secondary studies can be found in \textit{three} different research tracks \added{related to this topic}.

\subsubsection{SIPS Users Compliance} 
Compliance of SIPS users is mainly concerned with questions related to compliance to norms imposed on (1) SIPS users directly, (2) business processes of SIPS users (as one of the essential elements of the SIPS operational environment), or (3) the operational environment of SIPS. Secondary studies in this category provide insights into methods for processing regulation but do not make an explicit connection between such methods and SE. Some studies in this category (e.g., Ghanavati et al.~\cite{fame}) consider methods and tools for business process compliance as applicable for RE; still, they do not discuss an explicit connection between business process compliance and implementation of compliance of SIPS.

Mustapha et al.~\cite{new} reported on several research directions towards some of the challenges faced in managing compliance of business processes. The systematic review identified BPCM (Business Process Compliance Requirements Management) approaches, the most addressed constraint types, environments where business processes have been more implemented, and a comparative analysis of several works that have been carried out in this field of research. Mustapha et al.~\cite{new} considered two classes of the application domain: traditional and cloud environments. Still, such categorization does not allow a sufficiently granular analysis of differences between implementing compliance in different domains.

Another relevant study was conducted by Hashmi et al.~\cite{Hashmi2017AreWD}. Their survey aims to understand the current state of affairs in the compliance of business processes, summarise the weaknesses of existing techniques for business process compliance, and identify areas for further work.

Fellmann et al.~\cite{Fellmann} provided an overview of the focus and distribution of current compliance approaches for business processes by conducting a literature review.

Becker et al.~\cite{Becker} conducted a literature review to analyze existing business process compliance and to check approaches according to (1) their applicability to arbitrary modeling techniques and (2) their ability to address a preferably wide range of possible compliance rules. 

Shamsaei et al.~\cite{goalInd} suggest goal-oriented compliance management using Key Performance Indicators (KPIs) to measure the compliance level of organizations as an area that can be further developed. They undertook a systematic literature review to investigate their hypothesis, querying four major search engines and performing manual searches in related workshops and citations.

Ghanavati et al.~\cite{fame} reported on a systematic literature review focusing on goal-oriented legal compliance of business processes. Eighty-eight studies were selected out of nearly 800 unique studies extracted from five search engines, with manual additions from the Requirements Engineering Journal and four relevant conferences.

\subsubsection{SIPS Developers Compliance}
Secondary studies in this category mainly focus on implementing regulatory compliance in SIPS development. These studies are usually related to regulations specifically imposing regulatory demands on the SIPS development process (e.g., safety and security standards); they cover only a part of our approach to compliance of SIPS.

Ardila et al.~\cite{castellanos2022compliance} conducted a systematic literature review on compliance checking of SE processes published in 2022. They identified 41 studies and characterized the methods for automatic compliance-checking of software processes, including used techniques, potential impacts, and challenges. The authors considered challenges to compliance of SIPS development processes but identified only five categories of abstract challenges (``the use of software process modeling languages'', ``language suitability for addressing normative requirements'', ``towards a generic and domain-agnostic method'', ``increase the level of automation and tool support'', ``application in practice'') that were not directly mapped to the information extracted from publications.

Moyon et al.~\cite{moyon2020security} report on a systematic mapping study where they addressed security compliance in agile software development. They described the maturity of the field and domains where security-compliant agile SE was investigated. The authors considered PPs for regulatory compliance related to agile software development.

Naira et al.~\cite{exslr} developed a taxonomy that classifies the information and artifacts considered as evidence for safety. They reviewed 218 studies and the existing safety evidence structuring and assessment techniques and further studied the relevant challenges that have been the target of investigation in the academic literature.

\subsubsection{SIPS Compliance}
Few previous secondary studies have tried to address SIPS compliance with a broader research perspective that would cover both the compliance perspective of SIPS developers, SIPS users, and SIPS as a system. This way, studies having such scope could potentially cover not only compliance in SIPS development processes but also compliance of SIPS as a standalone product or service and SIPS in use by a particular type of organization. Such works were characterized by the efforts to consider all the tasks throughout the legal or regulatory compliance process~\cite{goal}, focusing on SIPS compliance and users of software~\cite{mubarkoot2021software}.

Studies in this research track usually have a specific scope, such as applying a particular type of modeling PPs (e.g., goal-oriented methods~\cite{goal}).

Further, some studies applying a broader approach often lack a clear scope of SIPS compliance and cover business processes or other ``types'' of compliance along with SIPS compliance. As a result, only a few works provide data sufficient to analyze the implications of regulatory compliance on SE.

The secondary study with the scope closest to ours is Mubarkoot et al.~\cite{mubarkoot2023software}. In this study, the authors have conducted a systematic literature review to investigate software compliance requirements, factors, policies, and the challenges they address.
The study considered human- and technology-related challenges to compliance but did not provide their clear and granular categorization.
The study primarily focused on policies as a means of addressing challenges. Also, the study suggested classifying policies based on the type of compliance challenges they address. This classification includes two large classes, which are as follows:  technological challenges (software certification, regulation-oriented architecture, model-driven development, applying most restrictive laws, outsourcing, runtime security auditing) and human challenges (SETA, promote organizational and social bounds, incorporate appropriate responses, reward and punishment, investigate workarounds, internal control, and auditing, establish codes of ethics). Still, the granularity of classification suggested in the publication does not allow further analysis useful for implementing regulatory compliance in SE.
This study identified the following types of users considered in publications: architects, auditors, developers, domain experts, end users, Information System professionals, legal experts, managers, requirements engineers, safety engineers, and service providers. Still, reported data does not allow the identification of concrete trends regarding stakeholders' involvement.

Mubarkoot \& Altmann~\cite{mubarkoot2021software} investigated frameworks used for managing the compliance of software and software services and their applications across different industries in their systematic literature review published in 2021. The study was developed based on industry classification of the frameworks' specific needs, business requirements, and the context of compliance in 63 identified studies.

Kempe et al.~\cite{kempe2021regulatory} published a systematic literature review in 2021 and analyzed 20 studies to research regulatory and security standard requirements addressed throughout the SDLC.

Negri‑Ribalta et al.~\cite{negri2024understanding} conducted a systematic mapping study to discover the trends of PPs proposed in requirement engineering to achieve GDPR compliance. The paper also identified the different stages of RE the primary studies focused on, the concrete GDPR norms studies covered, and if the authors of the primary studies belonged to different disciplines. The study found that most studies approached GDPR compliance as a whole without focusing on specific ``elements to achieve compliance''. Also, this study claims that interdisciplinary research in this area is essential as legal knowledge is required and that GDPR requirements affect a range of engineering disciplines, not only requirements engineering.

Abdullah et al.~\cite{Syed} undertook a review of the current research on compliance management topics in the Information Systems research, with the ultimate goal of carrying out a gap analysis between research-based solutions and the current needs of compliance management professionals. 

Cleven et al.~\cite{Cleven} examined the state-of-the-art compliance research in Information Systems using a comprehensive literature analysis.

Finally, Otto et al.~\cite{Otto7} surveyed research efforts over the past 50 years in modeling and using legal texts for system development. They identified the strengths and weaknesses of each approach and proposed a broad set of requirements for tool support that would aid requirements engineers and compliance auditors.

Akhigbe et al.~\cite{goal} explored how goal-oriented and non-goal-oriented modeling methods have been used for legal and regulatory compliance and identified their main claimed benefits and drawbacks based on the kind of compliance tasks they perform. Using a systematic literature mapping approach, they evaluated 103 studies. Akhigbe et al.~\cite{goal} considered compliance challenges but only in the context of the application of goal modeling (e.g., the challenge of applicability of goal models for only comparing models generated from legal texts and not the legal texts themselves). Also, the authors considered the target audience for applying goal-oriented and non-goal-oriented modeling methods for legal or regulatory compliance. However, only three categories of the target audience were identified, such as regulated parties (considered in 82\% of studies), regulators (considered in 12\% of studies), and both regulated parties and regulators (in 6\% of studies).

Considering previous related work, our research aims to complement the existing state of research in several ways. In particular, in this study, we approach compliance of SIPS:
\begin{itemize}
\item independently of a particular regulation or field of regulation (e.g., GDPR);
\item holistically by considering the contributions of RE to all the main process areas (PAs) of the SIPS life cycle;
\item as involving both SIPS users' and SIPS developers' compliance perspectives;
\item independently of a type of software system or SIPS (e.g., safety-critical systems);
\item focusing specifically on regulatory compliance, and excluding other types of compliance (e.g., compliance with contracts, internal policies);
\item with a broader approach to compliance that does not only consider the compliance of requirements but rather the compliance of the software systems in general
\end{itemize}

\section{Study Design}\label{sec:SD}
In this section, we describe the goal of this study and its research questions. Then, we describe the methodology we applied.

\subsection{Research Goal and Research Questions} \label{RQ}
This secondary review aims to comprehensively assess the current evidence regarding RE for regulatory compliance of SIPS. It focuses on identifying the challenges in this domain and explores how established PPs address those challenges. Our primary objective is to provide a detailed overview of the existing research landscape while offering valuable insights for future investigations, benefiting researchers and practitioners. To achieve these aims, we propose the following research questions (RQ):
\begin{enumerate}[label=\textbf{$RQ$\arabic*}]
    \item\label{rq1} What are the reported challenges in regulatory compliance of SIPS?
    \item\label{rq2} What is needed to address the challenges to the regulatory compliance of SIPS, and what principles and practices (PPs) are used/suggested?
    \item\label{rq3} Which stakeholders are involved in developing, applying and/or validating principles and practices for regulatory compliance of SIPS?
    \item\label{rq4} What are the main software process areas (PAs) involved in enabling the regulatory compliance of SIPS?
    \item\label{rq5} Which regulations and domains of application are the most prevalent in regulatory compliance of SIPS?
\end{enumerate}
These research questions have been conceptualized to address various critical aspects of regulatory RE in ensuring compliance with SIPS. The challenges associated with regulatory compliance in SIPS largely stem from the inherent complexity of regulations as a source of requirements. Furthermore, the development of effective PPs, alongside the active involvement of stakeholders, is closely tied to the RE process. Additionally, variations in applicable regulations and differences across application domains are particularly significant in shaping the RE process.

\subsection{Research Methodology} \label{RM}
For this study, we followed the systematic mapping study(SMS) procedure according to Kitchenham et al.~\cite{Kitchenham07guidelines}. We extracted and analyzed multiple interrelated categories of data (something similar to a systematic literature review (SLR) approach). Nevertheless, several factors characterize our study as an extensive SMS. The primary goal of our study was to embrace different RE-related aspects of SIPS regulatory compliance and their potential interrelationships. To this end, we collected and analyzed data on a high level of granularity. For example, we did not distinguish in detail how particular challenges were reported in primary studies, but we calculated the occurrence of mentions of challenges. Similarly, we did not conduct a detailed analysis of the involvement of different stakeholders. 

\begin{figure*}
    \centering
    \includegraphics[scale=0.36]{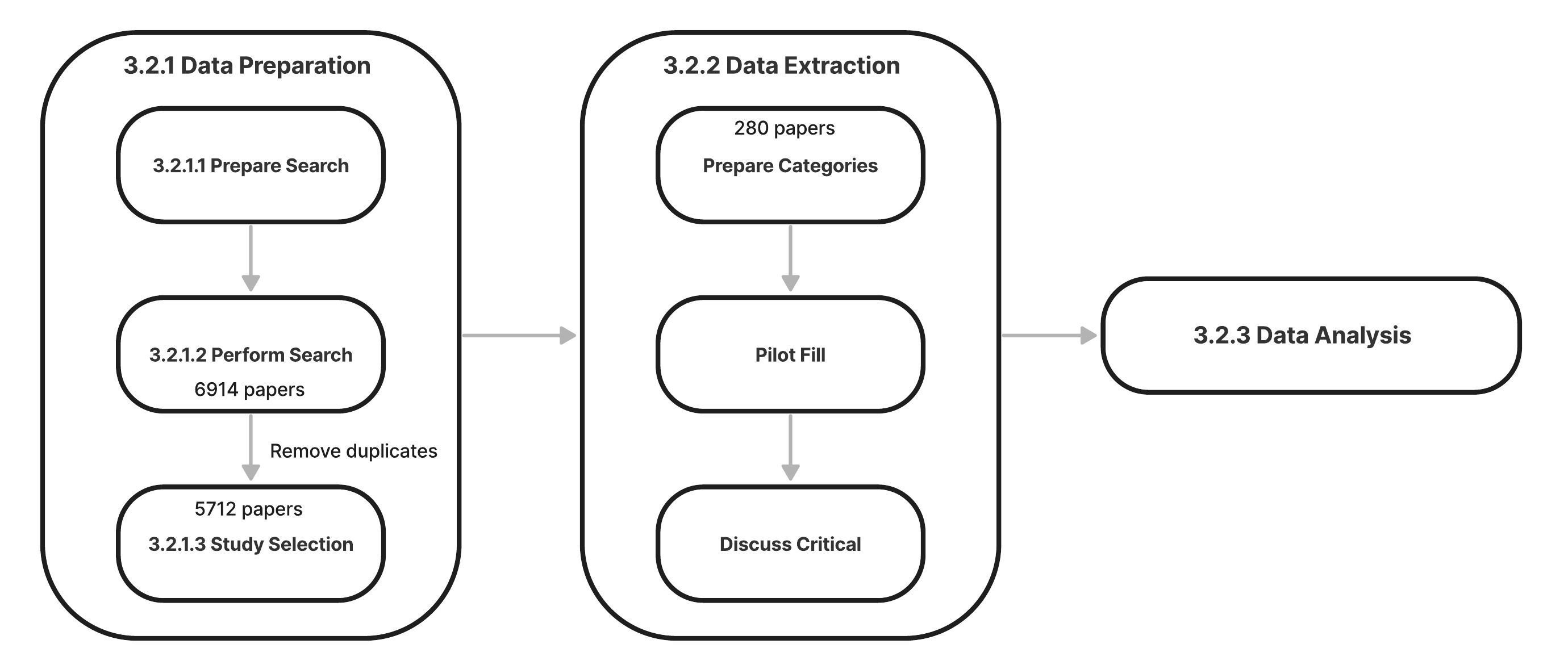}
    \caption{Study Setup}
    \label{fig:setup}
\end{figure*}

The level of analysis we selected is sufficient to identify general trends and make high-level conclusions about the relationships between the categories of extracted data. Subsequently, we encourage other researchers to undertake detailed literature reviews to further analyze and validate the trends identified in this mapping study. Figure~\ref{fig:setup} summarises the study setup, which is split into three major phases---data preparation, extraction, and analysis. 

\subsubsection{Data Preparation}\label{sec:dp}

During the data preparation phase, the first two authors were responsible for collecting the primary studies for data extraction. Initially, we developed a search string guided by the aforementioned research questions. This involved a careful process beginning with the creation of a comprehensive set of keywords, which were thoroughly discussed and refined in collaboration with all co-authors. Following this, we conducted a pilot search to test and optimize the query. We then applied the finalized search string across a predefined set of databases, systematically screening the identified studies based on their titles and abstracts (see Figure~\ref{fig:setup}). 

Based on the RQs, we also prepared a dataset containing a list of known studies relevant to our review. We searched the most important SE databases to ensure we did not miss any relevant studies concerning the regulatory compliance of SIPS. We selected these sources employing the recommendations of Dyba et al.\cite{4343750}, using ACM Digital Library \footnote{\href{https://dl.acm.org/search/advanced}{https://dl.acm.org/search/advanced}}, IEEE Xplore\footnote{\href{https://ieeexplore.ieee.org/search/advanced/command}{https://ieeexplore.ieee.org/search/advanced/command}}, and an indexing system (Scopus \footnote{\href{https://www.scopus.com/search/form.uri?display=advanced}{https://www.scopus.com/search}}). We extended their proposal to include Elsevier (Science Direct \footnote{\href{https://www.sciencedirect.com/search}{https://www.sciencedirect.com/search}}) as well~\footnote{It is among the databases like ACM and IEEE that host the major journals and conference proceedings related to SE~\cite{10.1145/3444689}}.

\paragraph{Preparation of the Search String}
As an initial step, we developed the search string following the guidelines provided by Brereton et al.~\cite{10.1016/j.jss.2006.07.009}. To ensure the search results remained relevant to our study's scope, we began by incorporating the term  ``software'' followed by ``requirement''. Next, we selected additional keywords related to regulatory compliance. After updating the search string, we conducted a pilot search to test its effectiveness. For example, one of the tried search queries was \textit{((``Requirement* engineering'' OR ``Software engineering'' OR ``Software system'') AND (``Regulatory compliance'' OR ``Legal compliance'' OR ``Regulation''))}--- returned twice as many search results from a single database during preliminary investigations.

While conducting trial searches, we observed studies focusing on regulatory RE and compliance of systems other than SIPS (e.g., hardware, embedded systems) and compliance with regulations focused on hardware. We argue that regulatory RE and compliance of SIPS are different from non-SIPS systems. For example, it is well-known that SIPS development teams actively apply agile methods, while other engineering teams (e.g., mechanical engineering teams) only rarely apply agile methodologies~\cite{kasauli2020charting}. In addition and, according to some of the studies in the domain of medical devices, regulations on medical devices created without specifically considering software as a medical device are considered to be problematic and hardly applicable to ``software as a medical device'' and software-intensive medical devices (e.g.,~\cite{granlund2020medical}). To exclude the studies that were not focused on SIPS from the search results set, we added the keyword ``software'' with a boolean operator ``AND''. We tested this search string and found out that it allows us to reduce the number of primary studies focused on non-SIPS requirements and compliance significantly while not losing studies we deemed relevant. This change of the search string also allowed \added{the removal of} a massive number of studies not focused on RE and compliance of any systems, but rather requirements applicable only to organizations, business processes, etc.

In the process of experimenting with various search strategies, we have also tried variations of the same words (e.g., “regulation”, “regulatory”) and the application of wildcards (e.g., “regulat*”). The results of such variations turned out to yield inconsistent results sets across the databases. For example, we observed that the number of results in Scopus and IEEE increased by about 200 hits each while the number of results in ACM decreased by about 200 hits. Furthermore, the search in title, abstract, and keywords with the application of wildcards was not possible for ScienceDirect. To address this issue, we have reviewed the first 100-200 search results with and without wildcards for Scopus and IEEE. Our review showed that the number of hits mainly increased due to the increase of the number of hits in other disciplines not related to SIPS engineering (e.g., regulated medicines consumption) and we were not able to identify any primary studies relevant to our research among the reviewed search results. Taking this into account, we therefore decided to apply the search string without wildcards across all the databases. The final search query is:

\textbf{software AND requirement AND (regulation OR regulatory OR compliance OR law OR legal OR certification)}

\paragraph{Performing Search} 
We adapted the search string to each database (depending on the allowed syntax) and performed the search for authors' keywords, titles, and abstracts. Table~\ref{tab:results} shows the search query and the number of results for each database. After removing the duplicates, we collected 5712 studies, which we divided for screening among the first two authors by reading the title and abstract.

\begin{table*}
\footnotesize
\centering
	\caption{Search strings for each data source and number of results}\label{tscriptsizeab:results}
        \setlength\extrarowheight{-6pt}
	\begin{tabular}{p{2.5cm}p{10cm}p{1.5cm}l}\toprule  
		\thead[l]{Database}                   & \thead[l]{Entries} & \thead[l]{Results} \\\midrule  
		\emph{Scopus}    & TITLE-ABS-KEY ( software AND requirement AND ( regulation OR regulatory OR compliance OR law OR legal OR certification ) ) AND PUBYEAR > 2016 AND PUBYEAR < 2024         & 4295     \\
		\emph{IEEE Xplore}    & ("Abstract":software AND requirement AND (regulation OR regulatory OR compliance OR law OR legal OR certification)) OR (”Author Keywords”: software AND requirement AND (regulation OR regulatory OR compliance OR law OR legal OR certification)) OR (”Publication Title”: software AND requirement AND (regulation OR regulatory OR compliance OR law OR legal OR certification)). Filters Applied: 2017 - 2023         & 1246     \\
		\emph{ACM Digital Library} & Title:(software AND requirement AND (regulation OR regulatory OR compliance OR law OR legal OR certification)) OR Abstract:(software AND requirement AND (regulation OR regulatory OR compliance OR law OR legal OR certification)) OR Keyword:(software AND requirement AND (regulation OR regulatory OR compliance OR law OR legal OR certification)) AND [E-Publication Date: (01/01/2017 TO 12/31/2023)] & 987     \\
		\emph{Science Direct}    & Title, abstract, keywords: software AND requirement AND ( regulation OR regulatory OR compliance OR law OR legal OR certification. Filters Applied: 2017 - 2023 & 386      \\ \midrule
		Total:                &     & 6914    \\
		Without duplicates:    &    & 5712    \\
        After first assessment:       &   & 341     \\
		Primary studies selected:       &   & 280     \\		
		\bottomrule
	\end{tabular}
\end{table*}

After searching, we have also tried to apply the snowballing procedure as per the guidelines by Wohlin~\cite{wohlin2014guidelines} for further identification of primary studies. Applying the snowballing procedure for our study turned out to be challenging. Firstly, the main scope of our study was primary studies considering the contribution of RE to SIPS regulatory compliance. Many primary studies focus on other process areas such as software design or quality assurance, but still include aspects of RE to support these process areas. However, identifying such studies via snowballing without applying keyword search is impossible. For example,~\cite{ayala-rivera_grace_2018} is one of the primary studies we have identified while executing our search query and is relevant to our study. One of the studies referring to it is~\cite{portillo2019towards}. This study is relevant from the perspective of the regulatory compliance of SIPS as its main goal is to ensure regulatory compliance of log data processing with regulations such as GDPR. However, this study fully omits aspects related to RE required to achieve their goal. Therefore, the only option we deemed to be systematic and effective in identify\added{ing} the relevance of such studies is to search for the keyword ``requirement''. Next, our systematic mapping is limited to the most recent publications which limits the effectiveness of snowballing. Finally, we intentionally searched for publications belonging to different communities and disciplines to include different perspectives making it particularly challenging to account for all the communities in a starter set. While RE is the main research community of interest for our study, it is known that existing RE studies do not provide an overview of how regulatory requirements are approached throughout the SDLC~\cite{kempe2021perspectives}.

\paragraph{Study Selection}
The first two authors screened the studies based on their titles and the abstract, then scanned the entire text only when necessary to reach a judgment. In this phase of the study selection process, we used an initial set of \hyperlink{inclusion}{inclusion} and \hyperlink{exclusion}{exclusion} criteria to identify the relevant studies. \added{Specifically, we used the following inclusion criteria}:

\begin{itemize}
	\item The study must be in English.
    \item The study published in the period between January 1st, 2017 and December 31st, 2023.
	\item The context of the study is only SE - defined as ``a systematic approach to the analysis, design, assessment, implementation, test, maintenance and re-engineering of software, that is, the application of engineering to software'' (Laplante~\cite{lap}) or software systems. This also includes SE in the context of other types of systems and products (e.g., automotive, mechatronics).
	\item The study is focused or relevant to regulatory compliance.
	\item The study must be published in a peer-reviewed scientific journal, conference, symposium, or workshop.
\end{itemize}

We also decided to select studies published in a particular period. Regulation of software-intensive technologies evolved worldwide in the recent decade. And regulators have been proactively trying to respond to technological developments~\cite{digitalRegulation}. As a result, new regulations are enacted, and new regulatory approaches and frameworks are being tried out. GDPR marked the introduction of such a new framework aimed at responding to the most recent advancements in ICTs~\cite{rotenberg2013updating, muller2024european}. In recent years, regulators have drafted or enacted multiple new regulations (e.g., the Cyber Resilience Act, AI Act proposed in 2021). As these new regulations incorporate the most recent regulatory ideas, we have decided to focus on regulatory RE and compliance of SIPS with such new regulations. Therefore, our study focuses on the most recent primary studies, starting from 2017 - the year after the General Data Protection Regulation (GDPR) was enacted in April 2016 and one year before GDPR became enforceable in May 2018. We considered the following exclusion criteria during our study:

\begin{itemize}
	\item The study is not peer-reviewed.
	\item The study is a secondary study.
	\item The study is a doctoral thesis, master's thesis, or opinion report, or a book chapter.
	\item The study's reference to regulatory compliance is tangential to the overall purpose or contributions. 
	\item The study is extended in another study (e.g., a journal extension of a conference paper).
        \item The study is already mentioned in another database (duplicated Studies).
     \item The study can not be accessed. 
\end{itemize}

At the end of the first phase of the selection process, we conducted an additional synchronization involving the first two authors and the fifth author. The first two authors identified 183 ``borderline studies'' selection of which using the defined selection criteria was challenging. It was decided to analyze these studies. Both researchers checked the studies together until they reached an agreement. After this assessment round, the number of studies was reduced from 5712 to 341. To improve the selection process and as the result of synchronization between the researchers, the initial list of exclusion criteria was refined with additional exclusion subcriteria that are as follows:

\begin{itemize}
\item studies considering the regulatory compliance of other types of systems or systems containing software without specifically addressing software component compliance (e.g., radio equipment~\cite{mueck2017radio});
\item studies considering business process compliance without explicitly considering business process modeling as an RE method or other application for SIPS engineering (e.g.,~\cite{arogundade2020algorithm});
\item studies considering the regulatory compliance of IT infrastructure of organizations, without focusing specifically on the implementation of regulatory compliance at the level of a software system (e.g., development of communication protocols);
\item studies considering compliance other than to general obligatory regulations or standards (e.g., contracts~\cite{fantoni2021text});
\item regulatory compliance of software system or PPs applicable for regulatory compliance is not a primary focus of the study (e.g.,~\cite{ponsard2018helping} reported GDPR compliance as one of the challenges that small and medium enterprises encounter, but did not focus on GDPR compliance);
\item studies related to the application of software as a tool for compliance rather than an object of compliance (e.g., studies on RegTech solutions not intended for SIPS~\cite{ryan2020gdpr});
\item studies on the regulatory compliance of software development organizations without an explicit connection to regulatory compliance of SE processes or SIPS (e.g., \cite{ponsard2018helping} presented a set of concerns underlying the design of a business continuity plan for companies that have certified invoicing software);
\item studies considering regulations related to software systems, without specific focus on the implementation of compliance of SIPS (e.g., Cybersecurity Maturity Assessment Framework in compliance with NIS Directive~\cite{drivas2020nis}).
\end{itemize}

In the next phase of the selection process, each researcher applied additional (refined) exclusion criteria to studies included after the first phase (341 studies). Like the previous phase, the two primary authors scanned the abstract, and in unclear cases, the full text was examined. After this final assessment round, 280 final primary studies were identified.
For having a mutual agreement between two researchers while doing the first round of assessment and study selection based on the title and abstract, Cohen's kappa\cite{cohen} was calculated. Many studies (5712) were divided between the first two researchers, and ensuring that both had the same understanding of the analysis was important. Fifty studies were randomly selected from the databases and analyzed by the researchers separately for the inclusion or exclusion of the study. The Cohen's Kappa was 0.9404, which means 95.8 percent agreement or almost perfect agreement. Both raters agreed that the process and understanding were aligned enough to proceed with an entire run.

\begin{table}
\footnotesize
\centering
	\caption{Extraction Categories}\label{tab:category}
         \setlength\extrarowheight{-8pt}
	\begin{tabular}{p{1cm}p{4.25cm}p{1.25cm}l}\toprule  
		\thead[l]{RQ}   &  \thead[l]{Extraction Category}  &  \thead[l]{Coding Style}  \\\midrule  
            G   & Author(s) & O\\
                & Affiliation(s) & O\\
                & Venue & O\\
                & Publication Year & O\\
		RQ1 & Challenges to compliance & O \\ 
	    RQ2 & Principles and Practices Process & O \\
                & Principles and Practices Type & L \\ 
                & Automation Process & O \\ 
                & Automation Level & L \\
            RQ3 & Involved Stakeholders & O \\
                & Phase of Involvement & O \\
            RQ4 & PA & L \\
            RQ5 & Regulations Considered & O \\
                & Field of Regulation & L \\
		      & Domain of Application & L \\
             Q  & Rigor & L \\
                & Relevance & L \\\bottomrule
	\end{tabular}
 \\[1.5pt] 
  General Questions (G), Open (O), Labeled (L), Quality of Study (Q)
\end{table}

\subsubsection{Data Extraction} \label{sub:dataext}
First, a set of categories was prepared for the data extraction phase based on the research questions at hand. Then, two researchers filled the categories by reviewing each study. We tried to answer the RQs with the extracted data, and after two rounds of pilot extractions, we refined the categories. There was a discussion between the authors to reach an agreement on unclear assignments. The extraction categories, the representative RQ, and the coding styles are listed in Table~\ref{tab:category}. The ``open'' coding style was used for 10 out of 17 questions, while the ``labeled'' coding style was employed for the remaining seven questions. In the case of the ``labeled'' coding style, a predetermined set of possible labels was established before the data extraction phase. However, this list of labels was subject to evolution as new information was uncovered during the extraction process. Any alterations made to the labels necessitated the re-labeling of all previously coded studies. Moreover, to address some general findings regarding the research area, we extracted meta-data categories like ``author(s)'', ``affiliation(s)'', ``venue'' and ``publication year''. All the information referred to is available either in the primary study or on the publisher's website.

To address \textit{RQ1}, we extracted the reported challenges concerning the SIPS. The coding style for this category is open, and in the data analysis, we categorized those challenges. 

\begin{table*}
\footnotesize
\centering
	\caption{PPs Types (inspired by Offermann et al. \cite{PPtype})}\label{tab:pp}
        \setlength\extrarowheight{-10pt}
	\begin{tabular}{p{2.5cm}p{4.5cm}p{6.5cm}l}\toprule  
		\textbf{PP Type}         &  \textbf{Sub-types} &  \textbf{Structure}  \\\midrule  
            Methodology     & Guidelines, E-learning programs, Methods and Techniques, Processes and General Approaches  & A systematic framework that defines and organizes a collection of methods and principles to guide the process of creating or interacting with a system\\
		Metrics and Measurements & -- & A mathematical model that supports measuring aspects of systems or methods  \\ 
	    Model & Lexicon, Taxonomy & A structure, or behavior-related description of a system, commonly employing some formalism (e.g. UML) and possibly text, is a generally formalized system used to formulate statements representing parts of reality\\
            Pattern & Architecture  & Definition of reusable elements of design with its benefits and application context \\
            Requirements & -- & Description, Statement about System (A system or service of type X should have some property Y [because of Z]) \\
            Tool & -- &  A program or a set of programs, which is running on a computer that may be used to develop, modify, test, analyze, produce, or modify another program, its data, or its documentation\cite{Leannatool}\\\bottomrule
	\end{tabular}
\end{table*}

To address \textit{RQ2}, we extracted the ``principles and practices'' to better understand the implemented PP process. Moreover, we extracted the ``principles and practices type'', the type of PP implemented to address the mentioned challenges. Six categories were predefined for PP types (Table~\ref{tab:pp}).  Inspired by the work of Offermann et al.~\cite{PPtype}, we developed our categorization of 'Principles and Practices (PP) Types' tailored to the context of our study. Although we drew inspiration from Offermann’s artifact types, we decided to refine and adapt these categories to better align with our research's specific focus and scope. We used the term 'methodology' instead of 'method' to encapsulate a broader range of practices, including guidelines, e-learning programs, methods and techniques, processes, and general approaches. We believe 'methodology' is a more comprehensive term covering various workflows and overarching frameworks rather than just individual methods. We followed several steps: initially, we extracted the categories and statements of the authors from the primary studies, focusing on the proposed designs. The first two authors independently formulated their categories and interpretations based on the gathered information. The final step involved consolidating these categories into a unified taxonomy, which was then discussed and validated with a broader group of authors.
Each sub-type within our categories was carefully considered to reflect specific aspects of the PPs. For example, 'guidelines' refer to sets of recommendations or instructions; 'methods and techniques' are the definition of activities on how to create or interact with a system; 'processes' and 'general approaches' denote systematic series of actions or overall strategies; 'architecture' within 'pattern' refers to the design and structure of systems; and 'lexicon' or 'taxonomy' within 'model' indicates a structured vocabulary or classification system.

Lastly, in the case of any automation, we extracted the ``Automation'' and ``Automation Level''. For this purpose, we considered the definition and automation types described in Parasuraman et al.~\cite{Automation}. They have suggested four levels of automation in their research: 
\begin{enumerate}
    \item information acquisition: Automation in acquiring information applies to sensing and recording input data. 
    \item information analysis: Automation in information analysis incorporates cognitive functions, such as working memory and inferential processes.
    \item decision and action selection: The third stage, which involves decision and action selection, requires choosing from various alternatives. Automating this stage encompasses varying degrees of enhancing or substituting human decision-making with machine-based decision-making processes.
    \item action implementation: The last stage, action implementation, pertains to the actual execution of the chosen action. Automating this stage involves varying degrees of machine-based execution of the selected action, typically substituting human manual or vocal execution.
\end{enumerate}

To address \textit{RQ3}, we extracted ``involved stakeholders'' and ``phase of involvement'' to better understand if any experts were involved in the process of developing the PP, and if yes, at which phase.

To address \textit{RQ4}, we extracted for which software development main process area (PA) the PP was developed. We used a predefined label for extraction: 

\begin{itemize}
    \item \textit{RE (Requirements Engineering)}
    \begin{itemize}
        \item REE (Elicitation)
        \item REA/M (Analysis/Modeling)
        \item RES (Specification)
        \item REVV (Verification and Validation)
        \item REM (Management)
    \end{itemize}
    \item \textit{SD (Design)}
    \item \textit{SDev (Development)}
    \item \textit{SQA (Quality Assurance)}
    \item \textit{SDep (Deployment)}
    \item \textit{SM (Maintenance)}
\end{itemize}

To address \textit{RQ5}, we extracted which regulations were considered and investigated under the ``regulations considered'' extraction category. Additionally, ``field of regulation'' and ``domain of application'' were extracted. We applied the predefined lists of fields of regulation and domains of application formulated in the previous secondary studies~\cite{goal}, \cite{mubarkoot2021software}. After finishing the data extraction using these categories, we have refined and adapted the categories to the data that we have obtained. Next, we provide the final list of fields of regulation and domain of application with short explanations.

\textit{Field of Regulation}: Accessibility, AI/ML (capturing any regulations applicable to AI/ML-based software systems), Business (capturing regulations applicable to both business processes and/or enterprise information systems across different industries), Privacy, Privacy\&Security (identified in case same regulation considers both Privacy and Security simultaneously), Quality, Safety, Security, Traffic law, User rights (capturing regulations applicable to SIPS users only, e.g., human rights, patient rights).

\textit{Domain of Application}: Automotive, Avionics, Cloud computing, E-commerce, Education, Energy (including nuclear energy, electricity distribution), Enterprise (capturing wide range industries (e.g., retail, food industries) and activities characteristic or enterprises (e.g., marketing, taxation)), Finances, Government, Healthcare, IoT, Manufacturing, Media, Metrology, Military, Smart home/city, Software development (capturing primary study which focused on SIPS product and outsourcing companies, not belonging to specific industry), Telecommunications, Transport.

Finally, to evaluate the quality of the studies, we extracted ``scientific rigor'' and ``industrial relevance'', which is a scoring system based on the work of Ivarsson and Gorschek~\cite{tonny}. We adopted the scoring of 0 (weak description) or 1 (medium and strong description) across the research question study design and validity for the scientific rigor. To evaluate the industrial relevance, we scored the sub-types: subject, context, scale, and research method.

\subsubsection{Data Analysis}
We used the extracted data to address each research question during the data analysis phase. The data underwent a thorough cleaning and organi\added{z}ing process, and various figures and tables were generated to illustrate the findings. Descriptive statistics and frequency analysis were employed to analyze and summarize the data, leading to the presentation of various figure types, including bar charts, stacked bar charts, and maps (also known as categorical bubble plots). Additionally, the ``challenges to compliance'' open-coded category was examined, and its categorization was established in this phase. The first two authors carried out the data mapping task with active collaboration and consultation with the other authors.

\begin{figure}
     \centering
		\begin{tikzpicture}
        \begin{axis}[
            ybar=7pt,
          width=1\linewidth,
            xlabel={Year Published},
		ylabel={Number of Primary Studies},
            y label style = {font = \scriptsize},
            x label style = {font = \scriptsize},
            symbolic x coords={2017, 2018, 2019, 2020, 2021, 2022, 2023},
            xtick=data,
            nodes near coords,
            x tick label style = {font = \scriptsize, text width = 1.5cm, align = center},
            y tick label style = {font = \scriptsize},
            axis x line*=bottom,
            axis y line*=left
          ]
            \addplot[ybar,fill=MyBlue] coordinates {
                (2017,   31)
                (2018,   35)
                (2019,   37)
                (2020,   41)
                (2021,   43)
                (2022,   45)
                (2023,   48)
            };
        \end{axis}
		\end{tikzpicture}\caption{Number of primary studies published per year}\label{fig:year}
\end{figure}
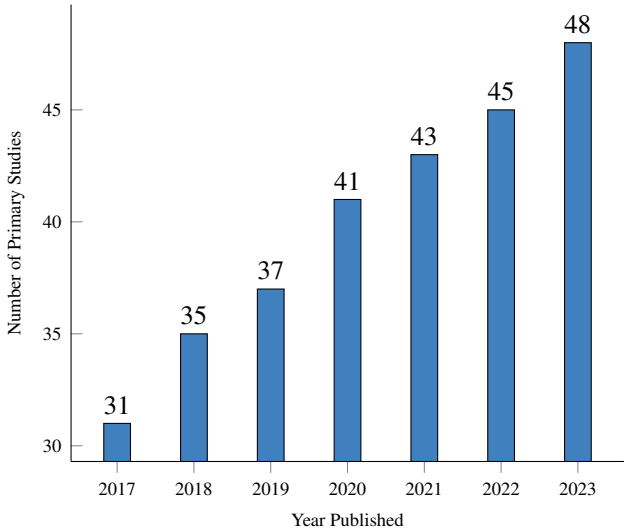

\section{Results and Discussions}\label{sec:results}
The following section presents the results and a discussion structured according to our research questions. We begin with an overview of general trends throughout the years, venues, and authors.
\subsection{Trends throughout years, venues, and authors} \label{sec:trends}
The study selection process resulted in 280 relevant studies.
Most selected studies were authored by researchers affiliated with academia 60.71\% (170/280). A smaller part, 33 studies (11.78\%), were from the authors affiliated with the industry, and 77 studies (27.5\%) were conducted through collaborations between academia and industry. Figure~\ref{fig:year} presents the number of studies published over the years. In this research, we targeted the last seven years' studies from 2017 (January 1) until 2023 (December 31). There has been a general increase in the number of studies over the years.

\begin{figure}
\centering
\begin{tikzpicture}
        \begin{axis}[ 
            ybar=7pt,
          width=1\linewidth,
		ylabel={Number of Primary Studies},
            symbolic x coords={REFSQ, PROFES, ICSE, FedCSIS, CEUR, Access, ISSREW, CCIS, RE, DASC},
            xtick=data,
            y label style = {font = \scriptsize},
            x label style = {font = \scriptsize},
            x tick label style = {font = \scriptsize, text width = 1.9cm, align = center, rotate = 70, anchor = north east},
            y tick label style = {font = \scriptsize},
            nodes near coords,
            axis x line*=bottom,
            axis y line*=left,
            xlabel style={yshift=-0.4cm},
            xlabel={Publication Venue}
          ]

            \addplot[ybar,fill=MyBlue] coordinates {
                (DASC,     13)
                (RE,       9)
                (CCIS,     7)
                (ISSREW,   7)
                (Access,   6)
                (CEUR,     5)
                (FedCSIS,  5)
                (ICSE,     5)
                (PROFES,   5)
                (REFSQ,    5)
            };
        \end{axis}
		\end{tikzpicture}\caption{Venues with the highest number of published studies}\label{fig:venue}
\end{figure}
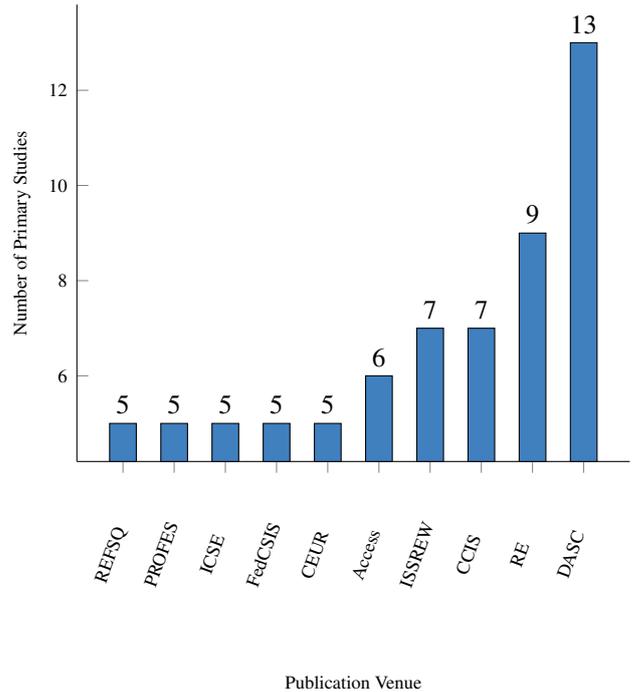

There are 181 unique venues where the primary studies were published. Figure~\ref{fig:venue} shows the list of the top venues with the most published studies (five or more). The venues are Digital Avionics Systems Conference (DASC), Requirements Engineering Conference (RE), Communications in Computer and Information Science (CCIS), International Symposium on Software Reliability Engineering Workshops (ISSREW), IEEE Access journal (Access), CEUR Workshop Proceedings (CEUR), Federated Conference on Computer Science and Information Systems (FedCSIS), International Conference on Software Engineering (ICSE), Product Focused Software Process Improvement Conference (PROFES), and Requirements Engineering: Foundation for Software Quality Conference (REFSQ). They account for 23.92\% (67/280) of all primary studies.

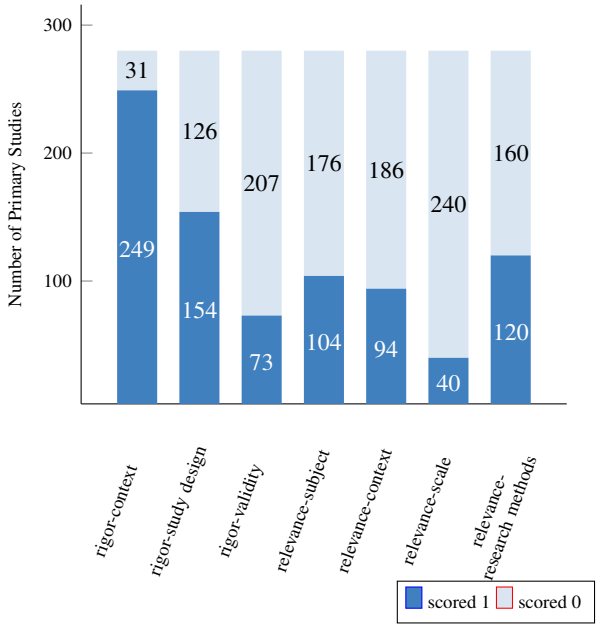
\begin{figure}
\centering
\begin{tikzpicture}
\begin{axis}[
    width=0.9\linewidth,
    ybar stacked,
    bar width=15pt,
    nodes near coords,
    axis x line*=bottom,
    axis y line*=left,
    enlargelimits=0.15,
    legend style={at={(0.65,-0.550)}, anchor=south west,legend columns=-1},
    y label style = {font = \scriptsize},
    x label style = {font = \scriptsize},
    nodes near coords style={font=\small},
    ylabel={Number of Primary Studies},
    symbolic x coords={rigor-context, rigor-study design, rigor-validity, relevance-subject, 
		relevance-context, relevance-scale, relevance-research methods},
    xtick=data,
    x tick label style={align = center, rotate = 70, anchor = north east, font = \scriptsize, text width = 2.6cm},
    y tick label style = {font = \scriptsize},
    ]
\addplot+[ybar,fill=MyBlue,draw=none,text=white] plot coordinates {(rigor-context,249) (rigor-study design,154) (rigor-validity,73) (relevance-subject,104) (relevance-context,94) (relevance-scale,40) (relevance-research methods,120)};
\addplot+[ybar,fill=NextBlue,draw=none,text=black] plot coordinates {(rigor-context,31) (rigor-study design,126) (rigor-validity,207) (relevance-subject,176) (relevance-context,186) (relevance-scale,240) (relevance-research methods,160)};
\legend{\scriptsize \strut scored 1, \scriptsize \strut scored 0}
\end{axis}
\end{tikzpicture}
\caption{Ratings of scientific rigor and industrial relevance for every study}\label{fig:rr}
\end{figure}

Finally, we also identified \added{the} top ten authors of the primary studies who are as follows Tommi Mikkonen (who is a co-author of 7 studies), Johnny Marques (6 studies), Fergal McCaffery (6 studies), Marko Esche (5 studies), Daniel Mendez (5 studies). Edna Dias Canedo (4 studies), Peng-Fei Gu (4 studies), Barbara Gallina (4 studies), John Mylopoulos (4 studies), Vlad Stirbu (4 studies). These authors authored 17.5\% (49/280) of the 280 primary studies. Ivarsson and Gorschek~\cite{tonny} outlined a method for evaluating scientific rigor and industrial relevance using several criteria (see Table~\ref{tab:category}). Figure~\ref{fig:rr} visualizes the rigor and relevance values extracted from the 280 primary studies. The three left bars of Figure~\ref{fig:rr} show scores for the scientific “rigor” of the primary studies (0 or 1). Most of the studies (249) received a score of 1 for rigor-context, with only 31 primary studies receiving a score of 0. For rigor-study design, more than 55\% (154/280) of the studies show an overall good quality. However, rigor-validity has almost 73.92\% (207/280) 0’s, which means many studies did not report any threats to validity. 
The right four bars of Figure\ref{fig:rr} show scores for the industrial “relevance” of the primary studies (either 0 or 1). Overall, there is a low industrial involvement in the analyzed primary studies. Relevance-research methods and relevance-subject show the highest values, as represented by 42.85\% (120/280) and 37.14\%(104/280) studies evaluated as 1 for them. Relevance-context had 33.57\% (94/280) 1’s over 0’s, and relevance-scale had the lowest overall score with 14.28\% (40/280).

\subsection{RQ1: What are the reported challenges in regulatory compliance of SIPS?} \label{RRQ1}
\begin{boxResults}
\textit{Highlights
\vspace{-2mm}
\begin{itemize}[leftmargin=*]
    \setlength\itemsep{2mm}
    \item We identified 14 categories of challenges to regulatory compliance of SIPS.
    \item Most considered categories of challenges are abstractness, conflicts of regulatory demands with SE practices, and demand for domain knowledge.
    \item There is a need for studies on specific challenge categories and the relations between them.
\end{itemize}}
\end{boxResults}
The existing body of literature contains information about a wide range of challenges to SIPS regulatory compliance related to RE. \added{255} studies out of 280 (\added{91.1}\%) mentioned at least one challenge to regulatory compliance. We observed that research of challenges has been increasing in recent years, however only some studies applied empirical methods to systematically research challenges or conduct in-depth research of concrete challenges (see section~\ref{sec:rq1Discussion} for details) and, as we have observed, the industrial relevance of the papers was low. Due to this, we have extracted any mention of challenges to regulatory compliance without introducing any additional qualifying criteria. Most of the analyzed primary studies mentioned challenges as motivation for their study or as a side result without elaborating on them in detail or providing a review of previous studies. Only some studies focused on challenges or explored them in detail. For example, Camilli et al.~\cite{dalpiaz_risk-driven_2021} described challenges in compliance assurance for collaborative AI systems, Peixoto et al.~\cite{madhavji_understanding_2020,peixoto2023perspective} explored how developers perceive and interpret privacy requirements, Canedo et al.~\cite{dias_canedo_perceptions_2020,canedo2021agile} described the perceptions of ICT practitioners towards software privacy in the context of the Brazilian General Data Protection Law, Galvez \& Gurses~\cite{galvez_odyssey_2018} have explored challenges to privacy threat modeling as required per GDPR in conditions of agile and service-oriented SE, and Li et al.~\cite{li_continuous_2019,li2022towards} described practical challenges to GDPR compliance in the context of small companies and Andrade et al.~\cite{andrade2023personal} who explored how organizations integrate personal data privacy into their software development processes. The only study that approached challenges to compliance specifically and systematically using empirical methods was Peixoto et al.~\cite{peixoto2023perspective} which mapped personal, behavioral, and external environment factors influencing the interpretation and perception of privacy. Li et al.~\cite{li2022towards} also reported multiple challenges identified in the course of their design study, but identified only three groups of challenges.
Most studies contained recurring challenges such as the abstractness of regulations, compliance with multiple regulations, etc. To analyze the challenges, we have grouped related challenges into categories that we describe next. Table~\ref{tab:challenges} contains the list of all the challenges, the number of primary studies that mention each category (each primary study can mention multiple challenges), examples of data about challenges extracted from studies for each category of challenges, and references to all primary studies mentioning challenges in each category.

\begin{table*} 
\resizebox{\textwidth}{!}{%
\footnotesize
\centering
	\begin{tabular}{p{2.5cm}P{1.1cm}p{5.4cm}P{10cm}} \toprule  
		\thead{Challenge \\ category} & \thead{\# of \\ studies} & \thead{Example of \\ a challenge} & \thead{References to \\ primary studies}  
            \\\midrule
            Abstractness of regulations & 96 & ``They find that the requirements of the standard are expressed at a high level.''~\cite{bujok_approach_2017} & \cite{bujok_approach_2017, madhavji_understanding_2020, granlund2020medical, ardila2017towards, ayala-rivera_grace_2018, fan_empirical_2020, antunes_operationalization_2020, martin_methods_2018, guaman_gdpr_2021, hjerppe_general_2019, farhadi2019compliance, alsaadi_minimizing_2019, garg_iterative_2021, alshammari_model-based_2018, grant_towards_2018, dewitte_comparison_2019, cook2020shamroq, odarushchenko_software_2020, dias_canedo_perceptions_2020, huth_appropriate_2019, paz2019requirements, hahnle_software_2019, gu2020discussion, guo_corba_2020, mustapha_systematic_2020, biscoglio2017certification, Usman, esche_formalization_2019, bartolini2019gdpr, netto_identifying_2019, patwardhan_moving_2018, jantti_studying_2020, velychko_testing_2019, jensen_towards_2019, patwardhan2018towards, sakul-ung_towards_2019, kosenkov2020towards, kneuper_translating_2020, massey2017modeling, strielkina2018cybersecurity, ghaisas2018resolving, wieland2018implementation, metayer2019modelling, romero2019adapting, bagheri2020synthesis, slavov2020requirements, zaitsev2018agile, barricelli202115, stirbu2021introducing, zieni2021transparency, poth2021lean, gomez2021towards, muram2021facilitating, durling2021certification, sovrano2018making, campanile2022towards, mashaly2022privacy, canedo2022guidelines, alhirabi2022privacy, li2022towards, peyrone2022formal, olukoya2022assessing, van2022recommendations, ludvigsen2022software, leite2022impact, barletta2022gdpr, kyhlstedt2022need, ryan2022support, illescas2022representing, kempe2022documenting, sarrala2022towards, breaux2022legal, ladkin2022assigning, tan2021application, mclachlan2021smart, price2017regulating, almada2023regulation, mueck2023upcoming, peixoto2023perspective, culley2023insights, tang2023helping, lucaj2023ai, li2023regulating, alhirabi2023parrot, rouland2023eliciting, makrakis2023wipp, durand2023formal, amalfitano2023documenting, ekambaranathan2023navigating, desolda2023materialist, prokhorenkov2023toward, sangaroonsilp2023taxonomy, agirre2023up2date, kosenkov2021vision, klymenko2022understanding, khan2022enhanced}\\
            Conflicts/changes to existing practices & 84 & ``Certain DO-178C objectives make it difficult to reuse software previously developed out of DO‐178C context.''~\cite{dmitriev_lean_2020} & \cite{bujok_approach_2017, dalpiaz_risk-driven_2021, granlund2020medical, wagner_metrics_2020-1, pierce_integrating_2020, moyon2018towards, guaman_gdpr_2021, dmitriev_lean_2020, galvez_odyssey_2018, huang_software-defined_2019, madhavji_methodology_2020, ozcan-top_hybrid_2018, garg_iterative_2021, xu_design_2021, laukkarinen_regulated_2018, dieudonne_rmc_2021, li_continuous_2019, dias_canedo_perceptions_2020, liu_reuse_2021, moyon_how_2020, paz2019requirements, istvan_software-defined_2021, guarro_formal_2017, xu_study_2019-1, engiel2017tool, zeller_integrated_2020, molina_design_2021, grace_identifying_2018, alsaadi2019investigating, yu_livebox_2019, keutzer_medical_2020, perez_optimization_2022, huth2020process, barbosa_re4ch_2018, grunbacher_specifying_2017, ozcan-top_what_2019, baron2021towards, reinhartz-berger_towards_2019, meis2017pattern, kearney2021bridging, campbell2018lessons, stirbu2018towards, ferrell2018mindful, wieland2018implementation, romero2019adapting, bagheri2020synthesis, zaitsev2018agile, alhazmi2021m, stirbu2021introducing, canedo2021agile, kempe2021perspectives, poth2021lean, gomez2021towards, mashaly2022privacy, canedo2022guidelines, zanca2022regulatory, elliott2022know, muller2022explainability, conte2022privacy, li2022towards, ardagna2022bridging, chhetri2022data, ludvigsen2022software, leite2022impact, schidek2022agilization, kempe2022documenting, breaux2022legal, alkubaisy2021framework, agyei2022impact, toivakka2021towards, price2017regulating, marques2023enhancing, cha2023software, culley2023insights, li2023regulating, barbareschi2023automatic, ottun2023one, rouland2023eliciting, andrade2023personal, sangaroonsilp2023empirical, khan2022enhanced, hubbs2023automating, martens2022admed}\\
            Demand for domain knowledge & 72 & ``SMEs can face the challenges with insufficient knowledge about regulations and standards''~\cite{bujok_approach_2017} & \cite{bujok_approach_2017,madhavji_understanding_2020,fan_empirical_2020,pierce_integrating_2020,antunes_operationalization_2020,valenca2020privacy,martin_methods_2018,joshi_integrated_2020,galvez_odyssey_2018,wirtz2019risk,farhadi2019compliance,garg_iterative_2021,alshammari_model-based_2018,dewitte_comparison_2019,cook2020shamroq,dias_canedo_perceptions_2020,huth_appropriate_2019,guo_semantically_2017,guo_corba_2020,mandal2017modular,zeni_nomost_2018,zeni_annotating_2017,mougiakou_based_2017,Usman,gharib_copri_2021,kosenkov2020towards,boltz2022model,kearney2021bridging,campbell2018lessons,ghaisas2018resolving,kasisopha2019applying,alhazmi2021m,stirbu2021introducing,canedo2021agile,kempe2021perspectives,zieni2021transparency,poth2021lean,gomez2021towards,stefanova2021privacy,sovrano2018making,campanile2022towards,canedo2022guidelines,zanca2022regulatory,tang2022assessing,ferreira2022poster,li2022towards,peyrone2022formal,olukoya2022assessing,drabiak2022leveraging,leite2022impact,barletta2022gdpr,sarrala2022towards,breaux2022legal,agyei2022impact,ladkin2022assigning,mclachlan2021smart,marques2023enhancing,diepenbrock2023analysis,peixoto2023perspective,culley2023insights,ardo2023implications,neitzke2023enhancing,tang2023helping,lucaj2023ai,rocha2023privacy,alhirabi2023parrot,andrade2023personal,ferreira2023rulekeeper,ekambaranathan2023navigating,desolda2023materialist,herwanto2023towards,kosenkov2021vision}\\
            Absence of principles\&practices & 71 & ``Although previously introduced  methods still lack efficient means for the representation of  attacker motivation and have no prescribed way of constructing  attack scenarios.''~\cite{esche_representation_2017} & \cite{esche_representation_2017,mayr-dorn_timetracer_2020,ayala-rivera_grace_2018,martin_methods_2018,guaman_gdpr_2021,galvez_odyssey_2018,morisio_integration_2020,esche_developing_2020,moyon_how_2020,paz2019requirements,paz_building_2018,sartoli2020compliance,mougiakou_based_2017,Usman,singh_conformance_2021,sherry_design_2021,ahmed_symbolic_2019,todde_methodology_2020,jensen_towards_2019,marques2017verification,birnstill2018identity,mourby2021transparency,saracc2019certification,metayer2019modelling,romero2019adapting,huang2019csat,kuwajima2019adapting,bagheri2020synthesis,alhazmi2021m,barricelli202115,stirbu2021introducing,kempe2021perspectives,gomez2021towards,muram2021facilitating,hernandez2021conflicting,mann2021radar,canedo2022guidelines,tang2022assessing,elliott2022know,muller2022explainability,ferreira2022poster,yuba2022systematic,li2022towards,chhetri2022data,ryan2022support,khurshid2022eu,russell2022modeling,sarrala2022towards,breaux2022legal,javed2021ontology,alkubaisy2021framework,toivakka2021towards,peixoto2023perspective,tang2023helping,lucaj2023ai,rocha2023privacy,alhirabi2023parrot,ottun2023one,feng2023towards,rouland2023eliciting,andrade2023personal,ferreira2023rulekeeper,milankovich2023delta,amalfitano2023documenting,ekambaranathan2023navigating,prokhorenkov2023toward,guber2023privacy,kosenkov2021vision,schuster2022certification,bao2023certification,esche2023risk}\\
            Complexity & 69 & ``The resultant legal requirements models usually contain the complexities inherited from the original texts.''~\cite{madhavji_methodology_2020} & \cite{bujok_approach_2017,granlund2020medical,ayala-rivera_grace_2018,kellogg2020continuous,larrucea2017supporting,galvez_odyssey_2018,alsaadi_minimizing_2019,madhavji_methodology_2020,alshammari_model-based_2018,hayrapetian_empirically_2018,grant_towards_2018,li_continuous_2019,dias_canedo_perceptions_2020,guo_semantically_2017,moyon_how_2020,mandal2017modular,zeni_nomost_2018,jha2017adopting,zeni_annotating_2017,Usman,adedjouma_model-based_2018,tsohou_privacy_2020,patwardhan2018towards,strielkina2018cybersecurity,campbell2018lessons,ferrell2018mindful,marques2018tailoring,chitnis2017enabling,anish2021automated,stirbu2021introducing,mubarkoot2021towards,canedo2021agile,gomez2021towards,stirbu2021extending,muram2021facilitating,durling2021certification,stefanova2021privacy,sovrano2018making,mann2021radar,alkubaisy2021confis,canedo2022guidelines,tang2022assessing,alhirabi2022privacy,conte2022privacy,li2022towards,ardagna2022bridging,chhetri2022data,drabiak2022leveraging,sarrala2022towards,javed2021ontology,agyei2022impact,klein2023general,tan2021application,mclachlan2021smart,thiele2021regulatory,price2017regulating,almada2023regulation,colloud2023evolving,culley2023insights,tang2023helping,alhirabi2023parrot,rouland2023eliciting,milankovich2023delta,ekambaranathan2023navigating,cepeda2022challenges,sangaroonsilp2023empirical,kosenkov2021vision,baron2023framework,bao2023certification}\\
            Interaction with experts & 58 & ``Data Protection by Design is a truly interdisciplinary effort that involves many stakeholders such as legal experts, requirements engineers, software architects, developers, and system operators.''~\cite{sion_architectural_2019} & \cite{wagner_metrics_2020-1,galvez_odyssey_2018,hjerppe_general_2019,morisio_integration_2020,dewitte_comparison_2019,wang_modelling_2019,moyon_how_2020,sion_architectural_2019,Usman,huth2020process,boltz2022model,kearney2021bridging,strielkina2018cybersecurity,ferrell2018mindful,metayer2019modelling,kuwajima2019adapting,alhazmi2021m,mubarkoot2021towards,canedo2021agile,kempe2021perspectives,poth2021lean,stirbu2021extending,campanile2022towards,canedo2022guidelines,zanca2022regulatory,tang2022assessing,li2022towards,ardagna2022bridging,olukoya2022assessing,van2022recommendations,drabiak2022leveraging,leite2022impact,feng2022research,ryan2022support,illescas2022representing,sarrala2022towards,breaux2022legal,ladkin2022assigning,mclachlan2021smart,price2017regulating,peixoto2023perspective,colloud2023evolving,culley2023insights,ardo2023implications,lucaj2023ai,zapata2023review,alhirabi2023parrot,feng2023towards,saraiva2023privacy,andrade2023personal,perera2021envisioning,milankovich2023delta,desolda2023materialist,herwanto2023towards,cepeda2022challenges,prokhorenkov2023toward,kosenkov2021vision,klymenko2022understanding} \\
            Resource intensity & 58 & ``Implementing international standards and certification is expensive and time-consuming.''~\cite{bujok_approach_2017} & \cite{bujok_approach_2017,ayala-rivera_grace_2018,pierce_integrating_2020,kellogg2020continuous,joshi_integrated_2020,dmitriev_lean_2020,wirtz2019risk,hjerppe_general_2019,hayrapetian_empirically_2018,dewitte_comparison_2019,odarushchenko_software_2020,li_continuous_2019,mills_towards_2017,guo_semantically_2017,arogundade2018specifying,guo_corba_2020,biscoglio2017certification,streitferdt2018complete,Usman,patwardhan_moving_2018,ferreyra2020pdp,jantti_studying_2020,jensen_towards_2019,patwardhan2018towards,castellanos2017towards,streitferdt2017component,birnstill2018identity,chitnis2017enabling,metayer2019modelling,kasisopha2019applying,bagheri2020synthesis,aberkane2021automated,anish2021automated,kempe2021perspectives,stirbu2021extending,muram2021facilitating,mashaly2022privacy,alhirabi2022privacy,li2022towards,ardagna2022bridging,feng2022research,illescas2022representing,bhamidipati2022risk,agyei2022impact,tan2021application,diepenbrock2023analysis,culley2023insights,ardo2023implications,tang2023helping,alhirabi2023parrot,rouland2023eliciting,perera2021envisioning,sahu2023web,milankovich2023delta,sangaroonsilp2023empirical,baron2023framework,rouland2023security,hubbs2023automating}\\
            Provability & 55 & ``The scale of the argument inevitably makes it difficult for regulators to comprehend them and to be confident that the evidence provided is of satisfactory quality.''~\cite{harrison_verification_2017} & \cite{granlund2020medical,ayala-rivera_grace_2018,pierce_integrating_2020,ozcan-top_hybrid_2018,vanezi_formal_2019,harrison_verification_2017,hahnle_software_2019,guarro_formal_2017,biscoglio2017certification,Usman,gannous_toward_2018,castellanos2017towards,mourby2021transparency,kearney2021bridging,campbell2018lessons,saracc2019certification,metayer2019modelling,bagheri2020synthesis,canedo2021agile,kempe2021perspectives,poth2021lean,stirbu2021extending,muram2021facilitating,durling2021certification,mashaly2022privacy,zanca2022regulatory,muller2022explainability,ardagna2022bridging,niemiec2022will,chhetri2022data,ryan2022support,schidek2022agilization,kempe2022documenting,russell2022modeling,breaux2022legal,agyei2022impact,price2017regulating,marques2023enhancing,mueck2023upcoming,peixoto2023perspective,culley2023insights,neitzke2023enhancing,tang2023helping,li2023regulating,rocha2023privacy,barbareschi2023automatic,andrade2023personal,durand2023formal,milankovich2023delta,amalfitano2023documenting,ekambaranathan2023navigating,cepeda2022challenges,baron2023framework,rouland2023security,hubbs2023automating}\\
		    Enforcement challenges & 52 & ``Some respondents do not intend to use privacy practices even recognizing their importance.''~\cite{madhavji_understanding_2020} &  \cite{bujok_approach_2017,madhavji_understanding_2020,kellogg2020continuous,galvez_odyssey_2018,farhadi2019compliance,huang_software-defined_2019,li_continuous_2019,guo_semantically_2017,gu2020discussion,mougiakou_based_2017,tsohou2017enabling,kosenkov2020towards,birnstill2018identity,mourby2021transparency,campbell2018lessons,saracc2019certification,romero2019adapting,zaitsev2018agile,alhazmi2021m,barricelli202115,stirbu2021introducing,zieni2021transparency,gomez2021towards,hernandez2021conflicting,sovrano2018making,ferreira2022poster,conte2022privacy,li2022towards,ardagna2022bridging,pantelic2022cookies,drabiak2022leveraging,barletta2022gdpr,kyhlstedt2022need,kempe2022documenting,khurshid2022eu,russell2022modeling,breaux2022legal,klein2023general,toivakka2021towards,mclachlan2021smart,thiele2021regulatory,almada2023regulation,peixoto2023perspective,culley2023insights,ardo2023implications,lucaj2023ai,rocha2023privacy,andrade2023personal,ferreira2023rulekeeper,ekambaranathan2023navigating,guber2023privacy,werthwein2023concept}\\
            Dynamics of software context & 37 & ``Such systems often run in dynamic and uncertain environments that make it difficult providing strong assurances of compliance.''~\cite{dalpiaz_risk-driven_2021} & \cite{dalpiaz_risk-driven_2021,chechik_uncertain_2019,anisetti_semi-automatic_2020,hjerppe_general_2019,garg_iterative_2021,alshammari_model-based_2018,cook2020shamroq,dias_canedo_perceptions_2020,hahnle_software_2019,sartoli2020compliance,Usman,kosenkov2020towards,boltz2022model,streitferdt2017component,campbell2018lessons,mubarkoot2021towards,canedo2021agile,kempe2021perspectives,gomez2021towards,durling2021certification,mashaly2022privacy,canedo2022guidelines,alhirabi2022privacy,yuba2022systematic,li2022towards,olukoya2022assessing,drabiak2022leveraging,breaux2022legal,ladkin2022assigning,thiele2021regulatory,almada2023regulation,tang2023helping,li2023regulating,rouland2023eliciting,sangaroonsilp2023empirical,kosenkov2021vision,henderson2022toward}\\
            Multiplicity of regulations & 35 & ``For some of the safety-critical domains, there is a need to implement more than one standard.''~\cite{bujok_approach_2017} &  \cite{bujok_approach_2017,ardila2017towards,makkar_automotive_2019,antunes_operationalization_2020,stolfa_lightweight_2017,valenca2020privacy,joshi_integrated_2020,park_association-based_2018,huang_software-defined_2019,alsaadi_minimizing_2019,garg_iterative_2021,cook2020shamroq,arogundade2018specifying,yilmaz_application_2020,li_poet_2019,wieland2018implementation,elliott2022know,yuba2022systematic,li2022towards,olukoya2022assessing,pantelic2022cookies,drabiak2022leveraging,schidek2022agilization,breaux2022legal,mclachlan2021smart,thiele2021regulatory,mueck2023upcoming,colloud2023evolving,culley2023insights,lucaj2023ai,li2023regulating,zapata2023review,rouland2023eliciting,sahu2023web,kosenkov2021vision}\\
            Dynamics of software systems & 33 & ``This challenge is related to the continuous training of AI models in post-marketing settings.''~\cite{zapata2023review} & \cite{granlund2020medical,mayr-dorn_timetracer_2020,wagner_metrics_2020-1,kellogg2020continuous,galvez_odyssey_2018,hahnle_software_2019,streitferdt2018complete,besker2019how,boltz2022model,streitferdt2017component,kearney2021bridging,strielkina2018cybersecurity,campbell2018lessons,stirbu2018towards,kuwajima2019adapting,mubarkoot2021towards,mann2021radar,canedo2022guidelines,li2022towards,feng2022research,sarrala2022towards,ladkin2022assigning,thiele2021regulatory,price2017regulating,colloud2023evolving,culley2023insights,li2023regulating,zapata2023review,milankovich2023delta,cepeda2022challenges,sangaroonsilp2023empirical,henderson2022toward,esche2023risk}\\
            Organizational challenges & 29 & ``We found a lack of a privacy strategy and a lack of a privacy culture.''~\cite{peixoto2023perspective} & \cite{li_continuous_2019,serban_standard_2018,ferrell2018mindful,alhazmi2021m,stirbu2021introducing,canedo2021agile,kempe2021perspectives,poth2021lean,campanile2022towards,canedo2022guidelines,zanca2022regulatory,alhirabi2022privacy,muller2022explainability,li2022towards,van2022recommendations,drabiak2022leveraging,ryan2022support,kempe2022documenting,sarrala2022towards,ladkin2022assigning,mclachlan2021smart,thiele2021regulatory,peixoto2023perspective,ardo2023implications,rocha2023privacy,alhirabi2023parrot,andrade2023personal,milankovich2023delta,ekambaranathan2023navigating}\\
            Regulatory gaps & 26 & ``SOUP software requires special considerations, as the developers’ obligations related to design and implementation are not applied to it.''~\cite{stirbu2021extending} & \cite{granlund2020medical,mourby2021transparency,kearney2021bridging,ferrell2018mindful,kuwajima2019adapting,zaitsev2018agile,kempe2021perspectives,stirbu2021extending,hernandez2021conflicting,canedo2022guidelines,zanca2022regulatory,elliott2022know,yuba2022systematic,van2022recommendations,ludvigsen2022software,kyhlstedt2022need,zinchenko2022methodology,breaux2022legal,toivakka2021towards,thiele2021regulatory,price2017regulating,mueck2023upcoming,lucaj2023ai,li2023regulating,durand2023formal,henderson2022toward}
			\\\bottomrule
	\end{tabular}}
 \caption{References to all o the primary studies which considered categories of challenges}\label{tab:challenges}
\end{table*}

\subsubsection{Discussion \nameref{RRQ1}}\label{sec:rq1Discussion}
To analyze the extracted data about challenges, we identified similar challenges to categorize them. Next, we calculated the number of studies mentioning a particular category of challenges (independently of the number of mentions of a particular challenge in that study). Many of the challenges were described as closely interconnected. For example, in Moyon et al.~\cite{morisio_integration_2020}, one of the challenges is that ``the SE field lacks methods to demonstrate compliance with security standards when applying DevOps''. In such cases, the challenge category was selected, taking into account the context of the extracted text. We identified 14 categories of challenges, which we explain next.

\paragraph*{Abstractness of regulations (96/\added{255} studies (\added{37.6}\%))} The challenges in this category are related to preventing the immediate application of regulations as software requirements~\cite{bartolini2019gdpr} and/or their translation into software requirements~\cite{alshammari_model-based_2018}. In particular, the primary studies described abstractness as related to (1) the language used in regulations and/or (2) the content of regulations. A broad range of terms was used to describe the language of regulations. For example, it was denoted as abstract~\cite{ayala-rivera_grace_2018,peyrone2022formal}, unclear~\cite{bujok_approach_2017, guo_corba_2020}, verbose, vague~\cite{ayala-rivera_grace_2018,illescas2022representing}, ambiguous~\cite{ayala-rivera_grace_2018,li2022towards}, abstruse~\cite{guo_corba_2020}). There was a lack of terminological consistency in the description of these challenges. For example, primary studies could mention that regulations as a source of requirements are ambiguous and vague~\cite{ayala-rivera_grace_2018} without providing a clear distinction between the two. On the one hand, many primary studies recognized that abstractness is inherent and purposeful in covering multiple situations(e.g.,~\cite{massey2017modeling,kempe2022documenting}). On the other hand, many primary studies suggest that more concrete operationalizable norms are needed (e.g.,~\cite{lucaj2023ai, alhirabi2023parrot, amalfitano2023documenting}).

In terms of the content of regulations studies mention such challenges as provision of general principles only (e.g.,~\cite{campanile2022towards, mashaly2022privacy}), absence of concrete procedures to be implemented (e.g.,~\cite{fan_empirical_2020,price2017regulating}), obscurity about the PPs to be applied (e.g.,~\cite{muram2021facilitating, prokhorenkov2023toward}), or system properties to be achieved (e.g.,~\cite{bagheri2020synthesis}). Another aspect of the abstractness of the content of regulations was that regulations cover different aspects and have different granularity. For example, regulations can address both high-level goals and low-level design constraints~\cite{breaux2022legal} and also tend to cover multiple compliance aspects in addition to technologies such as processes, people, strategy~\cite{zaitsev2018agile}, industries or domains~\cite{breaux2022legal}, governance and management~\cite{tan2021application}, businesses and organization~\cite{kosenkov2021vision}.

Existing studies also widely recognize that as the result of the abstractness of regulations their interpretation is required (e.g.,~\cite{garg_iterative_2021}). Such interpretation poses a challenge itself because of the possibility of multiple and different interpretations~\cite{ayala-rivera_grace_2018, alsaadi_minimizing_2019}. This, increases the risk of non-compliance~\cite{cook2020shamroq,li2022towards} and non-consistent interpretation in different cases~\cite{zieni2021transparency,culley2023insights}. It is also unclear for software engineers how to conduct such interpretation correctly~\cite{peixoto2023perspective}. Due to this, a gap between the intention of regulators and interpretation within the developing companies may emerge~\cite{ryan2022support}.

\paragraph*{Conflicts or changes to existing SE practices caused by regulations (84/\added{255} studies (\added{32.9}\%))} We also identified a category of challenges related to the fact that regulations directly or indirectly conflict, with existing SE practices or necessitate changes to them. Some primary studies reported that regulations cause changes throughout the SDLC life cycle (e.g.,~\cite{kempe2021perspectives, agyei2022impact}). The most general challenges in this category are that regulations harm the SIPS development (e.g., constraints of data transfers~\cite{guaman_gdpr_2021}, demand use of additional computational resources~\cite{istvan_software-defined_2021}, slow down SIPS release~\cite{galvez_odyssey_2018}, constraint the flexibility of SIPS design~\cite{galvez_odyssey_2018}). Another aspect is related to the fact that regulations introduce new requirements~\cite{madhavji_methodology_2020, garg_iterative_2021, barbosa_re4ch_2018, meis2017pattern} (e.g., the demand for extended maintenance and quality assurance in operations~\cite{zanca2022regulatory}). Another challenge was related to the emergence of conflicts between regulatory requirements and other business and/or user requirements~\cite{engiel2017tool, grace_identifying_2018, elliott2022know, conte2022privacy, kempe2022documenting, alkubaisy2021framework, sangaroonsilp2023empirical}, or difficulty in integrating regulatory requirements into format used for usual requirements (e.g., use stories~\cite{canedo2021agile}).

Primary studies also reported challenges with multiple software engineering practices which are as follows: (1) different aspects of agile development\cite{bujok_approach_2017, granlund2020medical, wagner_metrics_2020-1, moyon2018towards, galvez_odyssey_2018, ozcan-top_hybrid_2018, laukkarinen_regulated_2018, moyon_how_2020, alsaadi2019investigating, huth2020process, baron2021towards, stirbu2018towards, zaitsev2018agile, stirbu2021introducing, marques2023enhancing, barbareschi2023automatic, andrade2023personal, khan2022enhanced, martens2022admed}, (2) DevOps, continuous delivery and maintenance~\cite{laukkarinen_devops_2017,granlund2020medical,wagner_metrics_2020-1,toivakka2021towards}, (3) regular updates~\cite{li2023regulating, bagheri2020synthesis}, (4) software reuse~\cite{reinhartz-berger_towards_2019, dmitriev_lean_2020, xu_design_2021, dieudonne2021rmc, liu_reuse_2021}, (5) rapid prototyping~\cite{dmitriev_lean_2020}, (6) automation~\cite{dalpiaz_risk-driven_2021}. Studies often mentioned multiple aspects of such conflicts. Some of the examples of conflicts between regulatory compliance demands and agile software development are as follows: iterative and incremental development complicates risk analysis, focus on functionality neglects traceability, audits are impeded by dynamic changes of processes~\cite{moyon2018towards} or are absent at all in agile development~\cite{laukkarinen_devops_2017}, documentation required for compliance purposes can be missing~\cite{wagner_metrics_2020-1}, tracking of changes in requirements is challenging~\cite{ozcan-top_hybrid_2018}. 

According to some studies, conflicts between SE practices and regulations can force companies to change their practice. For example, if it can be challenging to establish the suitability of existing SE methods (e.g., agile software development) to regulatory demands, companies may switch to more conservative development methods like Waterfall~\cite{bujok_approach_2017}.

\paragraph*{Demand for domain knowledge (72/\added{255} studies (\added{28.2}\%))} 
Challenges in this category were mainly related to the need for different types of domain knowledge and expertise in regulatory RE and implementation of regulatory compliance of SIPS.  We have identified five main types of domain knowledge and expertise that were mentioned in the studies which are as follows: (1) legal and/or compliance knowledge, (2) privacy knowledge, (3) security knowledge, (4) technical/software engineering knowledge, (5) application domain knowledge.
Studies described a wide range of challenges related to legal and compliance knowledge. Some of these are as follows absence of awareness about regulations or regulatory norms(e.g.,~\cite{madhavji_understanding_2020,Usman,kasisopha2019applying,alhazmi2021m,peixoto2023perspective}, demand for the knowledge of ``legalese jargon''(e.g.,~\cite{joshi_integrated_2020, cook2020shamroq,zeni_annotating_2017,kempe2021perspectives}), knowledge of the structure of legal documents (e.g.,~\cite{zeni_annotating_2017}), unwritten rules that might influence legal reasoning (e.g., the context of the legal scenario~\cite{ayala-rivera_grace_2018}), knowledge and experience in implementing regulations (e.g.,~\cite{dias_canedo_perceptions_2020,stirbu2021introducing}), systematic understanding of compliance across different development activities (e.g.,~\cite{kempe2021perspectives}) or ``in-depth knowledge of all legal consequences'' of decisions~\cite{leite2022impact}.
Legal and compliance knowledge was mentioned as complementary technical/software engineering knowledge. Technical knowledge was mentioned as essential not only for technical roles to conduct their activities but also as important for legal experts (e.g., for conducting code reviews~\cite{culley2023insights}), external reviewers (e.g., to verify compliance~\cite{tang2023helping}), and regulatory bodies~\cite{lucaj2023ai}. For example,~\cite{ladkin2022assigning} elaborated that regulatory demands largely depend on the characteristics of software, and its potential failures. Some examples of the technical knowledge mentioned in studies are as follows demanded for ``low level knowledge about the system implementation''~\cite{galvez_odyssey_2018}, knowledge of latest development technologies~\cite{stirbu2021introducing}, technical knowledge required to feasibly react to potential changes~\cite{boltz2022model}.

Privacy and security knowledge were mentioned mainly in the context of compliance with privacy and security regulations. Challenges related to privacy knowledge included such challenges as insufficient knowledge of privacy~\cite{dias_canedo_perceptions_2020,gharib_copri_2021}, confusion between privacy and security~\cite{madhavji_understanding_2020,andrade2023personal}, lack of experience~\cite{peixoto2023perspective, rocha2023privacy}. While demand for security knowledge was mentioned in some studies~\cite{moyon_how_2020,farhadi2019compliance, kempe2021perspectives, barletta2022gdpr,ardo2023implications} the challenges related to it were described on a very general level.
The challenges of SIPS application domain knowledge were also considered important~\cite{zieni2021transparency, sovrano2018making, campanile2022towards}, especially in complex domains like healthcare~\cite{drabiak2022leveraging} and finances~\cite{culley2023insights}.

According to the primary studies, demand for expertise and domain knowledge also creates challenges for developing PPs for regulatory compliance implementation~\cite{guo_corba_2020}. A few studies mentioned that there is some form of a disconnectedness, or mismatch between the regulatory, engineering, and other perspectives involved in regulatory compliance implementation~\cite{martin_methods_2018, ayala-rivera_grace_2018, fan_empirical_2020, alshammari_model-based_2018, dewitte_comparison_2019, moyon_how_2020, olukoya2022assessing} with only limited understanding across the domains~\cite{hjerppe_general_2019}. Many publications described it as a ``semantic gap'' i.e., the difference in meaning between two different representations describing the same object. This, for example, includes terms mismatch~\cite{guo_semantically_2017, ghaisas2018resolving}, the difference between policymakers and software engineers in understanding the concept of privacy~\cite{alshammari_model-based_2018}, different approaches to risk assessment~\cite{dewitte_comparison_2019}.
Last, but not least, some primary studies mentioned different challenges related to knowledge management such as structuring knowledge~\cite{campbell2018lessons}, sharing knowledge~\cite{li2022towards}, and the importance of shared knowledge between software engineers and legal experts~\cite{olukoya2022assessing}. This highlights the connection of this category of challenges with the category of challenges concerned with the interaction between experts which we consider further.

\paragraph*{Absence or insufficiency of principles and practices (PPs) (71/\added{255} (\added{27.8}\%))} A significant number of the primary studies highlighted the absence or insufficiency of PPs to conduct regulatory RE and implement SIPS compliance. Many primary studies highlighted the absence of appropriate PPs for implementation of regulatory compliance(e.g.,~\cite{ahmed_formal_2018,velychko_testing_2019,guaman_gdpr_2021,jensen_towards_2019,dalpiaz_risk-driven_2021}), other studies emphasized the insufficiency and limitations of existing PPs to achieve regulatory compliance (e.g.,~\cite{paz2019requirements, esche_developing_2020, mayr-dorn_timetracer_2020,moyon_how_2020}) and in particular inefficiency~\cite{esche_representation_2017}, inflexibility~\cite{sarrala2022towards}, disconnectedness from the engineering practice~\cite{martin_methods_2018}, absence of the required details~\cite{galvez_odyssey_2018}), insufficient rigor~\cite{bagheri2020synthesis}. 
    
One challenge that emerged especially in the recent studies was the absence of integrated or unified PPs (e.g., ~\cite{alhirabi2023parrot}), i.e. incorporating compliance throughout the development life cycle~\cite{romero2019adapting, metayer2019modelling, kempe2021perspectives}, embracing the different aspects of system quality~\cite{kuwajima2019adapting}, combining multiple PPs~\cite{mann2021radar}, addressing multiple regulatory demands~\cite{chhetri2022data,khurshid2022eu}.
    
Some primary studies identified the absence of internal guidelines and learning resources as a challenge~\cite{ayala-rivera_grace_2018,alhazmi2021m,canedo2022guidelines}. Absence of the required automation or insufficient automation was also considered to be a challenge (e.g., absence of automation for handling data subject requests)~\cite{galvez_odyssey_2018, birnstill2018identity,metayer2019modelling, sartoli2020towards,galvez_odyssey_2018,canedo2022guidelines,feng2023towards}. Tool qualification was also mentioned in this context as, in many cases, it was unclear if a particular PP is eligible for achieving regulatory compliance~\cite{saracc2019certification,ottun2023one}.

\paragraph*{Complexity of regulations processing and implementation (69/\added{255} (\added{27}\%))} Many studies describe different complexities emerging in the process of regulatory RE. Some of the aspects of this category of challenges are: (1) complexity of regulations, (2) complexity of software systems, and other different types of complexity-related challenges.

The complexity of regulations (e.g.,~\cite{madhavji_methodology_2020}) arises from various challenges, such as multiple pre-conditions and exceptions~\cite{mandal2017modular}, text length~\cite{zeni_nomost_2018,rouland2023eliciting}, a complex structure of regulations (incl. variety of multiple formats and notation elements that can be used in standards)~\cite{adedjouma_model-based_2018}, the complexity of regulations modeling~\cite{madhavji_methodology_2020} and resulting models consisting of multiple concepts and relationships~\cite{zeni_nomost_2018}, trade-offs and conflicts of compliance requirements~\cite{Usman,alkubaisy2021confis,alkubaisy2021confis}, multiple regulatory concerns that need to be observed~\cite{stirbu2021introducing,li2022towards}, management of compliance~\cite{li2022towards}.

The complexity of software systems (e.g., large scale) is another aspect of complexity constraining regulatory RE and SIPS compliance. In particular, it is mentioned in the context of AI/ML systems~\cite{stirbu2021extending, conte2022privacy, price2017regulating, culley2023insights, cepeda2022challenges}, cloud systems~\cite{mann2021radar}, and IoT systems~\cite{strielkina2018cybersecurity, alhirabi2023parrot}.
Some studies mention that implementation of regulatory compliance is especially challenging in legacy systems~\cite{stefanova2021privacy, li2022towards, klein2023general, almada2023regulation, colloud2023evolving, sangaroonsilp2023empirical} and in the context of the growth of the complexity of the software supply chain~\cite{larrucea2017supporting,  mclachlan2021smart, culley2023insights, milankovich2023delta, ekambaranathan2023navigating, baron2023framework}.

Error-proneness emerging due to complexity is another aspect of complexity we included in this category of challenges~\cite{kellogg2020continuous, ayala-rivera_grace_2018,hayrapetian_empirically_2018, dias_canedo_perceptions_2020, zeni_annotating_2017, patwardhan2018towards, li2022towards}. Some studies mention the complexity of regulations along with resource intensity, which is often caused by such complexity. We elaborate on this challenge of resource intensity further.

We have added to this category all the challenges that mention complexity, error-proneness, or related challenges that are not connected directly to a need for legal knowledge or expertise. Challenges in this category are also relevant in the case of the availability of domain knowledge and expert human resources. Legal knowledge can help to address the complex structure of regulations, but it does not nullify the existence of regulations' complexity as a challenge.

\paragraph*{Interaction between experts (58/\added{255} studies (\added{22.8}\%))}
We also identified a category of challenges related to interaction between the different experts involved in implementing regulatory compliance. According to the primary studies, some regulatory demands like privacy by design are interdisciplinary in their nature~\cite{sion_architectural_2019}, and stakeholder interaction is required for their implementation~\cite{wagner_metrics_2020-1, morisio_integration_2020}. This includes interdisciplinary interaction between different domains of expertise (e.g., legal experts, software engineers and system operators in~\cite{sion_architectural_2019}, security and software engineering in~\cite{wang_modelling_2019}) and intradisciplinary interaction of different engineering roles (e.g., developers and verification team in~\cite{Usman}, designers and manufacturers in~\cite{ferrell2018mindful}). Primary studies in this category mention some of the following aspects of the interaction experts usually conduct their activities in isolation~\cite{sion_architectural_2019}, different perspectives (e.g., more holistic by security experts and more dispersed by software engineers~\cite{moyon_how_2020}), differences in the terms and concepts applied~\cite{boltz2022model,alhirabi2023parrot}, the absence of understanding of the benefits of interactions~\cite{tang2022assessing}, conflicting suggestions and conclusions~\cite{li2022towards}, cultural differences~\cite{breaux2022legal}.
	
According to the primary studies, the motivation for interaction between experts includes the demand to explore different aspects of compliance holistically~\cite{kuwajima2019adapting, kearney2021bridging} (such as technical, legal, application aspects), need to handle requirements at both development team and organizational levels~\cite{canedo2021agile}, assuring the transparency for external audits~\cite{poth2021lean}, addressing the complexity of business environment~\cite{sarrala2022towards}, achieving shared responsibility~\cite{poth2021lean}, balancing technical and legal objectives~\cite{ryan2022support,breaux2022legal}, or enabling some activities that are impossible without such an interaction (e.g., data protection impact assessment~\cite{tang2022assessing}). Some studies also emphasized the importance of collective shared understanding as a part of interaction~\cite{li2022towards,ryan2022support,sarrala2022towards}.

It is noteworthy that interaction with domain experts or users was especially highlighted in the context of regulatory compliance of AI/ML-based systems as such an interaction plays an essential role in quality assurance (~\cite{zanca2022regulatory, drabiak2022leveraging, feng2022research}).

\paragraph*{Resource intensity of processing and implementation of regulations (58/\added{255} studies (\added{22.8}\%))}
The primary studies describe the implementation of regulatory compliance as expensive~\cite{wirtz2019risk,pierce_integrating_2020, dmitriev_lean_2020,li2022towards,culley2023insights}, time-consuming~\cite{bujok_approach_2017,streitferdt2018complete,dmitriev_lean_2020,tang2023helping,milankovich2023delta} and laborious~\cite{kellogg2020continuous,joshi_integrated_2020,guo_corba_2020,birnstill2018identity,
bagheri2020synthesis}. In a few studies, the problem of resource intensity is mentioned, especially for small and medium enterprises~\cite{bujok_approach_2017,hjerppe_general_2019}. Primarily, resource intensity is related to the need to process regulations~\cite{ayala-rivera_grace_2018,joshi_integrated_2020}, conduct audits manually~\cite{kellogg2020continuous}, extra software development efforts~\cite{dmitriev_lean_2020}, expensiveness of access to legal knowledge and expertise, need to hire additional human resources~\cite{hjerppe_general_2019}, need to provide documentation to achieve compliance~\cite{jensen_towards_2019} or resources required to collect annotated data for the development of automation tools~\cite{patwardhan2018towards}. Another challenge is that these resources should be spent in a way that would maximize return on investment~\cite{wirtz2019risk}.

\paragraph*{Provability challenges (55/\added{255} studies (\added{21.6}\%))} Challenges in this category are mainly related to the development and provision of proofs and/or documentation of compliance overall~\cite{ozcan-top_hybrid_2018} or in a particular process area (e.g., DevOps~\cite{morisio_integration_2020}). This is an important aspect of SIPS compliance as without appropriate evidence regulators can assume non-compliance~\cite{kempe2021perspectives}. Among other challenges were provability challenges related to complexity and scale of compliance evidence~\cite{harrison_verification_2017} (which connects provability challenges to the aforementioned challenge of complexity), documentation for 3rd party components~\cite{granlund2020medical}, multiplicity of sources of evidence and data collection~\cite{hahnle_software_2019,metayer2019modelling} and trustworthiness of evidence~\cite{hahnle_software_2019}. Among the properties of the evidence that are challenging to achieve the following were discussed completeness~\cite{hahnle_software_2019,biscoglio2017certification}, appropriate granularity~\cite{campbell2018lessons}, sufficiency~\cite{bagheri2020synthesis}, evidence maintenance~\cite{ryan2022support}. Another challenge related to provability is traceability (e.g.,~\cite{Usman,bagheri2020synthesis,muram2021facilitating,durling2021certification,marques2023enhancing,durand2023formal}) as in some cases full traceability is required to demonstrate compliance~\cite{ozcan-top_hybrid_2018}. One provability challenge emerging only in a couple of recent primary studies is the need to monitor and evidence continuously compliance~\cite{kearney2021bridging, stirbu2021extending}.

\paragraph*{Enforcement challenges (52/\added{255} studies (\added{20.4}\%))} \added{This} category of challenges include\added{s} different heterogeneous challenges related to the enforcement of regulatory compliance, i.e. making sure that regulations are implemented as expected. Even with regulations \added{clearly specifying the required} PPs and with such PPs at hand, companies and SIPS development teams can struggle to implement regulatory compliance of SIPS (e.g., because of the need to integrate and consolidate PPs~\cite{bujok_approach_2017}). First of all, such challenges mention de facto intentional non-compliance, for example, when engineers ignore or do not intend to apply compliance practices~\cite{mougiakou_based_2017, madhavji_understanding_2020}. Another aspect of this category is the implementation of compliance only in a formal way (e.g., approaching compliance with a ``tick box'' approach without the consideration of SIPS compliance effectiveness~\cite{culley2023insights,lucaj2023ai}) or in a way undermining compliance usability or usefulness (e.g., by limiting other functionality in the context of execution of data subject rights~\cite{ekambaranathan2023navigating,birnstill2018identity})

In some cases, enforcement challenges are related to impossibility or special challenges in implementing compliance of SIPS (e.g., because of the impossibility \added{of} fully implement\added{ing} requirements in software code~\cite{almada2023regulation,conte2022privacy}), demand to additionally rely on human subjects to enforce or verify compliance (e.g., ~\cite{li2022towards,russell2022modeling}). Another challenge identified in the primary studies was the enforcement of regulatory compliance in systems behavior (e.g., data processing is GDPR compliant)~\cite{gu2020discussion,ferreira2022poster}. In particular, enforcement mechanisms can require extensive monitoring~\cite{galvez_odyssey_2018}, or other additional mechanisms~\cite{khurshid2022eu}, which are ``new, rare, or not well understood''~\cite{kempe2022documenting}. Moreover, such compliance enforcement can require appropriate ``regulatory infrastructure''~\cite{lucaj2023ai,galvez_odyssey_2018}. It is noteworthy that while the lack of enforcement is usually seen as problematic, ``compliance over-engineering''~\cite{stirbu2021introducing} is also seen as a challenge to address.

\paragraph*{Dynamics of software context and regulations (37/255 studies (\added{14.5}\%))} This category of challenges covers the challenges related to the changes in regulations and other aspects of software context. One of the most significant challenges in this category is change of regulations~\cite{garg_iterative_2021, hjerppe_general_2019, mubarkoot2021towards, mashaly2022privacy, olukoya2022assessing, rouland2023eliciting} or their interpretation~\cite{boltz2022model},
Some changes (e.g., the concept of privacy) can be linked to the changes in social norms, cultural context, and technology rather than with explicit changes~\cite{canedo2022guidelines,tang2023helping}. Li et al.~\cite{li2022towards} claim that such dynamics of regulations not only mean that SIPS engineers need to be ready to react to changes but also prepare for the upcoming regulations.

Another aspect of the dynamics of software context is the dynamic and uncertain environment of software systems. This creates challenges to completely specifying software functionality~\cite{chechik_uncertain_2019} and providing strong compliance assurance~\cite{dalpiaz_risk-driven_2021}. The challenge of the dynamic and uncertain environment is especially prominent for AI/ML-based systems. Primary studies mentioned that AI/ML-based systems often can show lower accuracy and efficiency when applied in real-world conditions~\cite{drabiak2022leveraging}, while it is challenging to account for software context during software testing.

\paragraph*{Multiplicity of regulations (35/\added{255} studies (\added{13.7}\%))}
The primary studies name a number of challenges emerging in connection to multiplicity of regulations that need to be processed and implemented concurrently. Multiple regulations can be applicable simultaneously due to diverse jurisdictions in which compliance is required~\cite{antunes_operationalization_2020, valenca2020privacy, garg_iterative_2021}, the multiplicity of sources in a particular field of law (e.g., court cases~\cite{cook2020shamroq,li2022towards}), multiplicity of regulators in a particular field of regulations~\cite{garg_iterative_2021}, or multiple levels of regulation (e.g., federal, state, and local levels in the USA~\cite{cook2020shamroq}). Some of the challenges that emerge in connection to such multiplicity are conflicts, overlaps, and differences between regulations that need to be resolved~\cite{garg_iterative_2021, pantelic2022cookies, colloud2023evolving, lucaj2023ai, zapata2023review, makkar_automotive_2019, joshi_integrated_2020}, demand to posses knowledge specific to each regulation~\cite{bujok_approach_2017}, preparation to the audit of compliance to multiple regulations~\cite{makkar_automotive_2019},
duplication of compliance processes and efforts~\cite{bujok_approach_2017,joshi_integrated_2020}.

\paragraph*{Dynamics of software systems (33/\added{255} studies (\added{12.9}\%))}
Another source of challenges to regulatory RE is dynamics and changes in software systems and software engineering artifacts. Primary studies recognize that overall changes in SIPS can have implications for regulatory compliance~\cite{kearney2021bridging}.

Some of the challenges in this category are related to the application of machine learning and artificial intelligence SIPS components and their evolution over time~\cite{dalpiaz_risk-driven_2021,thiele2021regulatory,li2023regulating,zapata2023review,cepeda2022challenges}, changes in the code~\cite{kellogg2020continuous}, or changes in the SIPS architecture~\cite{galvez_odyssey_2018} that can also result in a risk of non-compliance~\cite{larrucea2017supporting}. Run-time changes \added{in} software systems represent another distinct challenge that needs to be addressed~\cite{mann2021radar}. Continuous changes in software systems make it difficult to track and have an up-to-date regulatory compliance status of the SIPS~\cite{mubarkoot2021towards,li2022towards}. Changes can also result in the emergence of privacy threats and hence engineers need to be able to understand such new emergent privacy threats~\cite{sarrala2022towards}.

\paragraph*{Organizational challenges (29/\added{255} (\added{11.4}\%)} This category covers the challenges related to the way regulatory RE and implementation of SIPS compliance are organized. Nearly all primary studies in this category support the claim that the organizational environment affects the implementation of SIPS compliance~\cite{canedo2021agile,canedo2022guidelines}. Primary studies emphasize that SIPS compliance requires a number of organizational aspects which are often omitted. These are, for example, holistic organizational commitment~\cite{kempe2021perspectives}, compliance planning~\cite{kempe2022documenting}, and nurturing the required strategy and culture~\cite{peixoto2023perspective,ardo2023implications}.

The most prominent aspect in this category is related to the responsibility for SIPS compliance. This includes such challenges as engineers do not consider that they are responsible for compliance implementation~\cite{alhazmi2021m, peixoto2023perspective, rocha2023privacy, andrade2023personal}, it is unclear who is responsible for regulatory compliance~\cite{poth2021lean, mclachlan2021smart}, there is no clear approach to sharing responsibility with other roles or partner organizations~\cite{muller2022explainability, drabiak2022leveraging, thiele2021regulatory, li2022towards, ladkin2022assigning}, there is lack of investments~\cite{canedo2022guidelines,rocha2023privacy}. Other challenges in this category are, for example, as follows organizations or SIPS development teams do not prioritize compliance because of prioritization of functional requirements~\cite{alhazmi2021m}, compliance is considered as a necessary evil with little practical relevance~\cite{stirbu2021introducing}, or as not a mandatory concern~\cite{peixoto2023perspective}, or not a business concern~\cite{peixoto2023perspective}. Last but not least, a non-systematic approach to regulatory compliance is mentioned as a challenge in a few studies~\cite{ryan2022support,peixoto2023perspective}, including the absence and/or insufficiency of discussions of regulatory compliance within SIPS  teams and/or organization~\cite{alhirabi2022privacy, peixoto2023perspective}.

\paragraph*{Regulatory gaps (26/\added{255} studies (\added{10.2}\%))}  Some primary studies pointed to the regulatory gaps, i.e. absence of regulations applicable to SIPS. Such challenges contrast to abstractness challenges which point to the absence of concreteness of existing regulations. Some of the examples of challenges in this category are as follows absence of any regulations~\cite{yuba2022systematic}, absence of regulations for new technologies such as AI/ML-based systems~\cite{kuwajima2019adapting,zanca2022regulatory,kyhlstedt2022need,zinchenko2022methodology}, absence of regulations specifically targeting SIPS, but rather non-SIPS systems~\cite{granlund2020medical,kearney2021bridging}, demand for special consideration of some types of SIPS (e.g., software of unknown provenance~\cite{stirbu2021extending}), demand for more consistent and coordinated regulation~\cite{hernandez2021conflicting}. Also, one study mentioned that better involvement of regulators as stakeholders could be beneficial, but challenging~\cite{elliott2022know}.

\paragraph*{\added{Summary}} As we observed from the studies, many categories of challenges are closely related and our study has pointed to at least some of such relationships (e.g., challenges in developing PPs and demand for domain knowledge are caused by the absence of knowledge and expertise in domains involved in regulatory RE, domain knowledge related to the challenge of interaction between experts, complexity challenge leading to resource intensity, complexity as creating some of the provability challenges). However, we have not found any primary studies addressing such interrelationships. This emphasizes the importance of further systematic empirical research of challenges both for research and practical purposes.

Many of our findings align with earlier studies reporting regulatory compliance challenges. Next, we highlight some of such studies. The challenge of the conflicts between regulations and existing SE practices or \added{the} introduction of additional demands on them (like - explicit and structured conformance arguments) was, for example, identified by Graydon et al.~\cite{graydon2012arguing}. Kempe \& Massey~\cite{kempe2021regulatory}, in their secondary study, also identified general characteristics of regulations (e.g., ambiguity) as one of the top discussed challenges.

On the basis of the results of some of the empirical studies that we have identified, we can assume that challenges from different categories emerge simultaneously in practice. For example, Li et al.\cite{li2022towards} mentioned challenges belonging to 12 out of 14 challenge categories we have identified, and Peixoto et al.\cite{peixoto2023perspective} identified challenges belonging to 7 categories of challenges.

The major part of the research on challenges in the primary studies stays fragmented as studies often do not explore the essence of challenges and their interconnections. Only some studies have explored such challenges systematically but did not focus on them in detail (e.g., \cite{patwardhan2018towards,galvez_odyssey_2018, li_continuous_2019, wirtz2019risk}). Some studies reported challenges without the application of empirical methods (e.g., ~\cite{galvez_odyssey_2018,dalpiaz_risk-driven_2021,wirtz2019risk}. Only a few studies have rigorously identified multiple challenges with the application of empirical methods (e.g.,\cite{li2022towards, peixoto2023perspective}). However, the number of studies with high rigor and relevance that tried to investigate challenges with empirical methods (e.g., ~\cite{dias_canedo_perceptions_2020,madhavji_understanding_2020}) stays relatively low. Up to date, existing studies have not resulted in a systematic overview of challenges.
Further empirical evidence could be required to understand better such challenges as resource intensity and conflicts or changes caused by regulatory demands in SE practices. Kempe\&Massey~\cite{kempe2021regulatory}, in their secondary study, also expressed an opinion that technical implementation of regulatory compliance is related to the costs of development for the system as a whole. Nevertheless, in their systematic literature review, they did not identify any concrete quantitative assessments. We also found only scarce studies assessing the resources required to implement regulatory compliance of SIPS (e.g.,~\cite{dmitriev_lean_2020}). Studies exploring the impact of regulations on SE practices often do not employ a systematic approach that could be applied to other regulations or types of practices.

In our opinion, the absence of systematicity in the research of challenges potentially impacts the overall state of research on the topics of regulatory compliance of SIPS and regulatory RE. For example, early studies on the topic of regulatory compliance (that are outside of the scope of this study) suggested that regulations are often developed to be abstract intentionally~\cite{breaux2008analyzing} and suggested approaches to address unintended ambiguity~\cite{breaux2007systematic}. Our results confirm that abstractness remains a top challenge to regulatory compliance of SIPS until now.

\subsubsection{Primary studies in other disciplines}
Our selected primary studies also include studies belonging to the disciplines other than software engineering and computer science such as law~\cite{mourby2021transparency, ladkin2022assigning, mclachlan2021smart, thiele2021regulatory, price2017regulating}, compliance and regulation~\cite{culley2023insights}, medicine~\cite{kearney2021bridging, hernandez2021conflicting, zanca2022regulatory, muller2022explainability, van2022recommendations, niemiec2022will, drabiak2022leveraging, kyhlstedt2022need, zinchenko2022methodology, giordano2023medical}, risk regulation and management~\cite{ludvigsen2022software, almada2023regulation}. In one case the authors of one of the studies had a technical background, but the study was published in the legal venue~\cite{ladkin2022assigning}. Despite belonging to other disciplines these studies covered SIPS compliance and regulatory RE challenges in a way relevant to our study. For example, consider the integration of regulatory affairs professionals in software life cycle~\cite{kearney2021bridging}, post-market surveillance of software as a medical device~\cite{zinchenko2022methodology}. Challenges described in such studies fit into our existing categories of challenges. But often these challenges were described from a different perspective (e.g., legal studies considered the consequences of "hardcoding" compliance controls into SIPS and not updating them, such studies also approached regulatory compliance more ``holistically'' i.e. independently of particular SIPS lifecycle process areas). Medical studies emphasized a number of challenges specific to the use of AI/Ml-based SIPS which is related to the SIPS maintenance process areas.  Overall, challenges in the SIPS maintenance process area were under-considered in other primary studies. We consider that studies from other disciplines can contribute to regulatory RE and SIPS compliance and are important to consider. However, we suggest that a number of methodological issues should be addressed in the future to effectively review such studies and conduct interdisciplinary research on regulatory RE (e.g., studies in law have a different structure, different disciplines operate on different levels of abstraction, understanding of the expected results and contributions value vary in different disciplines, each discipline approach academic rigor differently, etc).

\subsection{RQ2: What is needed to address the challenges to the regulatory compliance of SIPS and what are the PPs used for that?}\label{RRQ2}

\begin{boxResults}
\textit{Highlights
\vspace{-2mm}
\begin{itemize}[leftmargin=*]
    \setlength\itemsep{2mm}
\item We have identified and categorized the principles and practices (six categories) used in the research to address different types of challenges of regulatory compliance of SIPS.
\item Many of the PPs addressed multiple challenges.
\item There is a lack of tool support for addressing the identified challenges.
\item Automation was utilized in 26.42\% of the primary studies to address challenges, with the most frequent application of automation seen in ``Conflicts/changes to existing practices'' and ``Complexity'' challenges.
\end{itemize}}
\end{boxResults}

The second research question aims to identify principles and practices (PPs) used to address the mentioned challenges and achieve SIPS regulatory compliance. As mentioned in Section~\ref{sub:dataext}, we used the categorization we have developed (see  Table~\ref{tab:pp}) to classify and analyze challenges. Table~\ref{tab:PPraw} shows the full list of six different types of PP and examples taken from the primary studies. \added{Some studies} did not develop any solution or PP but rather only identified challenges or envisioned \added{development and/or application of PPs in the future}. Since they are primarily suggestions rather than actual implemented approaches, we did not include them in the analysis for RQ2, as only implemented PPs addressing challenges were considered. For example, one study~\cite{madhavji_understanding_2020} investigates the personal factors affecting the developers’ understanding of privacy requirements during the vacancy period of a data protection law. Moreover, 23 studies suggested more than one type of PP. For example, one study~\cite{bujok_approach_2017} presented \textit{requirements} and \textit{methodology} as their solution types to address challenges. As a result, the total number of uniquely mentioned PPs in the studies is 24\added{9}. The most popular type of PP was \textit{methodology} with 69.47\% (17\added{3}/249). Creating \textit{tools} with 11.24\% (2\added{8}/249) and developing \textit{models} with 9.23\% (23/249) were the next most popular PP types for addressing the challenges. \textit{Patterns} 6.02\% (15/249) and \textit{requirements} 3.61\% (9/249) were among the least popular PP types. One study proposed \textit{metrics and measurements}. Full details are available in our published dataset.

\begin{table*} 
\footnotesize
\resizebox{\textwidth}{!}{%
\centering
\begin{tabular}{p{3cm}P{1.5cm}p{14.5cm}} \toprule  
	\textbf{Type of PP} & \textbf{\# of unique PPs} & \textbf{Exemplary Raw Text}
            \\\midrule
            Methodology & 173 & A semi-automatic methodology that assesses the security requirements of software systems concerning completeness and ambiguity, creating a bridge between the requirements documents and complying with standards.\cite{hayrapetian_empirically_2018} \\
             &  & A method to achieve Continuous Security Compliance by extending an agile process model to allow secure development and verifiable compliance with a standard.\cite{moyon2018towards} \\
             &  & A guideline summarized the references from the regulatory guide, codes, and standards that recommend the verification and validation methods that should be completed.\cite{maerani_developing_2021} \\      
             &  & A 6-step systematic approach (GuideMe) that supports the elicitation of solution requirements that link GDPR data protection obligations with the privacy controls that fulfill these obligations.\cite{ayala-rivera_grace_2018} \\ 
            Metrics and Measurements & 1 & Two metrics are proposed that can best measure the performance of each attribute, and an assessment of either agile tools, agile methods, or DevOps is provided as being best positioned to satisfy the regulated environment attributes of communications applications.\cite{wagner_metrics_2020-1} \\
            Model & 23 & Adopting the APDL model as a requirements model that can be expressed as a common language understood by those concerned with privacy and data protection and those responsible for developing and maintaining privacy-aware systems.\cite{alshammari_model-based_2018} \\
             &  & The definition of atomic expressions, as a prerequisite for creating the formal rules, can be considered a step towards a better interpretation of the standards requirements.\cite{ardila2017towards} \\
            Pattern & 15 & The pattern system guides software developers so that they can help users understand how their information system uses personal data.\cite{colesky_system_2018} \\
             &  & An architectural viewpoint for describing software architectures from a legal data protection perspective whose core modeling abstractions are based on an in-depth legal analysis of the EU General Data Protection Regulation.\cite{sion_architectural_2019} \\
            Requirements & 9 & Set of generic requirement-driven elements that can be applied to similar IoT-based architectures.\cite{antunes_operationalization_2020} \\
            Tool & 28 & The automated system, named HPDROID, bridges the semantic gap between the general rules of GDPR and the app implementations by identifying the data practices declared in the app privacy policy and the data-relevant behaviors in the app code.\cite{fan_empirical_2020} \\
    \bottomrule
	\end{tabular}
 }
 \caption{Six different types of PP with exemplary text from a primary study}
 \label{tab:PPraw}
\end{table*}

The PP types are not mutually exclusive, and we found that 5\added{0.2}\% (125/24\added{9}) of the primary studies address multiple challenges with their suggested PP. For example, one study~\cite{mayr-dorn_timetracer_2020} developed a tool for tackling the ``Absence of PPs'' and ``Dynamics of software systems'' challenges. Hence, the total number of times a PP addresses different challenges is 6\added{24}.

\begin{figure*} 
\scriptsize
    \centering
    \begin{tikzpicture}
    \begin{axis}[
    width=0.92\columnwidth,
    height=8cm,
    symbolic x coords = {NoChallenge, Regulatory, Organizational, SoftwareSystems, RegulationsMultiplicity, SoftwareContext, Enforcement, Provability, ResourceIntensity, Experts, Complexity, AbsenceOfPrinciples, DomainKnowledge, Conflicts, Abstractness},
    symbolic y coords = {Metrics, Requirements, Pattern, Model, Tool, Methodology},
    xticklabels={Regulatory gaps, Organizational challenges, Software dynamics, Multiple regulations, Context dynamics, Enforcement challenges, Provability, Resource intensity, Interaction with experts, Complexity, Absence of PPs, Domain knowledge, Changes to practices, Abstractness},
    yticklabels = {Metrics, Requirements, Pattern, Model, Tool, Methodology},
    x tick label style = {font = \scriptsize, text width = 4cm, align = right, rotate = 70, anchor = north east},
    y tick label style = {font = \scriptsize, text width = 2cm, align = right},
    xtick=data,
    ytick=data,
    xmajorgrids=true,
    ymajorgrids=true,
    grid style=dashed,
    axis x line*=bottom,
    axis y line*=left,
    ]
    \addplot[%
        scatter=true,
        only marks,
        point meta=\thisrow{color},
        fill opacity=0.1,text opacity=1,
        visualization depends on = {0.5*\thisrow{Val} \as \perpointmarksize},
        visualization depends on = {\thisrow{Val} \as \Val},
        scatter/@pre marker code/.append style={
        /tikz/mark size=\perpointmarksize, 
    },
        nodes near coords*={\contour{white}{$\pgfmathprintnumber\Val$}},
        nodes near coords style={text=black, font=\sffamily, font=\bfseries, font=\small, anchor=center}
    ] table [x={Group},y={Posttest}] {
Group  Posttest Val color

Regulatory	Metrics	0	0
Organizational	Metrics	0	0
SoftwareSystems	Metrics	1	0
RegulationsMultiplicity	Metrics	0	0
SoftwareContext	Metrics	0	0
Enforcement	Metrics	0	0
Provability	Metrics	0	0
ResourceIntensity	Metrics	0	0
Experts	Metrics	1	0
Complexity	Metrics	0	0
AbsenceOfPrinciples	Metrics	0	0
DomainKnowledge	Metrics	0	0
Conflicts	Metrics	1	0
Abstractness	Metrics	0	0

Regulatory	Requirements	0	0
Organizational	Requirements	0	0
SoftwareSystems	Requirements	0	0
RegulationsMultiplicity	Requirements	2	0
SoftwareContext	Requirements	0	0
Enforcement	Requirements	1	0
Provability	Requirements	0	0
ResourceIntensity	Requirements	0	0
Experts	Requirements	0	0
Complexity	Requirements	0	0
AbsenceOfPrinciples	Requirements	0	0
DomainKnowledge	Requirements	1	0
Conflicts	Requirements	1	0
Abstractness	Requirements	4	0

Regulatory	Pattern	0	0
Organizational	Pattern	1	0
SoftwareSystems	Pattern	1	0
RegulationsMultiplicity	Pattern	1	0
SoftwareContext	Pattern	2	0
Enforcement	Pattern	4	0
Provability	Pattern	2	0
ResourceIntensity	Pattern	5	0
Experts	Pattern	2	0
Complexity	Pattern	2	0
AbsenceOfPrinciples	Pattern	0	0
DomainKnowledge	Pattern	5	0
Conflicts	Pattern	5	0
Abstractness	Pattern	4	0

Regulatory	Model	0	0
Organizational	Model	0	0
SoftwareSystems	Model	1	0
RegulationsMultiplicity	Model	3	0
SoftwareContext	Model	1	0
Enforcement	Model	1	0
Provability	Model	4	0
ResourceIntensity	Model	7	0
Experts	Model	5	0
Complexity	Model	4	0
AbsenceOfPrinciples	Model	2	0
DomainKnowledge	Model	5	0
Conflicts	Model	1	0
Abstractness	Model	12	0

Regulatory	Tool	1	0
Organizational	Tool	6	0
SoftwareSystems	Tool	6	0
RegulationsMultiplicity	Tool	0	0
SoftwareContext	Tool	3	0
Enforcement	Tool	8	0
Provability	Tool	3	0
ResourceIntensity	Tool	7	0
Experts	Tool	7	0
Complexity	Tool	12	0
AbsenceOfPrinciples	Tool	14	0
DomainKnowledge	Tool	11	0
Conflicts	Tool	8	0
Abstractness	Tool	8	0

Regulatory	Methodology	14	0
Organizational	Methodology	14	0
SoftwareSystems	Methodology	15	0
RegulationsMultiplicity	Methodology	20	0
SoftwareContext	Methodology	19	0
Enforcement	Methodology	27	0
Provability	Methodology	29	0
ResourceIntensity	Methodology	35	0
Experts	Methodology	22	0
Complexity	Methodology	43	0
AbsenceOfPrinciples	Methodology	40	0
DomainKnowledge	Methodology	31	0
Conflicts	Methodology	52	0
Abstractness	Methodology	54	0
};
    \end{axis}
    \end{tikzpicture}
    \vspace{-15mm}\caption{Types of challenges mapped against the type of PPs}
    \label{fig:ppChall}
\end{figure*}
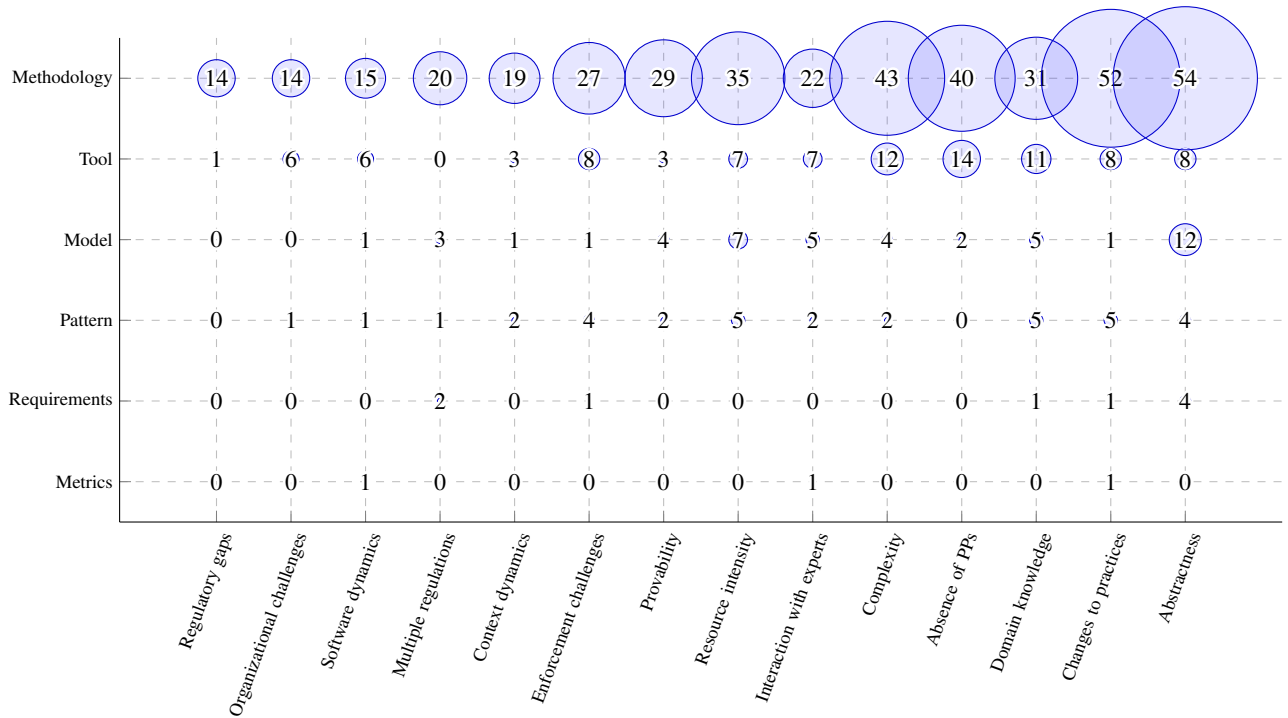

The bubble plot in Figure~\ref{fig:ppChall} shows the interplay of types of PP and the addressed challenge category. The X and Y axes have categorical values, producing a grid-like visualization. Another dimension, visualized as the size of each bubble, is the number of times when the intersection of the X and Y axes applies. For example, there are 5\added{2} times that a ``methodology'' was mentioned for addressing the ``conflicts/changes to existing practices'' challenge. 

We investigated to what degree the mentioned PPs are utilizing automation to address the challenges to the regulatory compliance of SIPS. Overall, 26.42\% (74/280) of the primary studies used some form of automation. With the proposed automation, several studies attempted to tackle multiple challenges. Consequently, automation was recommended to address challenges 88 times. The top two categories of challenges that were addressed by automation are ``Conflicts/changes to existing practices'' 18.18\% (16/88) and ``Complexity'' 14.77\% (13/88). ``Resource intensity'' and ``Absence of PPs'', each with 13.63\% (12/88) and 11.36\% (10/88), and ``Enforcement challenges'' and ``Abstractness'', with 10.22\% (9/88) and 7.95\% (7/88), are the next category of challenges that utilized automation. 

For example, Joshi et al.~\cite{joshi_integrated_2020} developed a comprehensive and semantically intricate knowledge graph encompassing a range of data compliance regulations. The knowledge graph is publicly available and can be used by enterprises to automate compliance workflows and formulate security protocols for their cloud-based operations. Zeni et al.~\cite{ZENI2018407} designed a semi-automated web-based system called GaiusT that automates the semantic annotation of legal documents. 

Additionally, as mentioned in Section~\ref{sub:dataext}, we classified the level of PP automation based on the categories suggested by Parasuraman et al.~\cite{Automation}. We found that 44.59\% (3\added{3}/74) of automation is at the \textit{information analysis} level, followed by 2\added{5}.67\% (1\added{9}/74) PPs that have automation at both \textit{information acquisition and information analysis} levels. Nine PPs include automation at \textit{information analysis and decision and action selection} while having only \textit{information acquisition} is 6.75\% (5/74). The combination of \textit{information acquisition, information analysis, decision and action selection, and action implementation} and only  \textit{decision and action selection} are each 4\% (3/74). Finally, two studies utilized automation at \textit{information acquisition, information analysis, and decision and action selection} level.\subsubsection{Discussion \nameref{RRQ2}}\label{sec:rq2Discussion}For some challenges like ``Domain knowledge and expert human resources'', there are different types of PP (e.g., models, tools, methodologies). Also, ten different tools exist to address this challenge. The problem space is not well analyzed, and the next step should involve a comparative evaluation of these PPs, examining the specific benefits and drawbacks of each approach in addressing the same challenge. This analysis would provide clearer guidance on which PP might be most effective in various contexts, aiding both researchers and practitioners in making informed decisions.

Moreover, looking at the results in Figure~\ref{fig:rr}, the number of studies in the ``relevance-scale dimension'' had a very low score, with 14.28\% (40/280). This suggests that the mentioned PPs were evaluated predominantly using
applications of unrealistic size (e.g., toy examples). This has two major implications. One, the basis for the developed/suggested PP might not be well anchored in cases in the industry, and two, and more importantly, the “solution” in the form of the PP was not tested out in a scalable and realistic manner. To bridge this gap, future research should focus on industry collaboration, where realistic examples and datasets are utilized to test the PPs. This approach would not only serve as a more rigorous test but also enhance industry confidence in adopting these PPs as viable solutions. Moreover, co-developing PPs with industry partners could lead to more practical and immediately applicable solutions.
Additionally, future research could focus on developing standardized frameworks or methodologies for evaluating PPs, ensuring that they are tested under comparable conditions, especially in terms of scalability and real-world applicability.

Finally, some categories of challenges are more widely covered by PPs. Looking at Figure~\ref{fig:ppChall}, ``Abstractness'' were addressed by a PP in 13.14\% (82/624) and ``conflicts/changes to existing practices'' in 10.89\% (68/624). However, challenges like ``organizational challenges'' 3.36\% (21/624) and ``regulatory gaps'' 2.4\% (15/624) are underrepresented by addressing PPs. Our analysis revealed a frequent lack of a clear linkage between the challenges identified in studies and the corresponding recommended PPs (for instance, studies that proposed various challenges did not explicitly state whether their suggested PP aims to address all the mentioned challenges or not). Future research should focus on establishing these linkages more explicitly, ensuring that proposed PPs are evaluated against all relevant challenges.
Lastly, as regulations evolve, so too must the PPs that support compliance. Future research could focus on developing adaptive PPs that can evolve in response to changes in regulations, ensuring continuous compliance without requiring extensive reworks. This would provide organizations with more flexible tools that can easily adapt to new regulatory requirements, reducing the cost and complexity of maintaining compliance.

\subsection{RQ3: Which stakeholders were involved in the development of PPs for regulatory compliance of SIPS?} \label{RRQ3}

\begin{boxResults}
\textit{Highlights
\vspace{-2mm}
\begin{itemize}[leftmargin=*]
\item Primary studies considered five stakeholders' categories: software engineers, legal experts, other experts, researchers, and other stakeholders.
\item 13.6\% of primary studies considered the involvement of both software engineering and legal experts.
\item The number of studies considering the involvement of stakeholders in the validation of principles and practices is relatively low.
\item The role of each category of stakeholders and the involvement of ``other stakeholders'' (e.g., suppliers, users) requires further research.
\end{itemize}}
\end{boxResults}

124/280 studies (44.3\%) considered the need for the involvement of one or more stakeholders in the development, application, and/or validation of suggested principles or practices. We extracted data about any mention of a need to involve certain stakeholders. The way in which the involvement of stakeholders was considered varied. In some cases, stakeholders were involved along with study authors; in other cases, their involvement was mentioned without any details, or their involvement was envisioned as future work.  For the purposes of answering this research question, we also counted the involvement in research activities, i.e., involvement of a role in research counts as participation in a PP development. We analyzed the data extracted from 124 primary studies mentioning the involvement of stakeholders and identified five main categories of stakeholders. As every study could consider the involvement of a few categories of stakeholders, we calculated the number of unique studies mentioning the involvement of each category of stakeholders one or more times.
90/124 (72.6\%) of studies considered the involvement of some SE roles (like developers, architects, requirements analysts, and/or product owners).
55/124 (44.4\%) studies considered the involvement of legal or compliance experts (compliance officers, personal data protection officers, regulators, and/or certification authorities). We also included privacy and personal data protection experts in this category as these roles are mainly related to compliance duties and mainly emerged after the enactment of GDPR. Noteworthy is that in some studies, regulators (e.g., certification authorities) were also involved in the development, application, and/or validation of PPs or corresponding research~\cite{anisetti_semi-automatic_2020, harrison_verification_2017,biscoglio2017certification}.
28/124 (22.6\%) studies considered the involvement of experts other than legal or compliance experts. Primarily, studies indicated the involvement of security and safety experts/engineers. In this context, it is important to mention that we assume that security and safety experts can also possess the knowledge and expertise required for compliance. Other experts involved were human factors specialists, process engineers, usability experts. 12/124 (9.7\%) studies considered the involvement of researchers. These were mainly computer science or SE researchers, but also researchers in linguistics and law were involved.
In \added{39}/124 (\added{31.5}\%) studies, authors considered the involvement of other stakeholders. Primarily, these were stakeholders external to the organization, such as suppliers, service providers, users, and data owners. In some cases also, internal stakeholders such as product management and sponsors were mentioned.
Among the 124 studies, \added{68} papers (\added{54.8}\%) mentioned the involvement of more than one category of stakeholders.

\begin{figure}
\centering
\begin{tikzpicture}
\begin{axis}[
    ybar stacked,
    bar width=15pt,
    height=0.9\linewidth,
    nodes near coords,
    y label style = {font = \scriptsize},
    x label style = {font = \scriptsize},
    y tick label style = {font = \scriptsize},
    axis x line*=bottom,
    axis y line*=left,
    enlargelimits=0.15,
    legend style={at={(0.65,-0.26)}, anchor=north west,legend columns=-1},
    nodes near coords style={font=\scriptsize},
    ylabel={Number of Primary Studies},
    symbolic x coords={Researchers, Other experts, Other stakeholders, Legal / compl. experts, Software engineers},
    xtick=data,
    x tick label style={rotate=45,anchor=east, font = \scriptsize},
    ]
\addplot+[ybar,fill=MyBlue,draw=none,text=white] plot coordinates {(Software engineers,35) (Legal / compl. experts,22) (Other experts,13) (Researchers,6) (Other stakeholders,16)};
\addplot+[ybar,fill=NextBlue,draw=none,text=black] plot coordinates {(Software engineers,52) (Legal / compl. experts,36) (Other experts,12) (Researchers,3) (Other stakeholders,19)};
\addplot+[ybar,fill=SecondBlue,draw=none,text=black] plot coordinates {(Software engineers,16) (Legal / compl. experts,9) (Other experts,8) (Researchers,3) (Other stakeholders,5)};
\legend{\scriptsize \strut dev.,
\scriptsize \strut appl.,
\scriptsize \strut valid.}
\end{axis}
\end{tikzpicture} \caption{Number of studies considering the involvement of stakeholders in principle or practice (1) development, (2) application, (3) validation. Note: each stakeholder can be involved in any of the three principle or practice life cycle stages simultaneously. The categories are sorted from the least mentioned on the left to the most mentioned on the right.} \label{fig:stakeholdersPhase}
\end{figure}
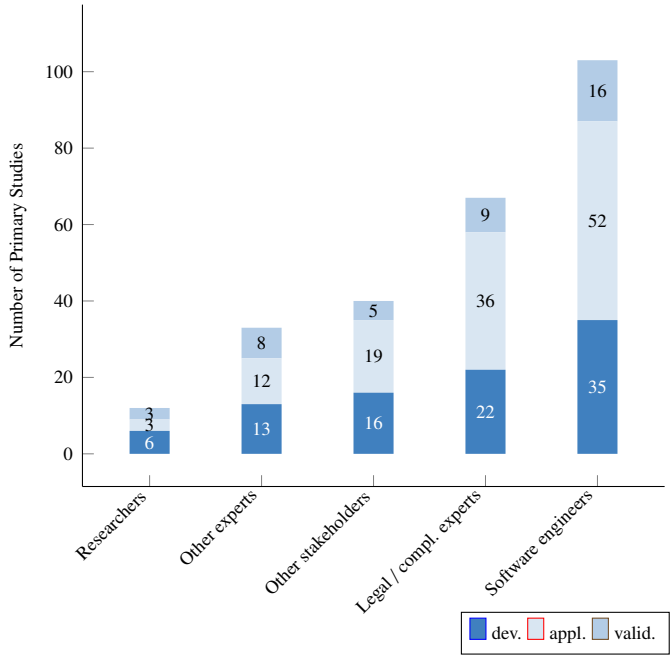

While extracting data from the primary studies, we have also observed that the involvement of stakeholders is usually considered for a particular phase of the PPs life cycle, such as the development, application, and validation of PP. To account for this difference, We also collected data about the involvement of stakeholders in any of these phases. Based on the information about stakeholder involvement reported in the paper, we assigned one of the three categories of PPs for each stakeholder considered in the paper. Figure~\ref{fig:stakeholdersPhase} illustrates the involvement of a particular category of stakeholders in a particular phase of the principles or practice life cycle. Some studies suggested the need for the involvement of the same category of stakeholders in different stages of the PP life cycle. \added{However,} the number of such studies was insignificant nine primary studies out of 124 (8.1\%)). Three primary studies~\cite{kellogg2020continuous, moyon_how_2020, hubbs2023automating} suggested that involvement of stakeholders is required simultaneously to application and validation of PPs, one study~\cite{sartoli2020towards} suggested that involvement is required to PP development and application, and four studies~\cite{marques2019set, gharib_copri_2021, alhirabi2023parrot, amalfitano2023documenting} suggested to involve stakeholders in development and validation of PPs). Only two studies~\cite{zeni_annotating_2017, alhirabi2023parrot} mentioned that the involvement of the same category of stakeholders is required for all three stages of the PP life cycle. This study also mentioned the involvement of multiple stakeholder categories: SE roles (in particular requirements analysts), legal experts (lawyers), other experts (domain experts), and researchers (researchers on linguistics).

\subsubsection{Discussion \nameref{RRQ3}}
Our findings go in hand with the previous study by Mubarkoot et al.~\cite{mubarkoot2023software} that identified that among all the primary studies they identified various stakeholders including legal experts, domain experts, and some external stakeholders. Nevertheless, their results show that domain experts, legal experts, and safety engineers were the least considered stakeholders. Our findings are similar to the secondary study of Akhigbe et al.~\cite{goal}, who found that concerns of regulators as stakeholders are also considered in about 12\% of studies. Earlier primary studies that are out of the scope of this study also pointed to the need to involve legal experts (e.g.,~\cite{massey2010evaluating}) or describe regulatory compliance implementation as an interdisciplinary activity (e.g.,~\cite{ojameruaye2014systematic}). Still, the number of studies that pointed to the involvement of both legal experts and software engineers in the PP life cycle stays relatively low, with 38 out of all 280 studies (13.6\%). The involvement of other experts and stakeholders on par with legal experts was an important discovery. The involvement of other experts (e.g., domain, security experts) was mentioned in some of the earlier primary studies known to the authors of this study (e.g.,~\cite{massey2010evaluating}) but did not receive significant attention. The involvement of other stakeholders, such as users and suppliers, seems to be relevant, taking into account that regulations can impact their interests. Still, to the best knowledge of the authors, this aspect of the implementation of regulatory compliance of SIPS was not explored in-depth in any previous studies. Future research could better explore the roles of stakeholders involved. For example, Ayala-Rivera\&Pasquale~\cite{ayala-rivera_grace_2018} envisioned the involvement of \added{stakeholders belonging to all five groups of stakeholders that we have identified} (IT professionals, data privacy experts, legal experts, governance, data processors) but did not provide any details about the role of all of the stakeholders.
According to the number of primary studies that have considered the involvement of researchers, we can assume that interdisciplinary research stays relatively limited (primary studies considered the involvement of researchers belonging to disciplines other than the discipline of the primary study).
The primary studies mainly indicated that stakeholders need to be involved in PP application. Less attention was paid to the involvement of stakeholders in the development of PPs. Furthermore, the involvement of all types of stakeholders in the validation of PPs was the least considered.
For a deeper analysis of stakeholder involvement, we conducted a mapping between the data on stakeholder involvement and (1) RE-related challenges to compliance (described in Section~\ref{RRQ1}), (2) PPs for regulatory compliance of SIPS (described in Section~\ref{RRQ2}), (3) fields of regulation (described in Section~\ref{RRQ5}).

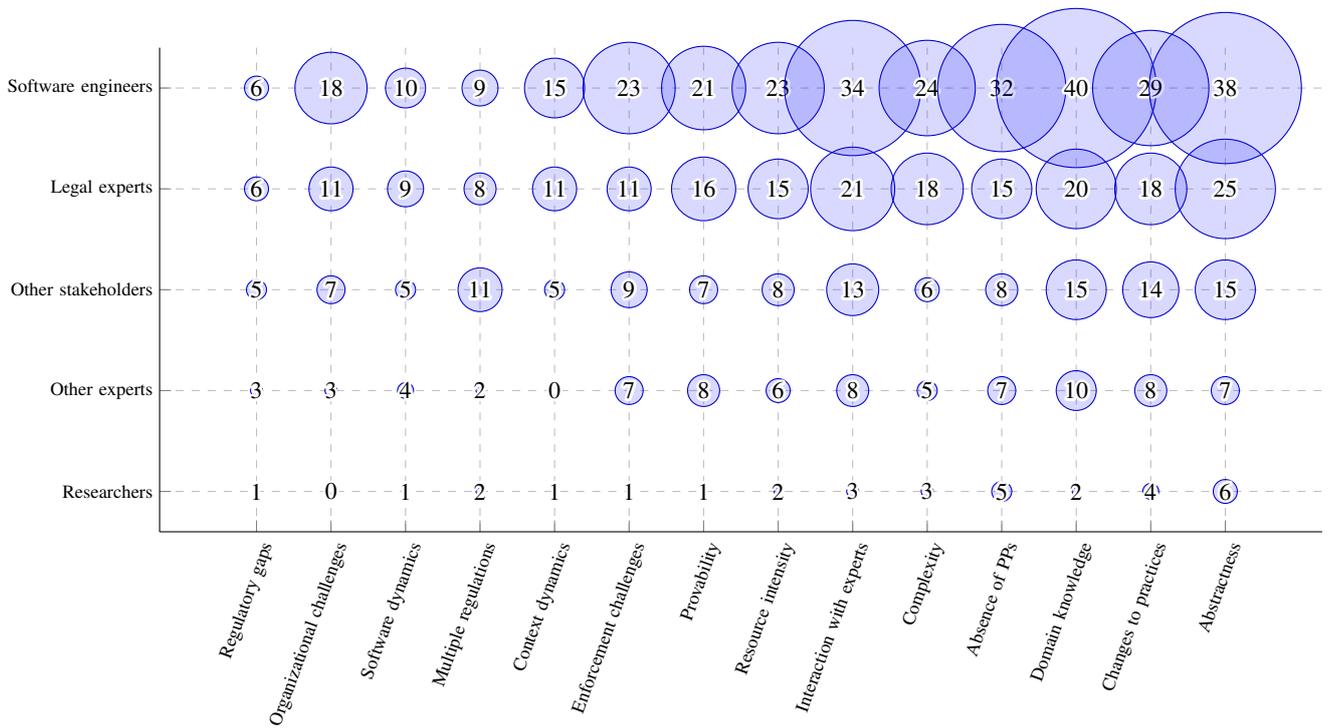
\begin{figure*}
\scriptsize
    \centering
    \begin{tikzpicture}
    \begin{axis}[
    width=0.92\columnwidth,
    height=8cm,
    symbolic x coords = {Regulatory, Organizational, SoftwareSystems, RegulationsMultiplicity, SoftwareContext, Enforcement, Provability, ResourceIntensity, Experts, Complexity, AbsenceOfPrinciples, DomainKnowledge, Conflicts, Abstractness},
    symbolic y coords = {researchers, otherExperts, otherStakeholders, LegalOrComplianceExperts, sotwareEngineering},
    xticklabels={Regulatory gaps, Organizational challenges, Software dynamics, Multiple regulations, Context dynamics, Enforcement challenges, Provability, Resource intensity, Interaction with experts, Complexity, Absence of PPs, Domain knowledge, Changes to practices, Abstractness},
    yticklabels = {Researchers, Other experts, Other stakeholders, Legal experts, Software engineers},
    x tick label style = {font = \scriptsize, text width = 4cm, align = right, rotate = 70, anchor = north east},
    y tick label style = {font = \scriptsize, text width = 2cm, align = right},
    xtick=data,
    ytick=data,
    xmajorgrids=true,
    ymajorgrids=true,
    grid style=dashed,
    axis x line*=bottom,
    axis y line*=left,
    ]
    \addplot[%
        scatter=true,
        only marks,
        point meta=\thisrow{color},
        fill opacity=0.15,text opacity=1,
        visualization depends on = {0.75*\thisrow{Val} \as \perpointmarksize},
        visualization depends on = {\thisrow{Val} \as \Val},
        scatter/@pre marker code/.append style={
        /tikz/mark size=\perpointmarksize
        },
        nodes near coords*={\contour{white}{$\pgfmathprintnumber\Val$}},
        nodes near coords style={text=black,font=\sffamily, font=\bfseries, font=\small, anchor=center},
    ] table [x={Group},y={Posttest}] {
Group  Posttest Val color
Regulatory	researchers	1	0
Organizational	researchers	0	0
SoftwareSystems	researchers	1	0
RegulationsMultiplicity	researchers	2	0
SoftwareContext	researchers	1	0
Enforcement	researchers	1	0
Provability	researchers	1	0
ResourceIntensity	researchers	2	0
Experts	researchers	3	0
Complexity	researchers	3	0
AbsenceOfPrinciples	researchers	5	0
DomainKnowledge	researchers	2	0
Conflicts	researchers	4	0
Abstractness	researchers	6	0

Regulatory	otherExperts	3	0
Organizational	otherExperts	3	0
SoftwareSystems	otherExperts	4	0
RegulationsMultiplicity	otherExperts	2	0
SoftwareContext	otherExperts	0	0
Enforcement	otherExperts	7	0
Provability	otherExperts	8	0
ResourceIntensity	otherExperts	6	0
Experts	otherExperts	8	0
Complexity	otherExperts	5	0
AbsenceOfPrinciples	otherExperts	7	0
DomainKnowledge	otherExperts	10	0
Conflicts	otherExperts	8	0
Abstractness	otherExperts	7	0

Regulatory	otherStakeholders	5	0
Organizational	otherStakeholders	7	0
SoftwareSystems	otherStakeholders	5	0
RegulationsMultiplicity	otherStakeholders	11	0
SoftwareContext	otherStakeholders	5	0
Enforcement	otherStakeholders	9	0
Provability	otherStakeholders	7	0
ResourceIntensity	otherStakeholders	8	0
Experts	otherStakeholders	13	0
Complexity	otherStakeholders	6	0
AbsenceOfPrinciples	otherStakeholders	8	0
DomainKnowledge	otherStakeholders	15	0
Conflicts	otherStakeholders	14	0			
Abstractness	otherStakeholders	15	0

Regulatory	LegalOrComplianceExperts	6	0
Organizational	LegalOrComplianceExperts	11	0
SoftwareSystems	LegalOrComplianceExperts	9	0
RegulationsMultiplicity	LegalOrComplianceExperts	8	0
SoftwareContext	LegalOrComplianceExperts	11	0
Enforcement	LegalOrComplianceExperts	11	0
Provability	LegalOrComplianceExperts	16	0
ResourceIntensity	LegalOrComplianceExperts	15	0
Experts	LegalOrComplianceExperts	21	0
Complexity	LegalOrComplianceExperts	18	0
AbsenceOfPrinciples	LegalOrComplianceExperts	15	0
DomainKnowledge	LegalOrComplianceExperts	20	0
Conflicts	LegalOrComplianceExperts	18	0
Abstractness	LegalOrComplianceExperts	25	0

Regulatory	sotwareEngineering	6	0
Organizational	sotwareEngineering	18	0
SoftwareSystems	sotwareEngineering	10	0
RegulationsMultiplicity	sotwareEngineering	9	0
SoftwareContext	sotwareEngineering	15	0
Enforcement	sotwareEngineering	23	0
Provability	sotwareEngineering	21	0
ResourceIntensity	sotwareEngineering	23	0
Experts	sotwareEngineering	34	0
Complexity	sotwareEngineering	24	0
AbsenceOfPrinciples	sotwareEngineering	32	0
DomainKnowledge	sotwareEngineering	40	0
Conflicts	sotwareEngineering	29	0
Abstractness	sotwareEngineering	38	0
};
    \end{axis}
    \end{tikzpicture}
    \vspace{-15mm}\caption{Categories of challenges that were mentioned in the papers (from least mentioned on the left to the most mentioned on the right except for ``No challenges'') mapped against categories of stakeholders involved (from the least mentioned on the bottom to the most mentioned on the top)}
    \label{fig:stakeholdersChallenges}
\end{figure*}

From the mapping of categories of challenges against categories of involved stakeholders, shown in Figure~\ref{fig:stakeholdersChallenges}, we can observe the categories of challenges in which stakeholders may need to be involved. Three categories of stakeholders (software engineers, legal experts, other experts, and other stakeholders) were primarily involved in studies that mentioned the following challenges: abstractness of regulations, domain knowledge and expert human resources, absence of PPs, interaction between experts, conflicts with existing practices, and resource intensity. Legal experts only were mainly involved in studies considering the abstractness of regulations, interaction with experts, domain knowledge, and conflicts with SE practices. It is noteworthy that both SE roles and legal experts were considered for involvement almost equally in studies considering the following challenges: dynamics of software context (SE roles: 15 and legal experts: 11), regulations' multiplicity (SE roles: 9 and legal experts: 8), dynamics of SIPS systems (SE roles: 10 and legal experts 9). The involvement of other types of experts was mentioned in studies reporting domain knowledge, conflicts with software engineering PPs, interaction with experts, and provability. The involvement of other stakeholders was also noted in publications reporting the challenges of abstractness, conflicts with SE PPs, domain knowledge, and interaction with experts. Studies reported that all categories of stakeholders should be involved in connection to such challenges as such challenges as abstractness, absence of PPs, changes to SE practices, interaction with experts, and complexity.
It is also noteworthy that despite the challenges of regulations' multiplicity, dynamics of systems, organizational challenges, and regulatory gaps are among the least considered in the primary studies, these studies relatively often mention the involvement of SE roles, legal experts, other experts, and other stakeholders.
Last but not least, the role of experts other than legal and compliance experts seems to be important in addressing domain knowledge-related challenges to compliance of SIPS. Overall, we see that all categories of stakeholders except for researchers and other experts were considered for involvement across all categories of challenges. Still, there are some differences between the involvement of stakeholders in studies mentioning different categories of challenges.

\begin{figure*}
\scriptsize
    \centering
    \begin{tikzpicture}
    \begin{axis}[
    width=0.92\columnwidth,
    height=6.5cm,
    symbolic x coords = {TrafficLaw,Accessibility,UserRights,Business,PrivacySecurity,AIML,Security,Safety,Quality,Privacy},
    symbolic y coords = {researchers, otherExperts, otherStakeholders, LegalOrComplianceExperts, sotwareEngineering},
    xticklabels={Traffic Law, Accessibility,User Rights, Business, Privacy Security, AI/ML, Security, Safety, Quality, Privacy},
    yticklabels = {Researchers, Other experts, Other stakeholders, Legal experts, Software engineers},
    x tick label style = {font = \scriptsize, text width = 4cm, align = right, rotate = 70, anchor = north east},
    y tick label style = {font = \scriptsize, text width = 2cm, align = right},
    xtick=data,
    ytick=data,
    xmajorgrids=true,
    ymajorgrids=true,
    grid style=dashed,
    axis x line*=bottom,
    axis y line*=left,
    ]
    \addplot[%
        scatter=true,
        only marks,
        point meta=\thisrow{color},
        fill opacity=0.15,text opacity=1,
        visualization depends on = {0.75*\thisrow{Val} \as \perpointmarksize},
        visualization depends on = {\thisrow{Val} \as \Val},
        scatter/@pre marker code/.append style={
        /tikz/mark size=\perpointmarksize
        },
        nodes near coords*={\contour{white}{$\pgfmathprintnumber\Val$}},
        nodes near coords style={text=black,font=\sffamily, font=\bfseries,
        font=\small,
        anchor=center},
    ] table [x={Group},y={Posttest}] {

Group  Posttest Val color
TrafficLaw	researchers	0	0
Accessibility	researchers	0	0
UserRights	researchers	0	0
Business	researchers	0	0
PrivacySecurity	researchers	2	0
AIML	researchers	1	0
Security	researchers	1	0
Safety	researchers	2	0
Quality	researchers	2	0
Privacy	researchers	5	0

TrafficLaw	otherExperts	0	0
Accessibility	otherExperts	0	0
UserRights	otherExperts	1	0
Business	otherExperts	2	0
PrivacySecurity	otherExperts	2	0
AIML	otherExperts	3	0
Security	otherExperts	8	0
Safety	otherExperts	3	0
Quality	otherExperts	7	0
Privacy	otherExperts	6	0

TrafficLaw	otherStakeholders	0	0
Accessibility	otherStakeholders	0	0
UserRights	otherStakeholders	0	0
Business	otherStakeholders	2	0
PrivacySecurity	otherStakeholders	3	0
AIML	otherStakeholders	2	0
Security	otherStakeholders	6	0
Safety	otherStakeholders	5	0
Quality	otherStakeholders	8	0
Privacy	otherStakeholders	19	0

TrafficLaw	LegalOrComplianceExperts	0	0
Accessibility	LegalOrComplianceExperts	0	0
UserRights	LegalOrComplianceExperts	1	0
Business	LegalOrComplianceExperts	3	0
PrivacySecurity	LegalOrComplianceExperts	2	0
AIML	LegalOrComplianceExperts	2	0
Security	LegalOrComplianceExperts	8	0
Safety	LegalOrComplianceExperts	7	0
Quality	LegalOrComplianceExperts	9	0
Privacy	LegalOrComplianceExperts	28	0

TrafficLaw	sotwareEngineering	0	0
Accessibility	sotwareEngineering	1	0
UserRights	sotwareEngineering	1	0
Business	sotwareEngineering	3	0
PrivacySecurity	sotwareEngineering	7	0
AIML	sotwareEngineering	1	0
Security	sotwareEngineering	10	0
Safety	sotwareEngineering	8	0
Quality	sotwareEngineering	16	0
Privacy	sotwareEngineering	50	0

};
    \end{axis}
    \end{tikzpicture}
    \vspace{-25mm}
    \caption{\added{Fields of regulation that were addressed in the papers mapped against categories of stakeholders involved}}
    \label{fig:stakeholdersFields}
\end{figure*}
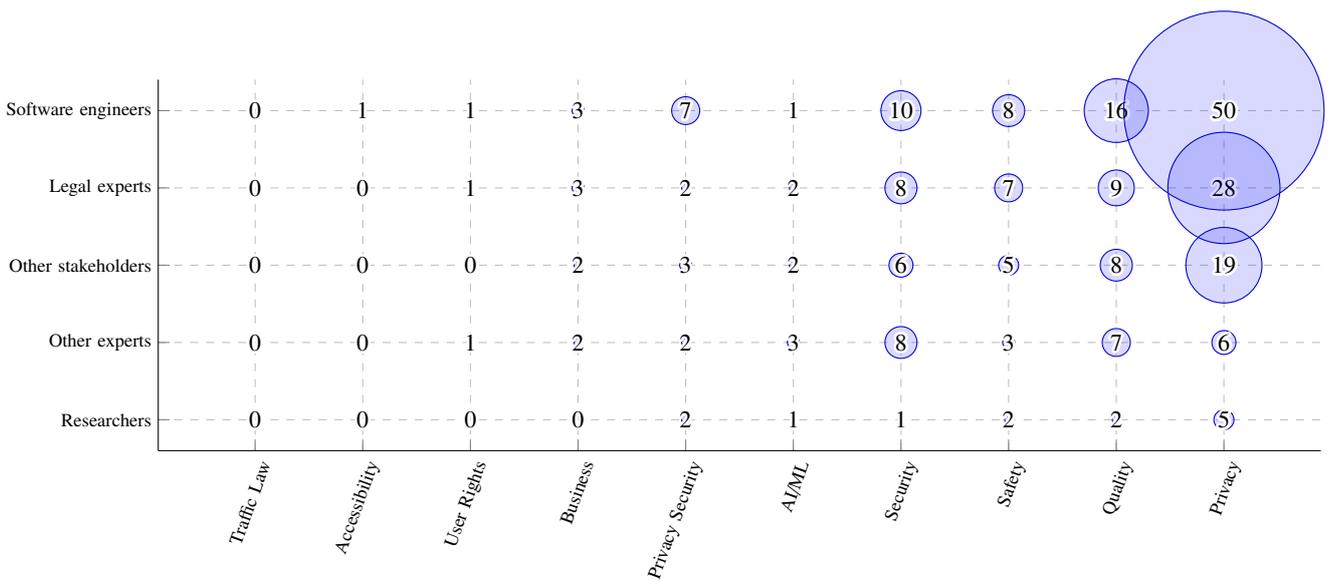

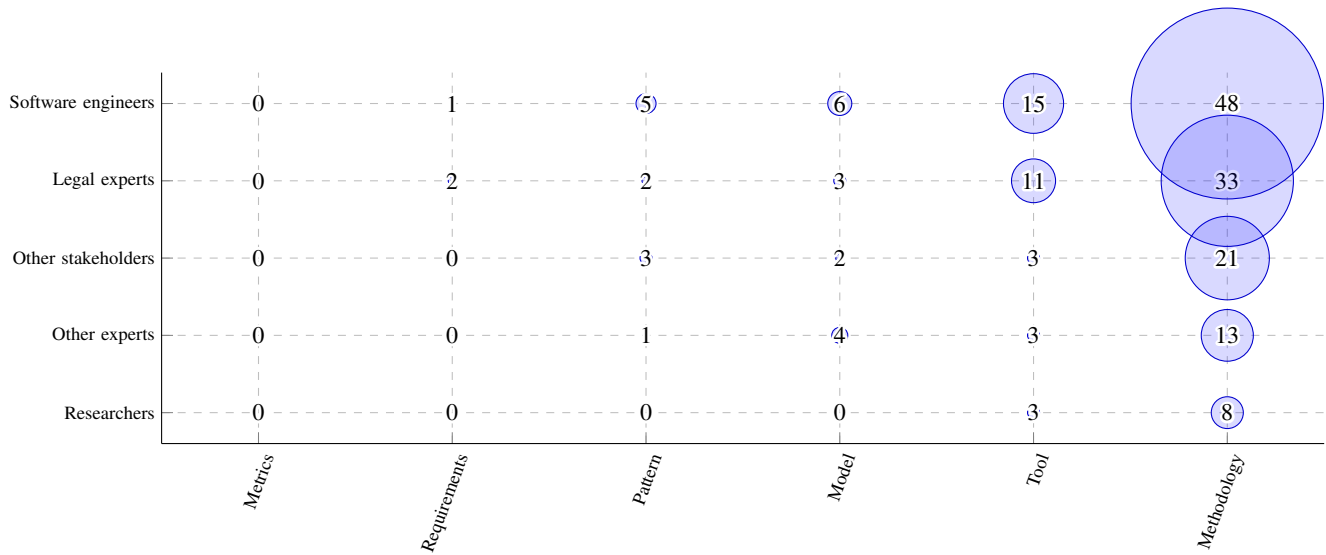
\begin{figure*} 
\scriptsize
    \centering
    \begin{tikzpicture}
    \begin{axis}[
    width=0.92\columnwidth,
    height=6.5cm,
    symbolic x coords = {Metrics, Requirements, Pattern, Model, Tool, Methodology},
    symbolic y coords = {researchers, otherExperts, otherStakeholders, LegalOrComplianceExperts, sotwareEngineering},
    xticklabels={Metrics, Requirements, Pattern, Model, Tool, Methodology},
    yticklabels = {Researchers, Other experts, Other stakeholders, Legal experts, Software engineers},
    x tick label style = {font = \scriptsize, text width = 4cm, align = right, rotate = 70, anchor = north east},
    y tick label style = {font = \scriptsize, text width = 2cm, align = right},
    xtick=data,
    ytick=data,
    xmajorgrids=true,
    ymajorgrids=true,
    grid style=dashed,
    axis x line*=bottom,
    axis y line*=left,
    ]
    \addplot[%
        scatter=true,
        only marks,
        point meta=\thisrow{color},
        fill opacity=0.15,text opacity=1,
        visualization depends on = {0.75*\thisrow{Val} \as \perpointmarksize},
        visualization depends on = {\thisrow{Val} \as \Val},
        scatter/@pre marker code/.append style={
        /tikz/mark size=\perpointmarksize
        },
        nodes near coords*={\contour{white}{$\pgfmathprintnumber\Val$}},
        nodes near coords style={text=black,font=\sffamily, font=\bfseries,
        font=\small, anchor=center},
    ] table [x={Group},y={Posttest}] {
Group  Posttest Val color
Metrics	researchers	0	0
Requirements	researchers	0	0
Pattern	researchers	0	0
Model	researchers	0	0
Tool	researchers	3	0
Methodology	researchers	8	0

Metrics	otherExperts	0	0
Requirements	otherExperts	0	0
Pattern	otherExperts	1	0
Model	otherExperts	4	0
Tool	otherExperts	3	0
Methodology	otherExperts	13	0

Metrics	otherStakeholders	0	0
Requirements	otherStakeholders	0	0
Pattern	otherStakeholders	3	0
Model	otherStakeholders	2	0
Tool	otherStakeholders	3	0
Methodology	otherStakeholders	21	0

Metrics	LegalOrComplianceExperts	0	0
Requirements	LegalOrComplianceExperts	2	0
Pattern	LegalOrComplianceExperts	2	0
Model	LegalOrComplianceExperts	3	0
Tool	LegalOrComplianceExperts	11	0
Methodology	LegalOrComplianceExperts	33	0

Metrics	sotwareEngineering	0	0
Requirements	sotwareEngineering	1	0
Pattern	sotwareEngineering	5	0
Model	sotwareEngineering	6	0
Tool	sotwareEngineering	15	0
Methodology	sotwareEngineering	48	0
};
    \end{axis}
    \end{tikzpicture}
    \vspace{-25mm}\caption{\added{Principles and practices suggested in studies mapped against categories of stakeholders involved}}
    \label{fig:ppsStakeholders}
\end{figure*}

Figure~\ref{fig:stakeholdersFields} maps the involvement of stakeholders concerning the fields of regulation that studies reported. SE roles, legal experts, and other stakeholders were most often mentioned in studies focusing on privacy \added{and quality}. Other experts were most often considered in studies focusing on security and quality as fields of regulation. We also observed that SE roles, legal experts, other stakeholders \added{and other experts} were considered for involvement in \added{seven} fields of regulation (Privacy, Quality, Safety, Security, \added{AI/ML, Privacy\&Security, and Business}). \added{Researchers were most often considered for involvement in studies considering privacy.}

Figure~\ref{fig:ppsStakeholders} maps the involvement of stakeholders concerning the category of PPs considered in primary studies. We observed that the involvement of software engineers and legal experts is considered for almost all PPs. The involvement of other experts \added{and other stakeholders} was considered for developing methodologies, tools, models \added{and patterns}, but not requirements. The only study that considered the development of metrics and measurements did not consider the involvement of any stakeholders.

\subsection{RQ4: What are the main software processes areas (PAs) involved in enabling the regulatory compliance of SIPS?}\label{RRQ4}

\begin{boxResults}
\textit{Highlights
\vspace{-2mm}
\begin{itemize}[leftmargin=*]
    \setlength\itemsep{2mm}
    \item Primary studies reported RE-related challenges to regulatory compliance of SIPS throughout almost all process areas of the SDLC.
    \item Requirements engineering as s standalone process area was considered in 74/\added{254} studies (29.1\%). In \added{58}/\added{254} studies (\added{22.8}\%) primary studies RE was considered in intersection with other process areas mainly SQA (19 studies) and SD (17 studies).
\end{itemize}}
\end{boxResults}

\begin{table*}
    \begin{minipage}{.45\textwidth}
      \centering
      \scriptsize
      \begin{tabular}{p{4.5cm}P{1.75cm}l}\\
      \hline
    \thead{Process \\ area(s)} & \thead{Primary \\ studies} \\
    \hline
RE*          			& 74\\
SDLC		            & 51\\
SQA             	    & 20\\
RE* \& SQA         	    & 19\\
SD                	    & 17\\
RE* \& SD          	    & 17\\
SDev              	    & 16\\
SQA \& SM         	    & 5\\
RE* \& SD \& SDev  	    & 4\\
RE* \& SD \& SQA   	    & 4\\
RE* \& SDev        	    & 3\\
RE* \& SDep        	    & 3\\
SD \& SDev        	    & 3\\
RE* \& SM          	    & 2\\
RE* \& SD \& SQA \& SM   & 2\\
SDev \& SDep      	    & 2\\
SM                	    & 2\\
RE* \& SD \& SDep 	    & 1\\
RE* \& SDev \& SDep	    & 1\\
RE* \& SDev \& SM 	    & 1\\
RE* \& SQA \& SM 	    & 1\\
        \bottomrule
      \end{tabular}
      \end{minipage}
      \quad\quad\quad\quad
    \begin{minipage}{.4\textwidth}
      \centering
      \scriptsize
    \begin{tabular}{p{4.5cm}P{1.75cm}l}\\
    \hline
    \thead{Process \\ area(s)} & \thead{Primary \\ studies} \\
    \hline
REA/M           & 33 \\
RE              & 30 \\
REE \& REA/M        & 19 \\
REE             & 18 \\
REVV            & 10 \\
RES             & 6 \\
REA/M \& REVV   & 5\\
REA/M \& RES    & 3\\
REA/M \& RES \& REVV \& REM & 2\\
REE \& RES      & 2\\
REE \& REA/M \& RES & 2\\
REE \& REVV       & 1\\
REE \& REA/M \& RES \& REVV & 1\\
        \bottomrule
      \end{tabular}
      \end{minipage}
    \caption{(1) Number of primary studies contributing to particular process area of software development life cycle (mutually exclusive). \emph{Process areas are as follows: RE - Requirements Engineering, SD - Design, SDev - Development, SQA - Quality Assurance, SDep - deployment, SM - Maintenance, SDLC - whole life cycle}.\\(2) Number of primary studies considering particular RE (including studies focusing on RE only or RE in conjunction with other process areas). \emph{RE processes are as follows: REE - Elicitation, REA/M - Analysis/Modeling, RES - Specification, REVV - Verification\&Validation, REM - Management}.}\label{tab:processAreas}
\end{table*}

We identified a particular process area for \added{254} out of 280 studies (\added{90.7}\%). We identified PAs based on the PA explicitly mentioned in the paper. In case the paper did not mention concrete PA, we followed definitions of PAs provided in Section~\ref{sec:terms} to identify the corresponding PA. Studies could mention a few PAs simultaneously. For example, the category ``RE \& SQA'' includes studies that considered contributions to both of these PAs simultaneously. A detailed breakdown of studies according to the PAs they considered can be found in Table~\ref{tab:processAreas}.
Studies have contributed to different PAs with a major focus on RE (74/\added{254} (29\added{.1}\%)), SDLC (51/\added{254} (20\%)), SQA (20/\added{254} (7.9\%)), SD (17/\added{254} (6.7\%)), SDev (16/\added{254} (6.3\%)), and SM (2/\added{254} (0.8\%)). The whole SDLC was mainly addressed in the context of application of particular development methods (e.g., agile~\cite{ozcan-top_what_2019,al-momani_privacy-aware_2019,wagner_metrics_2020-1, dmitriev_lean_2020, schidek2022agilization}, Unified Software Development Process~\cite{gomez2021towards}), implementation of compliance controls throughout the SDLC (e.g., traceability~\cite{stirbu2021introducing}), systematic integration of Legal/compliance or other experts into SDLC (e.g.,~\cite{kearney2021bridging, zanca2022regulatory}), the implementation of model-based SE or application of formal methods~\cite{jha2017adopting, bahig_formal_2017, ahmed_formal_2018, adedjouma_model-based_2018}, or implementation of assessment methods. Also, consideration of the whole SDLC was characteristic for many research studies as they usually were not focusing on a particular SDLC process area (e.g.,~\cite{canedo2022guidelines}).

We decided to additionally analyze the studies that considered RE (both as a standalone process area (74 studies) and in conjunction with other SDLC process areas (5\added{8} studies)) (see Table~\ref{tab:processAreas} for general trends).

Four standalone RE processes most considered in the primary studies are Analysis/Modeling (3\added{3} studies), RE in general (\added{30} studies), Elicitation (18 studies), and Validation and Verification (10 studies). Requirements Specification was considered in \added{6} studies and no primary study focused on Requirements Management as a standalone RE process. In the general RE category we have aggregated both primary studies that have considered all RE processes (mainly in the context of integration into software development methods (e.g., agile~\cite{huth2020process, barbosa_re4ch_2018, campanile2022towards})), considered RE on a general level, and primary studies for which it was not possible to identify concrete RE processes. In many cases it was hard to identify concrete RE process as studies simply provided a list of requirements derived in an unclear, ad hoc way and/or derived on the basis of the authors' experience~\cite{farhadi_static_2018, kunz_edge_2020, li_carenet_2017, li2022towards}. For example,~\cite{farhadi_static_2018} have provided a list of requirements or ``technical safeguards'' which was then applied for evaluating open source applications. Herewith, the authors have not explained how they derived these requirements, nor they have suggested any other principle or practice for RE specifically.
It is noteworthy that among the categories involving two or more RE processes \added{simultaneously} only the Elicitation and Analysis/Modeling category was considered in multiple studies (19 studies), while all the other categories \added{were considered in five or fewer primary studies}.

\subsubsection{Discussion \nameref{RRQ4}}
Our finding that RE is the main process area contributing to regulatory compliance is consistent with the findings in some of the previous studies~\cite{kempe2021regulatory}. Our results diverge from the results of Negri-Ribalta et al.~\cite{negri2024understanding} as in their study \added{focusing on General Data Protection Regulation (GDPR)} authors found out that the Requirements Elicitation process was addressed in 47 studies, Specification in 35 papers, and Verification in 33 studies. This difference could emerge because the authors have not defined Analysis/Modeling as a separate RE process and used only five processes in their secondary study (namely Elicitation, Specification, Verification and Validation, Management, Documentation)).

SM received insignificant attention in primary studies despite maintaining compliance due to changes in regulations is recognized to be challenging. Also, integrating software into an existing IT landscape of organizations can be another aspect requiring attention in the future. As our results show SIPS deployment (which includes DevOps) was considered in only some primary studies, but we can expect an increase of this number in the near future.
Some studies considered mainly SQA or SD process in conjunction with RE activities. The number of such studies considering SQA and SD in conjunction with RE activities was almost equal to the number of priamry studies considering SQA or SD without RE contribution. It is noteworthy that only a few studies considered SDev in conjunction with RE~\cite{morisio_integration_2020, farhadi2019compliance}.
\added{In the primary studies considering the RE and SD phases in conjunction,} RE was used to support regulatory compliance in SD in multiple ways. For example, through the identification of architecture components~\cite{antunes_operationalization_2020} and design principles~\cite{sousa2018openehr}. In the primary studies considering both RE and SQA in conjunction, RE was used to support SQA mainly by identifying requirements that need QA (e.g., in ~\cite{farhadi_static_2018}).
The contribution of the RE process area to other process areas in the context of regulatory compliance was somewhat addressed in~\cite{larrucea2017supporting} on the example of contribution to SQA. This study stated that developing SQA methods starts with extracting requirements and their modeling. In~\cite{laukkarinen_regulated_2018}, a tighter connection between RE and SDev was also described as essential for regulatory-compliant development. This is in line with some of the previous studies that suggested that process areas beyond RE can have specific needs that should be addressed by the RE process area~\cite{kempe2021regulatory}. For example, SQA can require guidance on the nature of the required evidence, their sufficiency, and procedures for evidence auditing~\cite{corriveau2014requirements}.
Some primary studies suggested that regulatory compliance implementation should involve the interaction of different SE roles (e.g., requirements engineers, software architects, developers in~\cite{sion2019architectural}). Nevertheless, no details about such interaction were described.
None of the studies considered the implementation of regulatory compliance throughout the SDLC in a way that would cover a contribution required in each of the process areas.
We see an increasing trend towards consideration of multiple SDLC process areas in interaction. Still, the number of such studies stays relatively low. As mentioned before, some studies provided the requirements that they used without any information about the RE methods they applied~\cite{sousa2018openehr} and some reused requirements or RE methods from other studies~\cite{istvan_software-defined_2021,morisio_integration_2020}. Further research could focus on better disentangling the contribution of RE to implementing compliance in other process areas.

The number of primary studies considering two or more RE processes in conjunction stays relatively low (3\added{5} studies) and only some primary studies that considered the whole RE processes have considered RE processes in interaction. A better understanding of all the RE processes in their interconnection could be another direction for future work.

\subsection{RQ5: Which regulations and domains of application are the most considered in regulatory compliance of SIPS?} \label{RRQ5}

\begin{boxResults}
\textit{Highlights
\vspace{-2mm}
\begin{itemize}[leftmargin=*]
\item Most primary studies focused only on a few of the most considered domains of application, fields of regulation, and regulations.
\item Only some of the studies considered regulatory compliance in the context of more than one field of regulation or more than one regulation.
\item Our results suggest there are differences between different fields of regulation in terms of the challenges.
\item Some of the regulations (e.g., HIPAA) stay in the focus of researchers despite they were enacted or revised a relatively long time ago.
\item There is a significant range of regulations (including recently enacted regulations). However, only some were considered in more than 5 studies.
\end{itemize}}
\end{boxResults}

\subsubsection{Domain of application}
As mentioned in Section~\ref{sub:dataext}, we have a predefined list of 20 ``\hyperlink{DA}{domain of application}''. The initial list of domains of application was based on domains of application \added{used} in previous secondary studies (\cite{goal}, \cite{mubarkoot2021software}). However, after applying the predefined list of domains we refined it to better accommodate the results and derived the final list with 19 categories. We identified the domain of application according to the domain that was a focus of the primary study or domain for which the application of a particular principle or practice was envisioned. Here, it is important to mention that applying suggested PPs may not be limited to only the reported domain. Figure~\ref{tab:domain} shows the number of primary studies considering the application domains. Of the 280 identified primary studies, 232 addressed an application domain. Healthcare domain was considered in 71 studies (30.6\%), Software development in 40 studies (17.2\%), Avionics in 38 studies (16.4\%), Automotive domain was mentioned in 17 studies (7.3\%), Cloud computing in 16 studies (6.9\%), Finances in 14 studies (6\%), Manufacturing domain in 12 studies (5.2\%), and other application domains were mentioned in less than 10 studies (please see Table~\ref{tab:domain} for further details). 

Our results suggest that overall research is skewed towards a couple of domains of application, such as healthcare, software development and avionics, as these three domains of application constitute 53.2\% (149 out of 280). Noteworthy is that only \added{26} studies have considered the application of two or more domains of application. While some primary studies we have mapped mentioned that their results are generalizable to other domains, they have not provided further details. Only 27 primary studies have considered two or more domains of applications simultaneously. We suggest that further research on regulatory RE and SIPS compliance beyond the established domains can be of value to the community. Also, it would be beneficial to look into the applicability/replicability of the study results and/or PPs across different domains.

\begin{table}
\footnotesize
\centering
        \setlength\extrarowheight{-8pt}
	\begin{tabular}{p{2.85cm}P{2.5cm}l}\toprule  
		\textbf{Domain of application} & \textbf{Number of studies}  \\\midrule  
            Healthcare &    71 \\
            Software development & 40 \\
            Avionics & 38\\
            Automotive & 17\\
            Cloud computing & 16\\
            Finances & 14\\
            Manufacturing & 12\\
            IoT & 9\\
            Education & 8\\
            Enterprise & 7\\
            Metrology & 7\\
            Transport & 7\\
            Energy & 6\\
            Government & 6\\
            Media & 4\\
            E-commerce & 3\\
            Military & 3\\
            Telecommunications & 3\\
            Smart home/city &	2\\
            \\\bottomrule
	\end{tabular}
 \caption{Number of primary studies addressing domains of application}\label{tab:domain}
\end{table}

\subsubsection{Fields of regulation}
For \added{268}/280 (\added{95.7}\%) studies, we have identified the field of regulation (Table~\ref{tab:fields}). \added{Some primary studies mentioned a particular field of regulation, but have not mentioned a concrete regulation it was addressing.} The most considered regulation field was Privacy, with 101/\added{268} (37.\added{7}\%), whereas 68/\added{268} (25.\added{3}\%) studies focused on Quality. We have categorized 63/\added{268} studies (23.\added{4}\%) studies as considering software Safety. We introduced this category to capture studies concerned with the overall SIPS development processes and multiple properties of systems demanded by regulations. Many studies in this category are related to implementing medical device regulations (such as the EU Medical Device Regulation (EU MDR)) and metrological regulations (e.g., Measuring Instruments Directive). These studies are concerned with multiple aspects of software systems simultaneously (Quality, Security) and software development processes. Some other regulations addressed in the studies belonging to this category were ISO 62304 (software life cycle processes of medical devices), ISO 13485 (Medical devices quality management systems), and ISO 14971 (application of risk management for medical devices). Security was considered in 35/\added{268} (13\%) studies. In some cases, two fields of regulation were considered simultaneously. We also introduced a separate category for studies considering AI and ML systems compliance because as of now (2024) such regulations address multiple AI/ML-related concerns simultaneously. etc.). We have identified 21/\added{268} (7.8\%) studies belonging to this category. Privacy\&Security field of regulation was considered in 18/\added{268} primary studies (6.7\%), Business related regulations were addressed in 7/\added{268} studies (2.6\%), User rights field of regulation in 4/\added{268} primary studies (1.5\%), Accessibility in 3/\added{268} (1.1\%), and Traffic law in 2/\added{268} primary studies (0.7\%).

\begin{table}
\footnotesize
\centering
        \setlength\extrarowheight{-8pt}
	\begin{tabular}{p{3.5cm}P{2.5cm}l}\toprule  
		\textbf{Field of regulation} & \textbf{Number of studies}  \\\midrule  
            Privacy & 101 \\
            Quality & 68 \\
            Safety & 63\\
            Security & 35\\
            AI/ML & 21\\
            Privacy\&Security & 18\\
            Business & 7\\
            User rights & 4\\
            Accessibility & 3\\
            Traffic law & 2
            \\\bottomrule
	\end{tabular}
 \caption{Number of primary studies addressing fields of regulations}\label{tab:fields}
\end{table}

\begin{table*}
\footnotesize
\resizebox{\textwidth}{!}{%
\centering
	\begin{tabular}{p{4cm}P{2.55cm}P{2.5cm}P{2.5cm}P{2cm}P{2.5cm}l} \toprule  
		\thead{Regulation} & \thead{Field of \\ regulation} & \thead{Number of \\ studies} & \thead{Date of \\ enactment} & \thead{In force \\ since} & \thead{Date of \\ last changes} 
            \\\midrule
            GDPR & Privacy & 79/260 (30.4\%) & 27/04/2016 & 25/05/2018 & 23/05/2018 \\
            (1) DO-178C; (2) DO-330; (3) DO-331; (4) DO-332 & Safety & 34/260 (13\%) & (1,2,3,4)): 13/12/2011 & (1,2,3,4)): 13/12/2011 & (1):05/01/2012 (2)):16/02/2021 (3):16/02/2021 (4): - \\
            HIPAA & Privacy / security & 18/260 6.9\% & 21/08/1996 & 14/04/2003, 14/04/2005, 14/04/2006 & 26/04/2024 \\
            (1) EU MDR, (2) EU MDD & Quality & 16/260(6.1\%) & (1) 05/04/2017, (2) 14/06/1993 & (1) 25/05/2021(2024), (2) 01/07/1994 & (1) 20/03/2023 \\
            ISO 26262 & Safety & 14/260 (5.4\%) &  11/11/2011 & 11/11/2011 & 17/12/2018 \\
			ISO 62304 & Quality & 14/260 (5.4\%) & 17/05/2006 & 17/05/2006 & 01/11/2017 \\
			ISO 13485 & Quality & 9/260 (3.5\%) & 01/03/2016 & 01/03/2016 & 01/03/2016 \\
            LGPD & Privacy & 9/260 (3.5\%) & 18/09/2020 & 01/08/2021 & 25/10/2022 \\
			(1) AI HLEG Guidel.; (2) EU AI Act & AI/ML & 8/260 (3.1\%) & (1)08/04/2019; (2)21/04/2021 & (1)08/04/2019; (2)01/08/2024 & - \\
            ISO 14971 & Quality & 7/122 (2.7\%) & 01/10/1998 & 01/10/1998 & 10/12/2019 \\
            IEC 62443 & Security & 7/260 (2.6\%) & 2002 & - & 2023 \\
			(1)ISO 27000, (2) ISO 27001; (3) ISO 27002; (4) ISO 27005 & Security & 4/260 (1.5\%) & (2):12/10/2005; (3):15/06/2005; (4):04/06/2008 & - & (2):25/10/2022; (3):12/02/2022; (4):25/10/2022 \\
	        IEC 61508 & Safety & 6/260 (2.3\%) & 2000 & 2010 &	\\
            IEC 50128 & Safety & 5/260 (1.9\%) & 11/2001 & 11/2001 & 03/2012 \\
            EU MID & Quality & 5/260 (2.5\%) & 31/03/2004 & 30/10/2006 & 27/01/2015 \\
			\\\bottomrule
	\end{tabular}}
 \caption{Number of primary studies considering regulations and additional information about each regulation}\label{tab:regulations}
\end{table*}

\begin{figure*} 
\scriptsize
    \centering
    \begin{tikzpicture}
    \begin{axis}[
    width=0.92\columnwidth,
    height=8cm,
    symbolic x coords = {Regulatory, Organizational, SoftwareSystems, RegulationsMultiplicity, SoftwareContext, Enforcement, Provability, ResourceIntensity, Experts, Complexity, AbsenceOfPrinciples, DomainKnowledge, Conflicts, Abstractness},
    symbolic y coords = {TrafficLaw, PrivacySecurity, AIML, Security, Safety, Quality, Privacy},
    xticklabels={Regulatory gaps, Organizational challenges, Software dynamics, Multiple regulations, Context dynamics, Enforcement challenges, Provability, Resource intensity, Interaction with experts, Complexity, Absence of PPs, Domain knowledge, Changes to practices, Abstractness},
    yticklabels = {Traffic Law, Privacy Security, AI/ML, Security, Safety, Quality, Privacy},
    x tick label style = {font = \scriptsize, text width = 4cm, align = right, rotate = 70, anchor = north east},
    y tick label style = {font = \scriptsize, text width = 2cm, align = right},
    xtick=data,
    ytick=data,
    xmajorgrids=true,
    ymajorgrids=true,
    grid style=dashed,
    axis x line*=bottom,
    axis y line*=left,
    ]
    \addplot[%
        scatter=true,
        only marks,
        point meta=\thisrow{color},
        fill opacity=0.15,text opacity=1,
        visualization depends on = {1*\thisrow{Val} \as \perpointmarksize},
        visualization depends on = {\thisrow{Val} \as \Val},
        scatter/@pre marker code/.append style={
        /tikz/mark size=\perpointmarksize
        },
        nodes near coords*={\contour{white}{$\pgfmathprintnumber\Val$}},
        nodes near coords style={text=black,font=\sffamily, font=\bfseries, font=\normalsize, anchor=center},
    ] table [x={Group},y={Posttest}] {
Group  Posttest Val color

Regulatory	TrafficLaw	1	0
Organizational	TrafficLaw	2	0
SoftwareSystems	TrafficLaw	1	0
RegulationsMultiplicity	TrafficLaw	2	0
SoftwareContext	TrafficLaw	1	0
Enforcement	TrafficLaw	2	0
Provability	TrafficLaw	0	0
ResourceIntensity	TrafficLaw	0	0
Experts	TrafficLaw	1	0
Complexity	TrafficLaw	2	0
AbsenceOfPrinciples	TrafficLaw	0	0
DomainKnowledge	TrafficLaw	1	0
Conflicts	TrafficLaw	0	0
Abstractness	TrafficLaw	1	0

Regulatory	PrivacySecurity	0	0
Organizational	PrivacySecurity	1	0
SoftwareSystems	PrivacySecurity	2	0
RegulationsMultiplicity	PrivacySecurity	2	0
SoftwareContext	PrivacySecurity	2	0
Enforcement	PrivacySecurity	2	0
Provability	PrivacySecurity	1	0
ResourceIntensity	PrivacySecurity	5	0
Experts	PrivacySecurity	3	0
Complexity	PrivacySecurity	5	0
AbsenceOfPrinciples	PrivacySecurity	3	0
DomainKnowledge	PrivacySecurity	7	0
Conflicts	PrivacySecurity	2	0
Abstractness	PrivacySecurity	10	0

Regulatory	AIML	12	0
Organizational	AIML	4	0
SoftwareSystems	AIML	7	0
RegulationsMultiplicity	AIML	9	0
SoftwareContext	AIML	5	0
Enforcement	AIML	8	0
Provability	AIML	7	0
ResourceIntensity	AIML	3	0
Experts	AIML	10	0
Complexity	AIML	8	0
AbsenceOfPrinciples	AIML	5	0
DomainKnowledge	AIML	6	0
Conflicts	AIML	5	0
Abstractness	AIML	8	0

Regulatory	Security	2	0
Organizational	Security	3	0
SoftwareSystems	Security	4	0
RegulationsMultiplicity	Security	5	0
SoftwareContext	Security	3	0
Enforcement	Security	5	0
Provability	Security	6	0
ResourceIntensity	Security	11	0
Experts	Security	9	0
Complexity	Security	8	0
AbsenceOfPrinciples	Security	7	0
DomainKnowledge	Security	6	0
Conflicts	Security	7	0
Abstractness	Security	14	0

Regulatory	Safety	4	0
Organizational	Safety	4	0
SoftwareSystems	Safety	6	0
RegulationsMultiplicity	Safety	2	0
SoftwareContext	Safety	5	0
Enforcement	Safety	7	0
Provability	Safety	15	0
ResourceIntensity	Safety	12	0
Experts	Safety	3	0
Complexity	Safety	14	0
AbsenceOfPrinciples	Safety	14	0
DomainKnowledge	Safety	5	0
Conflicts	Safety	23	0
Abstractness	Safety	15	0

Regulatory	Quality	17	0
Organizational	Quality	6	0
SoftwareSystems	Quality	12	0
RegulationsMultiplicity	Quality	12	0
SoftwareContext	Quality	1	0
Enforcement	Quality	8	0
Provability	Quality	18	0
ResourceIntensity	Quality	7	0
Experts	Quality	13	0
Complexity	Quality	15	0
AbsenceOfPrinciples	Quality	14	0
DomainKnowledge	Quality	8	0
Conflicts	Quality	26	0
Abstractness	Quality	18	0

Regulatory	Privacy	4	0
Organizational	Privacy	17	0
SoftwareSystems	Privacy	7	0
RegulationsMultiplicity	Privacy	10	0
SoftwareContext	Privacy	15	0
Enforcement	Privacy	27	0
Provability	Privacy	17	0
ResourceIntensity	Privacy	21	0
Experts	Privacy	29	0
Complexity	Privacy	27	0
AbsenceOfPrinciples	Privacy	33	0
DomainKnowledge	Privacy	45	0
Conflicts	Privacy	30	0
Abstractness	Privacy	42	0

};
    \end{axis}
    \end{tikzpicture}
    \vspace{-15mm}\caption{Challenges to regulatory RE and compliance of SIPS and mapped against fields of regulation. Note: Fields of regulation Accessibility, User Rights, and Business have mentioned only a few challenges and thus were excluded from this mapping (See text for the details).}
    \label{fig:fieldsChall}
\end{figure*}
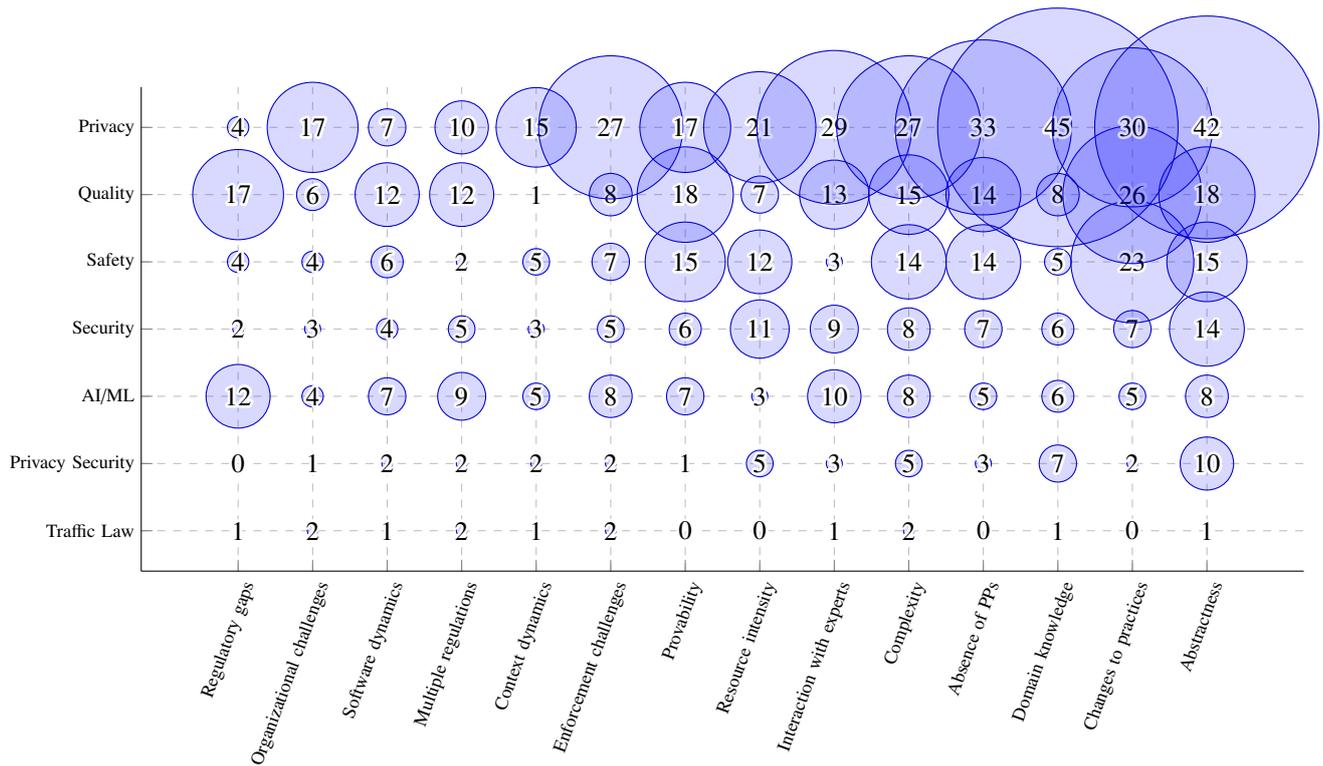

\subsubsection{Regulations}
260 out of 280 studies have mentioned at least one regulation in the context of their research. We summarize the regulations considered in the studies in Table~\ref{tab:regulations} together with additional information important for the discussion. In some cases primary studies described a wide range of different regulations from the same regulator, such as US National Institute of Standards and Technology (NIST) standards (mentioned in \added{11} studies) and US Food and Drug Administration (FDA) regulations (mentioned in \added{seven} studies). \added{Some of the regulations mentioned in less than five primary studies were as follows} Virginia Consumer Data Protection Act (US CDPA)~\cite{kempe2022documenting, sahu2023web}, Children's Online Privacy Protection Rule (US COPPA)~\cite{ekambaranathan2023navigating, breaux2022legal}, EU Cybersecurity Act~\cite{milankovich2023delta, khurshid2022eu}, the EU In Vitro Diagnostics Regulation (EU IVDR)~\cite{toivakka2021towards, muller2022explainability}, \added{US Health Information Technology for Economic and Clinical Health Act (US HITECH)}~\cite{li2020chainsdi, strielkina2018cybersecurity} US executive order 13960~\cite{ottun2023one}, US executive order 14028~\cite{khurshid2022eu}, EU Markets in Financial Instruments Directive 2014/65/EU (EU MiFID II)~\cite{culley2023insights}, US Sarbanes–Oxley Act (US SOX)~\cite{joshi_integrated_2020}. Some other rarely considered regulations were Chinese Cybersecurity Law~\cite{antunes_operationalization_2020}, US Federal Risk and Authorization Management Program (FedRAMP)~\cite{kellogg2020continuous}, US Federal Information Security Management Act of 2002 (US FISMA)~\cite{joshi_integrated_2020}. \added{While the majority of the studies mentioned the demand to be compliant with two or more regulations simultaneously, only some explicitly addressed this demand with corresponding PPs}. For example, Joshi et al.~\cite{joshi_integrated_2020} have considered multiple regulations (such as EU GDPR, US HIPAA, US FISMA, ISO 27001, ISO 27002 along with other non-regulatory types of sources of requirements like OAUTH, COBIT), but only within the RE process area. Strielkina et al.\cite{strielkina2018cybersecurity} suggested a hierarchical cybersecurity model for healthcare IoT systems encompassing relationships between different regulations and tool support for the corresponding decision-making. Anish et al.\cite{anish2021automated} used multiple regulations to extract penalty clauses and train their ML model for automated identification and deconstruction of penalty clauses.

\subsubsection{Discussion \nameref{RRQ5}}
There is an evident skew towards certain domains of application, fields of regulation, and regulations. The three most considered domains of application, Healthcare (71 primary studies), and Software development (40 primary studies), and Avionics (38 studies) make up \added{53.2}\% of all (280) primary studies. Fields of regulation of Privacy (101 primary studies) and Quality (68 primary studies) cover \added{60.4}\%  of all \added{primary studies}. The two most considered regulations, EU GDPR (79 primary studies) and DO-178C/330/331/332 avionics standards (34 primary studies), were considered in \added{40.4}\% primary studies. 

We have mapped fields of regulations and challenges (each study can contain more than one challenge) in Figure~\ref{fig:fieldsChall}. We excluded three fields of regulation (Accessibility, User rights, and Business) from the visualization as they reported only a few challenges. Studies focusing on Accessibility mentioned the following challenges, enforcement challenges (two primary studies) and conflicts with existing principles and practices. Primary studies focusing on the User rights field of regulations mentioned the following challenges enforcement challenges (one primary study), Regulatory gaps (one primary study), Complexity-related challenges (one primary study), and absence of principles and practices (one primary study). Studies considering the business field of regulation considered the challenges of multiplicity of regulations (one primary study), resource intensity (two primary studies), absence of PPs (one primary study), and abstractness of regulations (one primary study).

Some challenges are more considered in some fields of regulation. For example, challenges of \added{domain knowledge, abstractness, absence of PPs, and changes to software engineering practices are most prominent in the field of privacy. Simultaneously, in the field of safety, challenges related to changes to SE practices, abstractness, provability, complexity and absence of PPs are most considered. In the field of regulation of quality the most considered challenges were changes to SE practices, abstractness, provability, and regulatory gaps. AI/ML stands out in comparison to other fields of regulation with regulatory gaps, interaction with experts, multiplicity of regulations, and enforcement as the top-mentioned challenges.} This can result from differences in the maturity of research in different fields of regulation or related to basic differences between different fields of regulation. Unexpectedly, we found that the abstractness of regulations was mentioned as a challenge in such fields as safety and quality, in which regulations usually take place through technical standards.
Overall, for many fields of regulation, one of the 14 categories of challenges is mentioned at least \added{twice}.

It is noteworthy that many of the regulations considered in studies have been in force for a long time. The US Health Insurance Portability and Accountability Act (US HIPAA) was enacted in 1996 and last changed in 2013. Nevertheless, it stays among the most considered regulations even now. Many of the principles of the EU GDPR were based on the European Data Protection Directive~\cite{blanco2019using} enacted back in 1995. Still, GDPR, enacted based on the European Data Protection Directive, is the most considered regulation.
Only some studies considered compliance to a few regulations from different fields of regulation simultaneously (for example, compliance with privacy and safety regulations). This finding is in tune with the findings of the secondary study by Ardila et al.~\cite{castellanos2022compliance}, which also found that in almost all studies, regulations are addressed in isolation, reducing the possibility of generalizing the results.
Further research in particular fields of regulation is especially important, considering that multiple regulations are focused on the intersection between different fields of regulation (e.g., privacy and security). The understanding of privacy and security from regulatory and SE perspectives can differ, which demands addressing the gap between the two perspectives. Future research could focus on in-depth research of challenges, principles, and practices in the context of different regulation fields or on the intersection between regulation fields.

\section{Threats to Validity}\label{sec:threats}
We took inspiration from Montgomery et al. \cite{10.1007/s00766-021-00367-z} in structuring the threats to validity section of our SMS.


\subsection{Internal Validity}

We placed significant emphasis on ensuring the reliability of our research outcomes. To achieve this, we closely adhered to the research plan guidelines proposed by Kitchenham et al.~\cite{Kitchenham07guidelines} and Petersen et al.~\cite{Petersen20151}. Each step of our study was meticulously documented and followed according to the structured process outlined in Section~\ref{RM}. To enhance research reliability further, we conducted inter-rater reliability tests, including measuring Cohen's kappa, among the researchers involved in study selection, quality assessment, and data extraction. These measures aimed to maintain consistency in decision-making and reduce the potential for subjective biases. Throughout the study selection phase, the two first authors actively discussed any uncertainties regarding inclusion. In the event of disagreements, the default decision was to include the study, thus mitigating the risk of false negatives. The structure of this SMS adheres to the guidelines outlined in established literature for conducting an SMS within SE research. Our approach meticulously aligns with these recommendations. 

\subsection{External Validity}
The validity of our study may be influenced by the extent of coverage within the search results and potential biases in the selection of studies. It is plausible that certain pertinent studies may not have been included due to incomplete search strings or the omission of certain databases. To attain comprehensive evidence, our literature review search terms were deliberately broad. Our research was meticulously designed to unearth challenges related to the regulatory compliance of software systems without confining itself to specific regulations. It is plausible that our search parameters did not encompass studies narrowly focused on executing a singular regulation, omitting keywords like ``law'' or ``regulation''. Alternatively, studies might employ other terms, such as ``privacy'' and ``personal data protection'', or even reference specific regulations like GDPR or HIPAA instead of the more general keywords we employed. 
Researcher biases could potentially influence the assessment of studies. To counteract this possibility, a predefined set of inclusion and exclusion criteria was employed to guide the assessment process. We also incorporated studies that did not explicitly propose PP but still undertook relevant research, as these sources often contained valuable insights into prevailing challenges.
As we have considered the application of different keywords and wildcards across different databases, we conducted a trial selection from a subset of the search results to evaluate the relevance of the chosen keywords for our study.
Given the focus of this study -- regulatory RE as a process area contributing to regulatory compliance across the SIPS life cycle -- we assert that a keyword-based search offers the most systematic approach for selecting relevant studies. However, relying solely on keyword searches can introduce potential validity threats. To mitigate these risks, we reviewed the search results to identify key studies familiar to the authors, ensuring their inclusion. Additionally, we compared the outcomes of different trial searches to verify whether they encompassed these known studies and whether any new, significant primary studies emerged. The authors also discussed the breadth of coverage offered by the various trial searches. Finally, we evaluated and considered the methods and findings of related previous secondary studies.

\subsection{Construct Validity}
The principal challenge to construct validity lies within the phase of data extraction. Before commencing the comprehensive extraction process, the authors reached an agreement, and each researcher engaged in this task performed a preliminary extraction run. This preliminary run aimed to harmonize their comprehension of the extraction inquiries. Consequently, adjustments were made to the names, explanations of extraction questions, coding methodologies, and extraction procedures to synchronize with the researchers' interpretations.
Furthermore, low-confidence instances were identified and discussed between the first two authors until a consensus was reached.

\section{Conclusions and Further Research Directions}\label{sec:conclusions}
Implementation of regulatory compliance of SIPS remains an open and multifaceted goal for SE research and practice. This study contributes to disentangling regulatory compliance of SIPS by providing an overview of state of the art and identifying the most significant aspects of regulatory compliance of SIPS and regulatory RE.

Difficulties in implementing regulatory compliance of SIPS can be related to the 14 categories of challenges we have identified in the primary studies. Taking into account the number and interconnectedness of such challenges, further empirical research on particular challenges is required to create a sufficient understanding of the principles and practices that are required to address them.
This is essential as studies often do not clearly connect the challenges they identified and the principles and practices they suggested and/or evaluated in relation to such challenges. A better understanding of challenges could also contribute to developing tools and automation. For now, tools as a suggested principle and practice are significantly outnumbered by methodology as the most widespread type of principles and practice. Further research on principles and practices is also essential, as our results suggest that most existing principles and practices were not validated with sufficient rigor and, in many cases, had relatively low relevance for SE practice.

The involvement of categories of stakeholders other than software engineers in the development, application, and validation of PPs is an important finding for further SE research and practice. Our findings suggest that there may be different needs in terms of the involvement of stakeholders in different fields of regulation and in addressing different categories of challenges. Herewith, interdisciplinary research can contribute to a better analysis of such different needs.

Our results also show that regulatory compliance with SIPS may require a more systematic consideration throughout the SDLC. In particular, primary studies considered not only RE but also software design, quality assurance, etc. Many studies considered RE in conjunction with other process areas (e.g., conjunction of RE and SQA) which suggests that implementation of effective relationships between requirements engineering and other process areas is essential for the achievement of SIPS regulatory compliance.

Current research focuses mainly on the 2-3 most popular domains of application (healthcare, software development, avionics), fields of regulation (privacy, quality), and regulations (GDPR, DO-178/330 standards, HIPAA). This does not allow us to better explore and account for the difference between different regulations and domains of application. Our results suggest differences in challenges to compliance with SIPS among different regulation fields. Moreover, research on implementing requirements from some regulations stays popular even after such regulations were enacted or changed.

Future research on regulatory compliance of SIPS will need to take into account many of the aspects that we have identified, which can be summarized as follows:
\begin{itemize}
    \item the multiple categories of challenges to regulatory compliance of SIPS, which are interconnected;
    \item the interdependence of process areas in the implementation of compliance;
    \item role of requirements engineering in the implementation of regulatory compliance in subsequent process areas;
    \item the need for appropriate involvement of legal experts and, where necessary, domain experts;
    \item potential specificity of challenges across different fields of regulations (e.g., privacy, safety);
    \item the interdisciplinary nature of the topic of regulatory compliance of SIPS.
\end{itemize}

We suggest that regulatory requirements engineering practice for regulatory compliance of SIPS should better account for challenges, focus on more rigorous validation of principles and practices and involvement of the relevant groups of stakeholders of such principles and practices. Furthermore, differences in terms of SDLC process areas, regulations, and domains of application need to be considered.

Overall, requirements engineering for regulatory compliance of SIPS requires a more profound approach to more systematically ensure that there is a clear understanding of the challenges that need to be addressed and that developed principles and practices are relevant for the achievement of the final goal of ensuring regulatory compliance of software-intensive products and services.

\section{Appendix I}
Tables of contributions generated with \href{https://try-ctab.github.io/}{CTAB}\footnote{CTAB V0.1 can be accessed at: \url{https://try-ctab.github.io/}}.

\begin{table*}
    \begin{minipage}{.35\textwidth}
      \centering
      \scriptsize
\begin{tabular}{ r | m{0.145cm} | m{0.145cm} | m{0.145cm} | m{0.145cm} | m{0.145cm} | m{0.145cm} | m{0.145cm} | m{0.145cm} | m{0.145cm} | m{0.145cm} | m{0.145cm} | m{0.145cm} | m{0.145cm} |} 
\multicolumn{1}{c}{\textcolor{version}{\textcolor{background}{|}}} & \multicolumn{1}{c}{\rot{Conceptualization\textcolor{background}{|}}} & \multicolumn{1}{c}{\rot{Study design\textcolor{background}{|}}} & \multicolumn{1}{c}{\rot{Data collection\textcolor{background}{|}}} & \multicolumn{1}{c}{\rot{Data analysis*\textcolor{background}{|}}} & \multicolumn{1}{c}{\rot{Original draft*\textcolor{background}{|}}} & \multicolumn{1}{c}{\rot{Reviewing\&Editing\textcolor{background}{|}}} & \multicolumn{1}{c}{\rot{Supervision\textcolor{background}{|}}} \\ 
\hhline{~|-|-|-|-|-|-|-|} 
\textcolor{background}{|}O. Kosenkov*\textcolor{background}{|} & \cellcolor{C3}\textcolor{C3}{**|} & \cellcolor{C3}\textcolor{C3}{**|} & \cellcolor{C3}\textcolor{C3}{**|} & \cellcolor{C3}\textcolor{C3}{**|} & \cellcolor{C3}\textcolor{C3}{**|} & \cellcolor{C3}\textcolor{C3}{**|} & \cellcolor{C1}\textcolor{C1}{|} \\ 
\hhline{~|-|-|-|-|-|-|-|} 
\textcolor{background}{|}P. Elahidoost*\textcolor{background}{|} & \cellcolor{C2}\textcolor{C2}{*|} & \cellcolor{C3}\textcolor{C3}{**|} & \cellcolor{C3}\textcolor{C3}{**|} & \cellcolor{C3}\textcolor{C3}{**|} & \cellcolor{C3}\textcolor{C3}{**|} & \cellcolor{C3}\textcolor{C3}{**|} & \cellcolor{C1}\textcolor{C1}{|} \\  
\hhline{~|-|-|-|-|-|-|-|} 
\textcolor{background}{|}T. Gorschek\textcolor{background}{|} & \cellcolor{C2}\textcolor{C2}{*|} & \cellcolor{C2}\textcolor{C2}{*|} & \cellcolor{C1}\textcolor{C1}{|} & \cellcolor{C1}\textcolor{C1}{|} & \cellcolor{C1}\textcolor{C1}{|} & \cellcolor{C2}\textcolor{C2}{*|} & \cellcolor{C2}\textcolor{C2}{*|} \\ 
\hhline{~|-|-|-|-|-|-|-|} 
\textcolor{background}{|}J. Fischbach\textcolor{background}{|} & \cellcolor{C1}\textcolor{C1}{|} & \cellcolor{C1}\textcolor{C1}{|} & \cellcolor{C1}\textcolor{C1}{|} & \cellcolor{C1}\textcolor{C1}{|} & \cellcolor{C1}\textcolor{C1}{|} & \cellcolor{C3}\textcolor{C3}{*|} & \cellcolor{C2}\textcolor{C2}{*|} \\ 
\hhline{~|-|-|-|-|-|-|-|} 
\textcolor{background}{|}D. Mendez\textcolor{background}{|} & \cellcolor{C1}\textcolor{C1}{|} & \cellcolor{C1}\textcolor{C1}{|} & \cellcolor{C1}\textcolor{C1}{|} & \cellcolor{C1}\textcolor{C1}{|} & \cellcolor{C1}\textcolor{C1}{|} & \cellcolor{C2}\textcolor{C2}{*|} & \cellcolor{C3}\textcolor{C3}{**|} \\ 
\hhline{~|-|-|-|-|-|-|-|} 
\textcolor{background}{|}M. Unterkalmsteiner\textcolor{background}{|} & \cellcolor{C1}\textcolor{C1}{|} & \cellcolor{C1}\textcolor{C1}{|} & \cellcolor{C1}\textcolor{C1}{|} & \cellcolor{C1}\textcolor{C1}{|} & \cellcolor{C1}\textcolor{C1}{|} & \cellcolor{C3}\textcolor{C3}{**|} & \cellcolor{C3}\textcolor{C3}{**|} \\
\hhline{~|-|-|-|-|-|-|-|}
\textcolor{background}{|}D. Fucci\textcolor{background}{|} & \cellcolor{C1}\textcolor{C1}{|} & \cellcolor{C1}\textcolor{C1}{|} & \cellcolor{C1}\textcolor{C1}{|} & \cellcolor{C1}\textcolor{C1}{|} & \cellcolor{C1}\textcolor{C1}{|} & \cellcolor{C3}\textcolor{C3}{**|} & \cellcolor{C3}\textcolor{C3}{**|} \\
\hhline{~|-|-|-|-|-|-|-|} 
\textcolor{background}{|}R. Mohanani\textcolor{background}{|} & \cellcolor{C1}\textcolor{C1}{|} & \cellcolor{C2}\textcolor{C2}{*|} & \cellcolor{C1}\textcolor{C1}{|} & \cellcolor{C1}\textcolor{C1}{|} & \cellcolor{C1}\textcolor{C1}{|} & \cellcolor{C2}\textcolor{C2}{*|} & \cellcolor{C1}\textcolor{C1}{|} \\
\hhline{~|-|-|-|-|-|-|-|}
\end{tabular}
      \end{minipage}
      \quad\quad\quad\quad
    \begin{minipage}{.55\textwidth}
      \centering
      \scriptsize
\begin{tabular}{ r | m{0.145cm} | m{0.145cm} | m{0.145cm} | m{0.145cm} | m{0.145cm} | m{0.145cm} | m{0.145cm} | m{0.145cm} | m{0.145cm} | m{0.145cm} | m{0.145cm} | m{0.145cm} | m{0.145cm} |} 
\multicolumn{1}{c}{\textcolor{version}{\textcolor{background}{|}}} & \multicolumn{1}{c}{\rot{Introduction\textcolor{background}{|}}} & \multicolumn{1}{c}{\rot{Fundamentals\textcolor{background}{|}}} & \multicolumn{1}{c}{\rot{Related work\textcolor{background}{|}}} & \multicolumn{1}{c}{\rot{Study design\textcolor{background}{|}}} & \multicolumn{1}{c}{\rot{Results: General Trends\textcolor{background}{|}}} & \multicolumn{1}{c}{\rot{Results: RQ 1\textcolor{background}{|}}} & \multicolumn{1}{c}{\rot{Results: RQ 2\textcolor{background}{|}}} & \multicolumn{1}{c}{\rot{Results: RQ 3\& RQ4\textcolor{background}{|}}} &  \multicolumn{1}{c}{\rot{Results: RQ 5\textcolor{background}{|}}} & \multicolumn{1}{c}{\rot{Threats to validity\textcolor{background}{|}}} & \multicolumn{1}{c}{\rot{Conclusions\textcolor{background}{|}}} \\ 
\hhline{~|-|-|-|-|-|-|-|-|-|-|-|} 
\textcolor{background}{|}O. Kosenkov\textcolor{background}{|} & \cellcolor{C3}\textcolor{C3}{**|} & \cellcolor{C2}\textcolor{C2}{*|} & \cellcolor{C2}\textcolor{C2}{*|} & \cellcolor{C2}\textcolor{C1}{} & 
\cellcolor{C1}\textcolor{C1}{|} & 
\cellcolor{C3}\textcolor{C3}{**|} & \cellcolor{C1}\textcolor{C1}{|} & \cellcolor{C3}\textcolor{C3}{**|} & \cellcolor{C2}\textcolor{C2}{*} & \cellcolor{C1}\textcolor{C1}{|} & \cellcolor{C3}\textcolor{C3}{**|} \\ 
\hhline{~|-|-|-|-|-|-|-|-|-|-|-|} 
\textcolor{background}{|}P. Elahidoost\textcolor{background}{|} & \cellcolor{C1}\textcolor{C1}{|} & \cellcolor{C2}\textcolor{C2}{*|} & \cellcolor{C2}\textcolor{C2}{*|} & \cellcolor{C2}\textcolor{C3}{} & 
\cellcolor{C3}\textcolor{C3}{**|} & 
\cellcolor{C1}\textcolor{C1}{|} & \cellcolor{C3}\textcolor{C3}{**|} & \cellcolor{C1}\textcolor{C1}{|} & \cellcolor{C2}\textcolor{C2}{*|} & \cellcolor{C3}\textcolor{C3}{**|} & 
\cellcolor{C2}\textcolor{C2}{*} \\ 
\hhline{~|-|-|-|-|-|-|-|-|-|-|-|} 
\end{tabular}
      \end{minipage}
\caption{Contributions}\label{tab:contributions}
\end{table*}

\bibliographystyle{elsarticle-num-names} 
\bibliography{cas-refs}
\end{document}